\documentclass[useAMS,usenatbib]{mnras}
\usepackage{verbatim} 
\usepackage{natbib} 
\usepackage{amsmath} 
\usepackage{amsbsy}
\usepackage{amssymb}
\usepackage{mathrsfs} 
\usepackage{lscape} 
\usepackage{graphicx}
\usepackage{epstopdf}
\usepackage{deluxetable}
\usepackage{ctable} 
\usepackage{fixltx2e} 
\usepackage{url}

\usepackage{txfonts}
\usepackage[T1]{fontenc} 
\usepackage{aecompl}

\newcommand{\acknowledgments}{\begin{small}\section*{Acknowledgments}\end{small}}

\newcommand{\lcdm}{$\Lambda$CDM}
\newcommand{\Msun}{M$_{\odot}$}
\newcommand{\Msunyr}{M$_{\odot}$yr$^{-1}$}

\newcommand{\pathI}{./fig}

\title[Baryon Cycling and Galaxy Assembly on FIRE]{The Cosmic Baryon Cycle and Galaxy Mass Assembly in the FIRE Simulations}

\author[D. Angl{\'e}s-Alc{\'a}zar, et al.]{
\parbox[t]{\textwidth}{\vspace{-1cm}
Daniel Angl{\'e}s-Alc{\'a}zar$^{1}$\thanks{E-mail: anglesd@northwestern.edu},
Claude-Andr{\'e} Faucher-Gigu{\`e}re$^{1}$, 
Du\v{s}an Kere\v{s}$^{2}$,
Philip F. Hopkins$^3$,
Eliot Quataert$^4$,
Norman Murray$^{5,6}$}\vspace{0.1cm} \\
$^1$Center for Interdisciplinary Exploration and Research in Astrophysics (CIERA) and Department of Physics and Astronomy, \\ Northwestern University, 2145 Sheridan Road, Evanston, IL 60208, USA.\\
$^2$Department of Physics, Center for Astrophysics and Space Sciences, University of California, San Diego, \\ 9500 Gilman Drive, La Jolla, CA 92093, USA.\\
$^3$TAPIR, Mailcode 350-17, California Institute of Technology, Pasadena, CA 91125, USA.\\
$^4$Department of Astronomy and Theoretical Astrophysics Center, University of California Berkeley, Berkeley, CA 94720, USA.\\
$^5$Canadian Institute for Theoretical Astrophysics, 
60 St.\ George Street, University of Toronto, ON M5S 3H8, Canada.\\
$^6$Canada Research Chair in Astrophysics.}

\date{Submitted to MNRAS, October, 2016\vspace{-0.0cm}}
\pubyear{2016}

\begin{document}
\maketitle

\begin{abstract}
 
We use cosmological simulations from the FIRE (Feedback In Realistic Environments) project to study the baryon cycle and galaxy mass assembly for central galaxies in the halo mass range $M_{\rm halo} \sim 10^{10}$--$10^{13}$\,\Msun. By tracing cosmic inflows, galactic outflows, gas recycling, and merger histories, we quantify the contribution of physically distinct sources of material to galaxy growth. We show that in situ star formation fueled by fresh accretion dominates the early growth of galaxies of all masses, while the re-accretion of gas previously ejected in galactic winds often dominates the gas supply for a large portion of every galaxy's evolution. Externally processed material contributes increasingly to the growth of central galaxies at lower redshifts. This includes stars formed ex situ and gas delivered by mergers, as well as smooth \emph{intergalactic transfer} of gas from other galaxies, an important but previously under-appreciated growth mode. By $z=0$, {\it wind transfer}, i.e. the exchange of gas between galaxies via winds, can dominate 
gas accretion onto $\sim L^{*}$ galaxies over fresh accretion and standard wind recycling. Galaxies of all masses re-accrete $\gtrsim 50$\,\% of the gas ejected in winds and recurrent recycling is common. The total mass deposited in the intergalactic medium per unit stellar mass formed increases in lower mass galaxies. Re-accretion of wind ejecta occurs over a broad range of timescales, with median recycling times ($\sim 100-350$ Myr) shorter than previously found. Wind recycling typically occurs at the scale radius of the halo, independent of halo mass and redshift, suggesting a characteristic {\it recycling zone} around galaxies that scales with the size of the inner halo and the galaxy's stellar component.  
 
\end{abstract}


\section{Introduction}

The exchange of mass, energy, and metals between galaxies, their surrounding circumgalactic medium (CGM), and the intergalactic medium (IGM) represents an integral part of the modern paradigm of galaxy formation \citep{Dave2012,Lilly2013}.
Smooth accretion of fresh gas from the IGM, albeit difficult to detect, is required to sustain the observed star formation rate (SFR) in galaxies across cosmic time \citep{Erb2008,Tacconi2010,Putman2012,SanchezAlmeida2014}.  Stellar feedback processes power ubiquitous large scale outflows observed from local starbursts to high redshift galaxies \citep{Martin2005,Rupke2005,Veilleux2005,Weiner2009,Steidel2010,Martin2012,Newman2012,Heckman2015}.
Outflowing gas previously enriched in the interstellar medium (ISM) of galaxies must deposit heavy elements at large distances to explain the observed abundance of metals in the CGM \citep{Tumlinson2011,Werk2014}.  Nonetheless, observed wind velocities are such that a substantial portion of ejected material may recycle back onto galaxies \citep[e.g.][]{Rubin2012,Emonts2015,Pereira-Santaella2016}, providing a physically distinct source of gas infall \citep[``wind recycling";][]{Oppenheimer2010} that may fuel star formation at later times.  Much recent work has thus focused on improving our understanding of the main processes that constitute the {\it baryon cycle} in galaxy evolution.

Cosmological hydrodynamic simulations offer an ideal tool to model the cycling of baryons between galaxies and their surrounding gas, where intergalactic accretion feeds galaxies from the cosmic web \citep{Keres2005,Brooks2009,Dekel2009,Keres2009_ColdHot,Faucher-Giguere2011_BaryonicAssembly,vandeVoort2011}, powerful winds evacuate gas from galaxies \citep{Oppenheimer2008,Angles-Alcazar2014,Muratov2015,Muratov2016}, and outflowing gas often re-accretes back onto galaxies \citep{Oppenheimer2010,Christensen2016}. Notably, the balance between gas inflows, outflows, and recycling represents the core of galaxy equilibrium models, where simple analytic equations constrained by a limited number of free parameters are able to reproduce various global galaxy scaling relations \citep{Finlator2008,Bouche2010,Dave2012,Lilly2013,Mitra2015}. Semi-analytic models (SAMs ) built on dark matter halo merger trees also rely on baryon cycle processes \citep[e.g.][]{Somerville2008}, recently incorporating more detailed treatments of galactic winds and recycling \citep{Henriques2013,White2015}.

Galactic winds are indeed a key ingredient in all current successful galaxy formation models \citep{SomervilleDave2015}, required to regulate star formation in galaxies \citep{Angles-Alcazar2014,Hopkins2014_FIRE,Agertz2015_SF,Muratov2015}, produce disk galaxies with more realistic central baryonic distributions \citep{Governato2007,Angles-Alcazar2014,Christensen2014_bulges,Agertz2015_structure,Ma2016_StellarDisk}, and reproduce key observables including the galaxy stellar mass function \citep{Oppenheimer2010,Dave2011_MstarSFR,Vogelsberger2014}, the mass--metallicity relation \citep{Dave2011_GasMet,Torrey2014,Ma2016_Metallicity}, and the metal enrichment of the IGM \citep{Oppenheimer2006,Oppenheimer2008,Ford2013,Ford2014}.

Modeling realistic large scale winds has been difficult to achieve in cosmological hydrodynamic simulations, owing to relevant physical processes occurring at sub-resolution scales.  Successful ``sub-grid" models developed to circumvent this problem include parameterized kinetic winds temporarily decoupled from the dense ISM medium \citep{Springel2003_Multiphase,Oppenheimer2006,Oppenheimer2008,Angles-Alcazar2014}, the injection of thermal energy locally while gas cooling is temporarily disabled to avoid radiative losses \citep{Stinson2006,Christensen2016}, and the accumulation of feedback energy until individual thermal injections can be efficiently converted into kinetic energy \citep{DallaVecchia2012,Schaye2015}.
Recent improvements to the treatment of stellar feedback in high resolution galaxy-scale and cosmological ``zoom-in" simulations have allowed the generation of large scale winds without relying on hydrodynamic decoupling or delayed cooling techniques \citep{Hopkins2012_WindGeneration,Hopkins2014_FIRE,Agertz2015_SF,Agertz2015_structure,Keller2015,Muratov2015}.

In the Feedback In Realistic Environments (FIRE) cosmological zoom-in simulations\footnote{See the FIRE project website at:\url{http://fire.northwestern.edu}.} \citep{Hopkins2014_FIRE}, 
stellar feedback is modeled by injecting mass, energy, momentum, and metals on the scale of star-forming regions directly following stellar population synthesis models.
This combination of feedback processes together with enough resolution to begin to model the multi-phase structure of the ISM in galaxies generates powerful outflows self-consistently across a range of galaxy masses \citep{Muratov2015,Muratov2016}, while simultaneously reproducing a variety of observational constraints including the Kennicutt--Schmidt relation \citep{Hopkins2014_FIRE}, the stellar mass--halo mass relation \citep{Hopkins2014_FIRE,Feldmann2016}, the mass--metallicity relation \citep{Ma2016_Metallicity}, the HI covering fractions in Lyman-break galaxies and quasar-host halos \citep{Faucher-Giguere2015,Faucher-Giguere2016}, and the incidence and HI column density distribution of low-redshift Lyman limit systems \citep{Hafen2016}.

In this work, we use the FIRE simulations to link the local stellar feedback processes that shape the multi-phase structure of the ISM in galaxies with the large scale cycling of baryons.  Recently, \citet{Muratov2015,Muratov2016} presented a comprehensive analysis of the gusty, gaseous flows in the FIRE simulations based on the computation of instantaneous mass and metal fluxes through thin spherical shells around galaxies.  Here, we take advantage of the Lagrangian nature of smooth particle hydrodynamics (SPH) to perform a full particle tracking analysis of the FIRE simulations.  Particle tracking techniques are useful in cosmological simulations to trace the origin and evolution of individual parcels of gas \citep[e.g.][]{Keres2005,Oppenheimer2010,Genel2013,Ford2014,Ubler2014,Christensen2016,Nelson2015}.  By tracing back in time the gas and stellar components of galaxies, we characterize the main processes that constitute the baryon cycle in galaxy evolution.  We focus here on quantifying the contribution of physically distinct sources of material to the growth of central galaxies and the main statistical properties of wind recycling.

This paper is organized as follows.  We present our methodology in \S\ref{sec:meth}, including detailed descriptions of our particle tracking algorithms (\S\ref{sec:ana}) and the new definitions of galaxy growth components used in this work (\S\ref{sec:DissectGrowth}).  We report our results in \S\ref{sec:res}, where we describe the main sources of smooth gas accretion (\S\ref{sec:flows}), show representative galaxy growth histories (\S\ref{sec:paths}), analyze the halo mass dependence of galaxy growth components (\S\ref{sec:all}), and quantify the statistical properties of wind recycling (\S\ref{sec:winds}).  We discuss our findings and how they compare with previous work in \S\ref{sec:dis}. We summarize our conclusions in \S\ref{sec:con}.


\section{Methods}\label{sec:meth}

\begin{footnotesize}
\ctable[
  caption={{\normalsize Simulation parameters.}\label{tbl:sims}},center,star
  ]{lccccccc}{
\tnote[ ]{Parameters describing our simulations (units are physical): 
{\bf (1)} Name: simulation designation. 
{\bf (2)} $M_{\rm halo}$: mass of the central halo at $z=0$ (most massive halo in the high resolution region). 
{\bf (3)} $M_{*}$: stellar mass of the central galaxy at $z=0$.
{\bf (4)} $M_{\rm gas}$: ISM gas mass of the central galaxy at $z=0$.
{\bf (5)} $m_{\rm b}$: initial baryonic (gas and star) particle mass in the high-resolution region.
{\bf (6)} $\epsilon_{\rm b}$: minimum baryonic gravity/force softening (minimum SPH smoothing lengths are comparable or smaller); gas force softenings are adaptive.
{\bf (7)} $m_{\rm DM}$: dark matter particle mass in the high resolution region.
{\bf (8)} $\epsilon_{\rm DM}$: minimum dark matter force softening (fixed in physical units at all redshifts). 
}
}{
\hline\hline
\noalign{\vskip 0.5mm}
\multicolumn{1}{l}{Name} &
\multicolumn{1}{c}{$M_{\rm halo}(z=0)$ [\Msun]} &
\multicolumn{1}{c}{$M_{*}(z=0)$ [\Msun]} &
\multicolumn{1}{c}{$M_{\rm gas}(z=0)$ [\Msun]} &
\multicolumn{1}{c}{$m_{\rm b}$ [\Msun]} & 
\multicolumn{1}{c}{$\epsilon_{\rm b}$ [pc]} & 
\multicolumn{1}{c}{$m_{\rm DM}$ [\Msun]} & 
\multicolumn{1}{c}{$\epsilon_{\rm DM}$ [pc]} \\ 
\noalign{\vskip 0.5mm}
\hline
\noalign{\vskip 0.5mm}
{\bf m10} & 7.8e9 & 1.8e6  & 4.3e6  &  2.6e2 & 3 & 1.3e3 & 30 \\
{\bf m11} &  1.4e11 & 1.4e9  & 1.1e9  &  7.1e3 & 7 & 3.5e4 & 70 \\ 
{\bf m12v} &  6.3e11 & 1.7e10  & 1.0e9  &  3.9e4 & 10 & 2.0e5 & 140 \\ 
{\bf m12i} & 1.1e12 & 4.3e10  &  7.4e9  &  5.0e4 & 14 & 2.8e5 & 140 \\ 
{\bf m12q} & 1.2e12 &  1.4e10 & 2.7e9  &  7.1e3 & 10 & 2.8e5 & 140 \\ 
{\bf m13} &  6.1e12 & 7.6e10  & 3.2e9  &  3.7e5 & 21 & 2.3e6 & 210 \\
\hline\hline\\
}
\end{footnotesize}

\subsection{Simulations}\label{sec:sims}

We use the suite of FIRE cosmological zoom-in simulations presented in \citet{Hopkins2014_FIRE}.  All simulations are run using the ``P-SPH" mode of the GIZMO simulation code\footnote{A public version of GIZMO is available at: \url{http://www.tapir.caltech.edu/~phopkins/Site/GIZMO.html}.} \citep{Hopkins2015_Gizmo}, employing a pressure-entropy formulation of SPH that minimizes some of the major problems previous formulations of SPH have with fluid mixing instabilities \citep{Hopkins2013_PSPH,Saitoh2013}.
Gravitational forces are computed using a modified version of the tree-particle-mesh algorithm of the GADGET-3 code \citep{Springel2005_Gadget}, including adaptive gravitational softenings for the gas component.  We refer the reader to \citet{Hopkins2014_FIRE} for more details regarding the numerical elements of the gravity and hydrodynamic solvers.

Star formation and stellar feedback are modeled using the numerical models developed in \citet{Hopkins2014_FIRE}. 
Star formation occurs only in dense regions with hydrogen number density $n_{\rm H} \geq 5$--50\,cm$^{-3}$ from molecular, locally self-gravitating gas with a 100\,\% efficiency per local free-fall time.  In our simulations, the star formation efficiency on larger scales is regulated by the rate of formation and disruption of giant molecular clouds by feedback processes \citep[e.g.][]{Faucher-Giguere2013}, and it is fairly insensitive to the density threshold and other details of the small scale star formation law \citep{Hopkins2013_StarFormationLaw}.  
Once a star particle forms, stellar feedback is modeled by explicitly injecting energy, momentum, mass, and metals from stellar radiation pressure, HII photo-ionization and photo-electric heating, Type I and Type II supernovae, and stellar winds from AGB and O-type stars.  All feedback quantities and their time dependence are taken directly from stellar population synthesis modeling using STARBURST99 \citep{Leitherer1999}, where each star particle is treated as a single stellar population with known mass, age, and metallicity.
Metal enrichment is computed explicitly following each source of mass return individually, tracking the chemical abundances of nine metal species individually.  
We include radiative cooling from primordial gas \citep{Katz1996} in the presence of a uniform photo-ionizing background \citep{Faucher-Giguere2009}, as well as metal-line \citep{Wiersma2009}, fine-structure, and molecular cooling processes.

The simulations analyzed here use the zoom-in technique \citep[e.g.][]{Katz1993,Onorbe2014} to achieve the baryonic particle mass ($m_{\rm b}$) and force softenning ($\epsilon_{\rm b}$) required to begin to resolve the multi-phase structure of galaxies in a full cosmological setting.  The zoom-in regions are selected from large volume cosmological N-body simulations targeting typical halos in the mass range $10^{10}$--$10^{13}$\,\Msun~at $z=0$.  The resolution of each simulation is chosen to scale with the mass of the system to ensure that the formation of individual giant molecular clouds is resolved, with baryonic particle masses in the range $m_{\rm b} = 2.6\times10^2$--$3.7\times10^5\,$\Msun~and minimum force softenning lengths $\epsilon_{\rm b} = 3$--$21$\,pc for gas particles (force softenings are adaptive).
Following the notation adopted in \citet{Hopkins2014_FIRE}, we analyze the isolated dwarf irregular galaxy {\bf m10}, the massive dwarf spheroidal {\bf m11}, the three Milky Way-mass systems {\bf m12i}, {\bf m12q}, and {\bf m12v}, and the more massive early-type galaxy {\bf m13} (see Table~\ref{tbl:sims} for a list of simulation parameters).  Simulations {\bf m11, m12q, m12i}, and {\bf m13} were chosen to match a subset of initial conditions from the AGORA project \citep{Kim2014_AGORA}, while the initial conditions for simulation {\bf m12v} were studied in previous work \citep{Keres2009_ColdClouds,Faucher-Giguere2011_Filaments}.  We adopt a ``standard" flat \lcdm~cosmology with parameters $h \approx 0.7$, $\Omega_{\rm M} = 1-\Omega_{\rm \Lambda} \approx 0.27$, and $\Omega_{\rm b} \approx 0.046$ \citep[e.g.][]{Hinshaw2013,PlanckCollab2014}.

\begin{figure*}
\begin{center}
\includegraphics[scale=0.6]{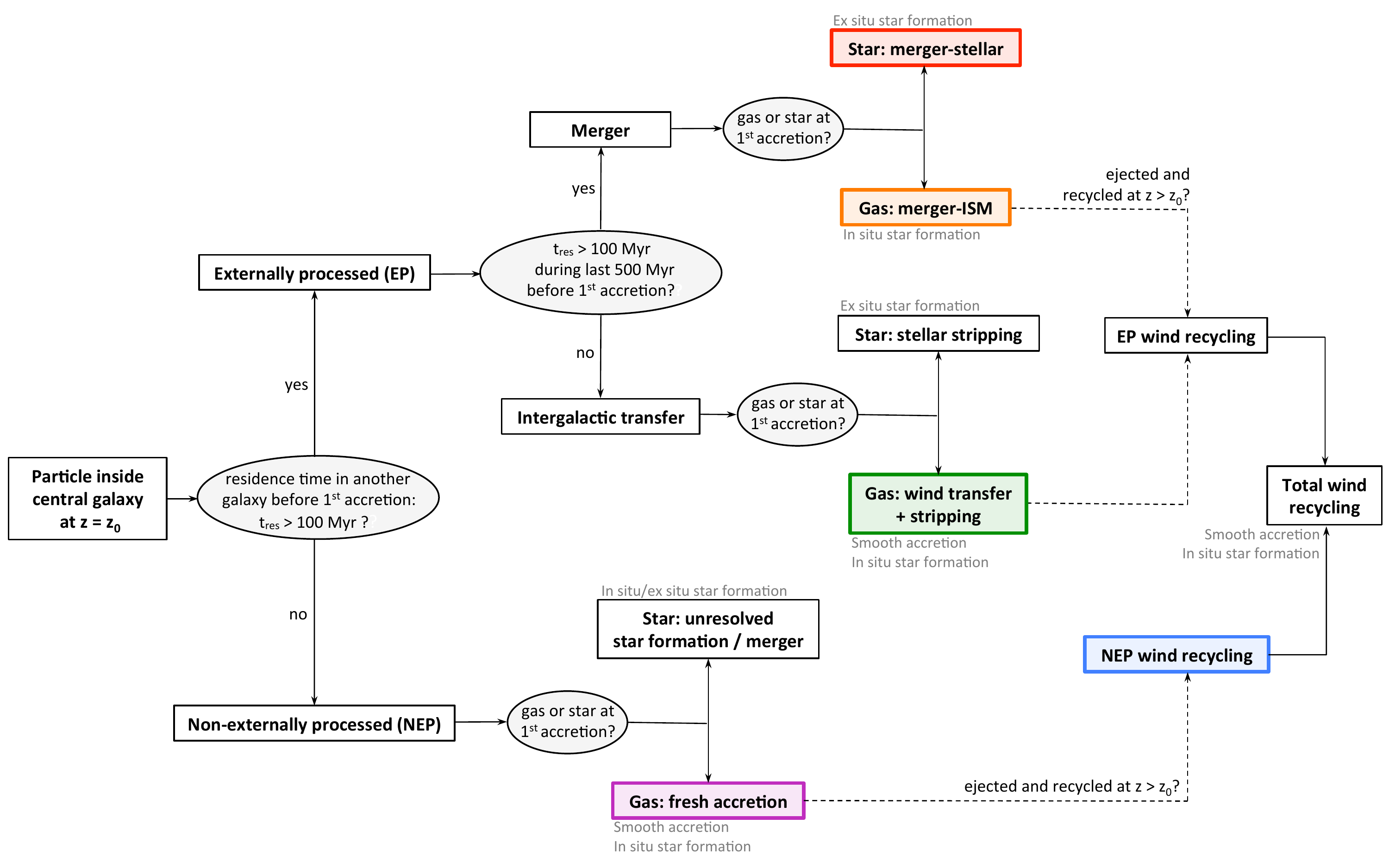}
\end{center}
\caption{Hierarchy of processes used throughout the paper to dissect the origin of the gas and stellar components of galaxies.  Particles (gas or star) inside of the central galaxy at $z=z_0$ are classified into different components based on their history prior to first accretion at $z>z_0$.  We first identify material preprocessed in another galaxy.  Externally processed material is further decomposed into mergers and intergalactic transfer.  The particle type at first accretion (gas or star) determines additional classifications into merger-stellar, merger-ISM, stellar stripping, and gas transfer. Non-externally processed material contains fresh gas accretion and stars from unresolved star formation events and mergers.  Gas ejected from the ISM of the central galaxy and recycled back before $z=z_0$ is further classified as wind recycling. 
Connections with previous definitions (in situ versus ex situ star formation and smooth accretion versus mergers) are indicated for each component.
Growth modes shown in Figures~\ref{fig:gmode} and \ref{fig:bars} are highlighted using the same color scheme. 
}
\label{fig:blocks}
\end{figure*}

\subsection{Analysis}\label{sec:ana}

We identify galaxies at each redshift as gravitationally bound collections of dense gas and star particles by means of {\sc skid}\footnote{
The original source code can be found at \url{http://www-hpcc.astro.washington.edu/tools/skid.html}. 
We use the modified version of {\sc skid} provided as part of the SPHGR package \citep[\url{https://bitbucket.org/rthompson/sphgr};][]{Thompson2015_SPHGR}.}. 
We use a linking length $\sim 0.2$ times the mean inter-particle separation and impose a minimum density threshold $n_{\rm H} \geq 0.1$\,cm$^{-3}$ for gas particles \citep[e.g.][]{Keres2005,Angles-Alcazar2014}.   
Dark matter halos are identified independently of the {\sc skid} galaxy definition using the Amiga Halo Finder \citep[AHF;][]{Gill2004_AHF,Knollmann2009_AHF}, where we use the evolving virial overdensity definition of \citet{Bryan1998}. Galaxies are linked to their host dark matter halos at each redshift by determining the AHF halo that contains the largest number of each galaxy's star particles.

In this work, we focus on the evolution of the central galaxy hosted by the most massive halo located near the center of the zoom-in region of each simulation at $z=0$, which we refer to as either the ``main" or ``central" galaxy throughout this paper.  
Starting at $z=0$, the full history of each galaxy is reconstructed back in time by identifying its most massive progenitor, which is defined as the {\sc skid} group in the previous redshift snapshot having the largest number of star particles in common. 
We require {\sc skid} groups to contain $\geq 100$ star particles; this resolution limit is imposed for all galaxy progenitors.
Global galaxy properties (e.g. stellar mass) are computed based on the particles that belong to the corresponding {\sc skid} group and that are within a radial distance to its center $r \leq 2\times R_{\rm eff}$, where the effective radius $R_{\rm eff}$ is defined as the aperture containing 50\,\% of the total stellar mass of the {\sc skid} group.  We define the CGM as the gas outside of the central galaxy but within the virial radius, while the IGM refers to gas outside the virial radius of dark matter halos.

Our particle tracking analysis begins by identifying every gas and star particle that belongs to the main progenitor galaxy at any redshift, using $\sim 440$ data snapshots per galaxy.  We track the full particle list over time (consisting of $\sim 4\times10^5$--$2\times10^6$ particles for different simulations), compiling information about their mass, position, velocity, and host galaxy/halo membership.  This is used to define the state of particles relative to the main progenitor galaxy at all times, starting from the redshift at which the galaxy is first resolved down to $z=0$, as well as to identify gas accretion and wind ejection events:
\begin{itemize}

\item Galactic {\it wind ejection events} are defined for gas particles that transition from being inside of the main galaxy to being outside of it from one redshift snapshot to the next.  There is no radial distance requirement in addition to the transition in {\sc skid} group membership, but wind particles are required to have a radial velocity $v_{\rm out} \ge 2\times V_{\rm c}$ by the time the ejection event is first identified, where $V_{\rm c}$ is the maximum circular velocity of the galaxy.
Additionally, wind particles must be outside of any other galaxy by the snapshot following the wind ejection event, thus removing possible contamination from any close galaxy fly-by that may be miss-identified by {\sc skid}.  The {\it last ejection event} of gas particles (if any) is of particular interest to quantify the amount of mass lost by galaxies in large scale winds.

\item Galaxy {\it accretion events} are defined for gas and star particles that transition from being outside of the main galaxy to being inside of it from one redshift snapshot to the next.  After the first gas accretion event, we select only the subsequent accretion events (if any) that are preceded by a wind ejection event.  This removes artificial accretion from gas loosely bound to the galaxy or due to small scale galactic fountains that could overestimate the accretion of gas from larger CGM scales. 
Thus, by construction, accretion and ejection events follow each other consecutively for gas particles that recycle multiple times. Throughout this paper, gas accretion rates ($\dot{M}_{\rm acc}$) refer to direct mass supply to the ISM of central galaxies. We compute $\dot{M}_{\rm acc}$ by adding the mass of all gas accretion events from one snapshot to the next, dividing by the time interval between snapshots.

\end{itemize}

\subsection{Classification of galaxy growth modes}\label{sec:DissectGrowth}

Figure~\ref{fig:blocks} summarizes how the different modes of galaxy growth are classified in our analysis.
We begin by separating the overall source of baryonic mass into {\it externally processed} and {\it non-externally processed} contributions, where the former refers to material that has been preprocessed inside another galaxy prior to contributing to the growth of the central galaxy.  {\it Mergers} contribute preprocessed material in the form of stars directly accreted onto the central galaxy as well as ISM gas from which new stars can later form.  Externally processed material also includes smooth {\it intergalactic transfer} of mass between galaxies.  This comprises gas and stars removed from satellites by e.g. ram pressure and tidal stripping processes, but also includes the {\it wind transfer} component, i.e. gas ejected from nearby galaxies in large scale winds that eventually accrete onto the central galaxy. Non-externally processed material contains primarily {\it fresh accretion} of gas from the IGM directly onto the ISM of the central galaxy. We also quantify each galaxy's own {\it wind recycling} material regardless of the original source of gas (i.e., externally and non-externally processed recycled gas).

\subsubsection{Externally processed material}\label{sec:ext}

For any arbitrary redshift $z=z_0$, we identify the first accretion event for every gas and star particle inside of the main galaxy.
We then compute the amount of time, if any, that these particles have resided in another galaxy prior to first accretion onto the main galaxy ($t_{\rm res}$).  We define a threshold preprocessing time $t_{\rm pro}$ such that particles that resided in another galaxy for $t_{\rm res} \geq t_{\rm pro}$ are classified as externally processed material, while particles with $t_{\rm res} < t_{\rm pro}$ contribute to the non-externally processed component.

Our fiducial threshold preprocessing time is comparable to typical galaxy dynamical times, $t_{\rm pro} = 100$\,Myr.  
This is chosen as a compromise between the typical time interval between snapshots (10--40\,Myr at $z > 1$ and 20--70\,Myr at $z < 1$) and shorter timescales minimizing the external effects on particles.
Non-externally processed material is dominated by {\it fresh accretion},   
corresponding to gas that never belonged to another galaxy for longer than $t_{\rm pro}$.
Nonetheless, non-externally processed material in principle also contains (1) stars that resided in galaxies below our mass resolution limit and (2) stars that formed inside of the central galaxy during the time interval between first accretion and the redshift snapshot following the gas accretion event.  In either case, the contribution of stars to the non-externally processed component is negligible.

\subsubsection{Distinguishing galaxy mergers from intergalactic transfer}\label{sec:mergtrans}

We look at each particle's past history for a time interval $t_{\rm m}$ prior to first accretion onto the main galaxy. 
We compute the residence time in another galaxy during this time interval, $t_{\rm res}(t_{\rm m})$, which is then compared against the threshold preprocessing time $t_{\rm pro}$.  Particles with $t_{\rm res}(t_{\rm m}) \geq t_{\rm pro}$ are defined to be part of the {\it merger} contribution, while accreted particles with $t_{\rm res}(t_{\rm m}) < t_{\rm pro}$ correspond to the {\it intergalactic transfer} component.  
The fiducial time interval for merger identification, $t_{\rm m} = 500$\,Myr, is chosen so as to easily accommodate the minimum residence time required for externally processed material ($t_{\rm pro} = 100$\,Myr) while short enough to capture the transfer of mass recently stripped or ejected from nearby galaxies. This ensures that merger-classified particles were indeed inside of the merging galaxy for at least $\sim 100$\,Myr during the $\sim 500$\,Myr immediately preceding the merger, while allowing us to identify the intergalactic transfer component as smoothly accreting particles that were removed from their host galaxy $\gtrsim 500$\,Myr prior to the accretion event.
We analyze the dependence of results on $t_{\rm pro}$ and $t_{\rm m}$ in Appendix~\ref{sec:appendix:num}.

Once the distinction between mergers and intergalactic transfer is made, we further identify the nature of the accreted material as either gaseous or stellar by the time of first accretion.  We thus separate galaxy mergers into the {\it merger-stellar} and {\it merger-ISM} gas contributions based on the type of particle accreted.  Likewise, the intergalactic transfer component is separated into {\it stellar stripping} and gas transfer, which occurs through {\it gas stripping} and (primarily) {\it wind transfer}. In this work, we show that the intergalactic transfer of mass between galaxies, previously largely neglected, may represent a substantial contribution to galaxy growth (up to $\sim 40$\,\% of the stellar content of Milky Way-mass galaxies at $z=0$).

\subsubsection{Wind recycling}\label{sec:windef}

Gas particles ejected from the central galaxy and recycled back before the redshift of interest ($z=z_0$) are classified as {\it wind recycling}.  Recycled gas can in principle remain in the galaxy, form stars, or be ejected in a wind again.  In each case, we track the cumulative number of wind ejection events as a function of redshift.  For particles inside of the main galaxy at $z=z_0$, the number of recycling times $N_{\rm REC}$ is defined as the cumulative number of ejection events prior to that redshift.  Star particles inherit $N_{\rm REC}$ from the recycling history of the gas particle progenitor prior to the star formation event, such that, e.g., the contribution of wind recycling to the stellar content of the main galaxy at $z=z_0$ can be easily computed based on particles with $N_{\rm REC} > 0$.

The galaxy growth contributions described above are defined based on the history of particles prior to first accretion onto the central galaxy, with the corresponding classification maintained at lower redshifts.  
Gas particles may be ejected from the main galaxy regardless of the original source of material.
Wind recycling introduces an additional classification level that depends on the evolutionary path of gas particles after their first accretion event (Figure~\ref{fig:blocks}).  
Throughout the paper, we illustrate the efficiency and overall contribution of wind recycling either only for the non-externally processed (NEP) component (i.e. NEP wind recycling; e.g. Figures~\ref{fig:gmode} and \ref{fig:bars}) or for the total (NEP+EP) wind recycling, including externally processed gas (e.g. Figure~\ref{fig:windrec}).
Note also that wind ejection events refer only to the main galaxy by definition,  
i.e. $N_{\rm REC} \equiv 0$ for all particles at the time of first accretion, so that the contribution of wind recycling reported for any property of the main galaxy corresponds to its own recycling process.

\subsubsection{Comparison to previous work}\label{sec:cwork}

Previous studies of galaxy mass assembly have typically considered either (1) the relative importance of smooth accretion versus mergers \citep{Murali2002}, (2) the contributions of ``in situ" versus ``ex situ" star formation to galaxy growth \citep{Oser2010}, or (3) the significance of wind recycling relative to other smooth accretion modes \citep{Oppenheimer2010}.
The classification scheme employed here attempts to incorporate most previous definitions into a single coherent representation of galaxy growth. The primary distinction between externally processed versus non-externally processed material establishes a fundamental difference between galaxy evolution in isolation and the role of external processes.  
This provides the basis for a more direct connection between galaxy mass assembly and baryon cycling, where traditional classifications into e.g. smooth accretion and mergers coexist with baryon cycle processes that explicitly consider wind recycling and the exchange of gas between galaxies in the overall mass budget.

Figure~\ref{fig:blocks} indicates the correspondence between previous definitions and the classifications adopted here. In situ star formation corresponds to stars formed inside of the central galaxy from gas provided by (1) fresh accretion, (2) wind recycling, (3) intergalactic transfer, and (4) the ISM of merging galaxies, while the ex situ component corresponds to stars formed outside of the central galaxy, i.e. stars accreted from (1) mergers and (2) stellar stripping of satellites.  Mergers are explicitly identified in our classification scheme while smooth accretion consists of (1) fresh gas, (2) wind recycling, and (3) intergalactic gas transfer.

\begin{figure*}
\begin{center}
\includegraphics[width=0.33\textwidth]{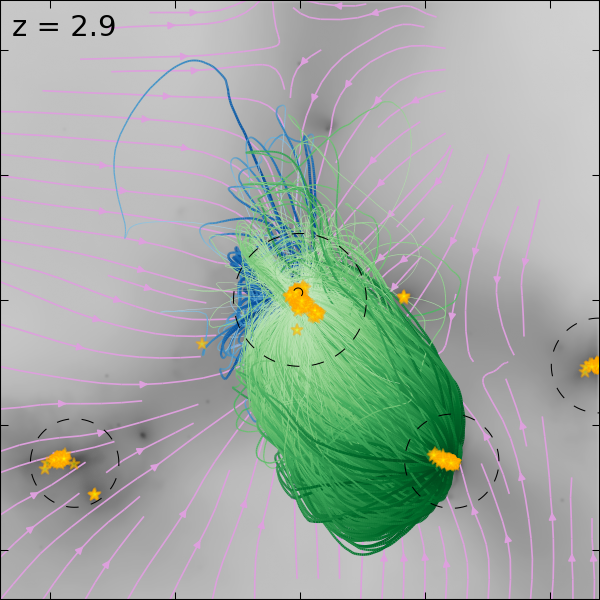}
\includegraphics[width=0.33\textwidth]{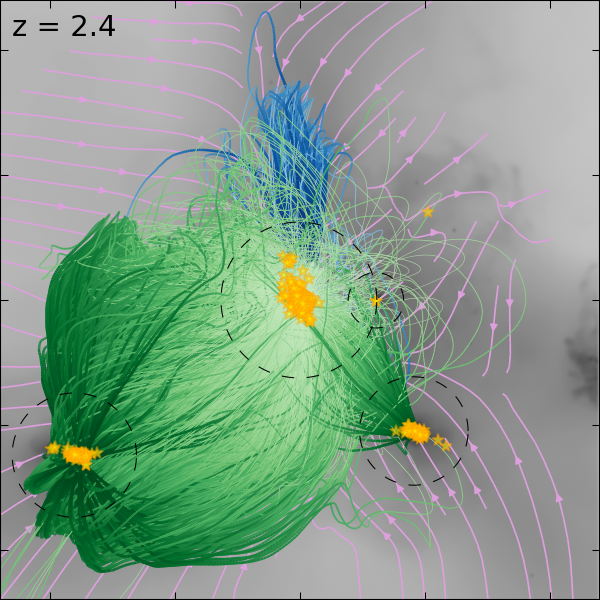}
\includegraphics[width=0.33\textwidth]{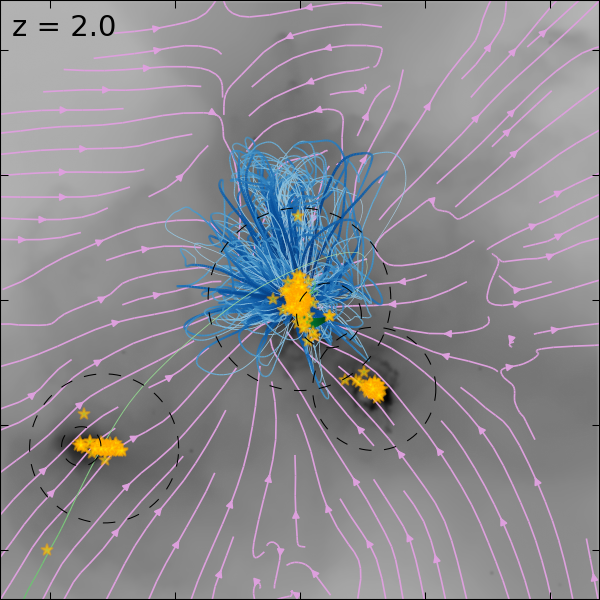}\\
\includegraphics[width=0.33\textwidth]{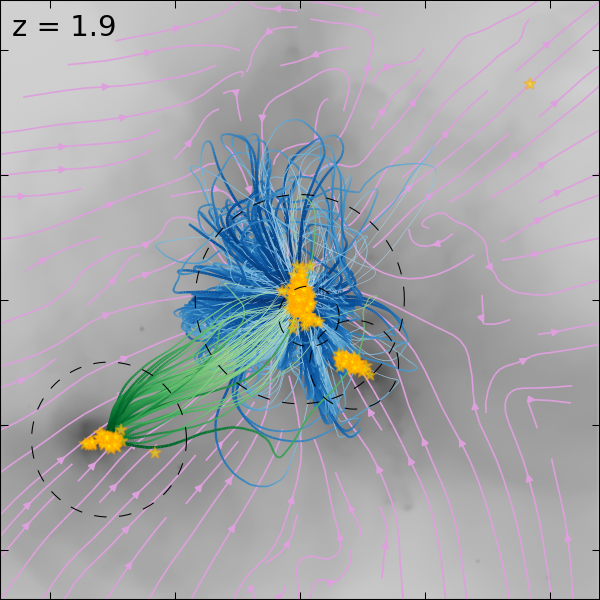}
\includegraphics[width=0.33\textwidth]{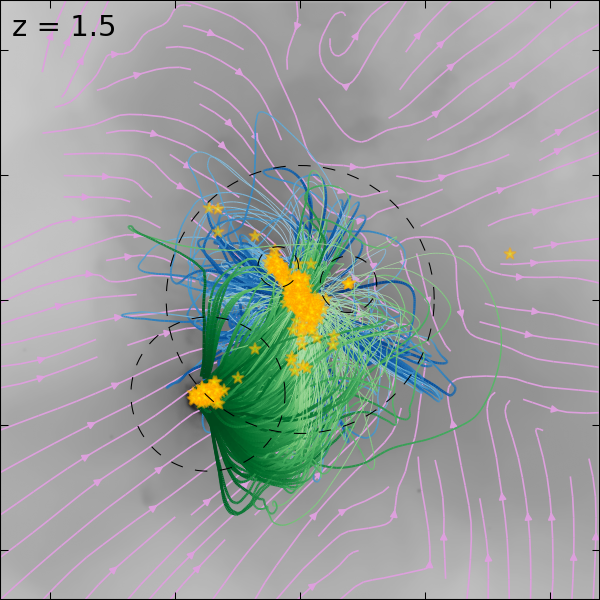}
\includegraphics[width=0.33\textwidth]{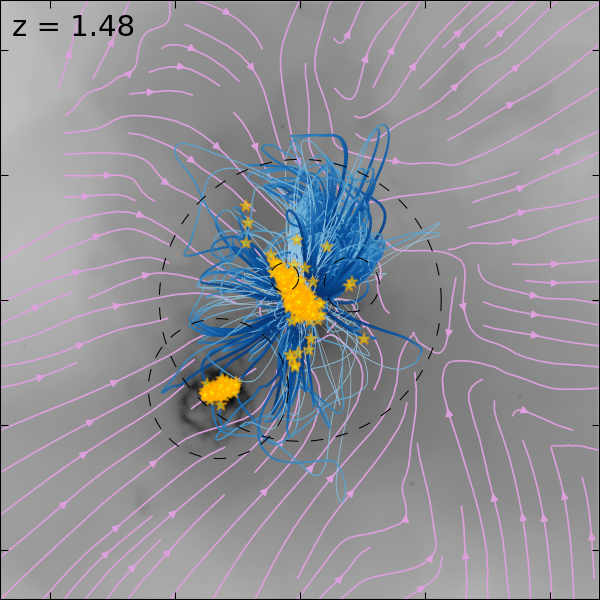}\\
\includegraphics[width=0.33\textwidth]{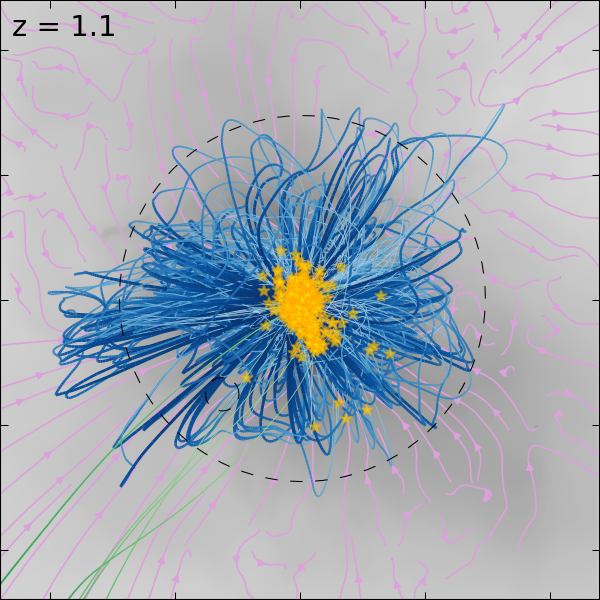}
\includegraphics[width=0.33\textwidth]{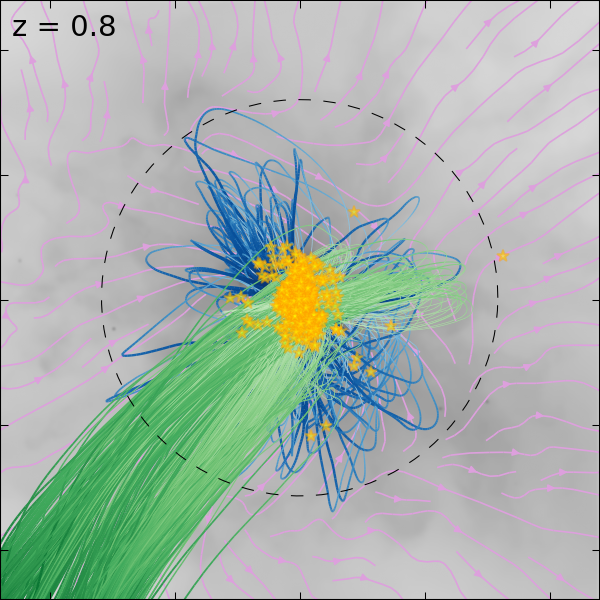}
\includegraphics[width=0.33\textwidth]{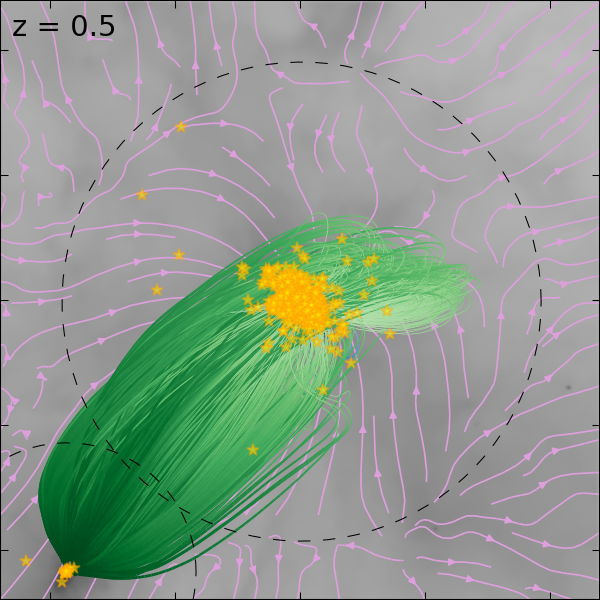}
\end{center}
\caption{Intergalactic gas flows in simulation {\bf m11}. The gray scale shows projected gas density distributions (logarithmically scaled) in a [240\,kpc]$^3$ volume (physical) centered on the main galaxy ($M_{\rm halo}(z=0) \approx 1.4\times10^{11}$\,\Msun) at different redshifts ($z \sim 3$--0.5), while the orange star symbols represent the stellar distribution within the same volume.
Purple lines indicate projected mean (mass-weighted) streamlines of fresh gas at each redshift.
Blue lines indicate the future trajectory of gas particles ejected from the central galaxy that will accrete back as part of the wind recycling component.
Green lines indicate the future trajectory of gas particles removed from the ISM of another galaxy that will smoothly accrete onto the central galaxy as part of the intergalactic transfer component.
Wind recycling and intergalactic transfer trajectories proceed in the direction of decreasing darkness and thickness of the blue and green lines, respectively.
In each panel, recycling/transfer trajectories are shown only for gas particles ejected within $\sim 50$--100\,Myr. 
Black dashed lines show the virial radius $R_{\rm vir}$ of each identified dark matter halo.
Note that streamlines of fresh gas (1) represent the bulk motion at each redshift but not the amount of fresh gas accreting onto the central galaxy and (2) may point away from the central galaxy owing to outflowing material (e.g., at $z=0.8$), reversing at later times.
Wind recycling occurs primarily within $R_{\rm vir}$ of the host halo.
The large volume covered by intergalactic transfer trajectories in some panels (e.g., at $z=2.9$ and $z=2.4$) illustrates the important contribution of galactic winds to this growth mode.  
}
\label{fig:flows}
\end{figure*}


\section{Results}\label{sec:res}

\subsection{Intergalactic gas flows}\label{sec:flows}

We begin by illustrating the primary sources of smooth gas accretion defined in \S\ref{sec:DissectGrowth}.  Figure~\ref{fig:flows} shows representative snapshots of our dwarf galaxy {\bf m11} (located at the center), where we indicate streamlines of fresh gas at various redshifts ($z\sim3$--0.5) and the future trajectories followed by wind recycling material and intergalactic transfer. 
Intergalactic gas flows can be rather complicated, with galaxies moving non-trivially relative to their surrounding medium and galactic outflows interacting with accreting material; individual gas particle trajectories have been smoothed for clarity, convolving their redshift-dependent coordinates with a flat window equivalent to $\sim 0.7$\,Gyr in width.  
Three filaments dominate the large scale distribution of gas surrounding the central galaxy at early times ($z\sim 3$), two of which contain the primary galaxies merging onto {\bf m11} at $z\lesssim 1.5$.
Fresh accretion traces the large scale motion of gas that will accrete directly onto the central galaxy, i.e. without prior contact with other galaxies. 
Note that streamlines of fresh accretion can be temporally directed radially outward from the central galaxy owing to the interaction of accreting gas with outflowing material (e.g. $z=0.5$ and $z=0.8$), delaying accretion onto the central galaxy.

Green lines in Figure~\ref{fig:flows} show intergalactic transfer trajectories originated at each redshift, connecting the ``source" galaxy (i.e. the current location of transfer material) to the accretion event onto the central ``destination" galaxy at later times.    
Galactic winds removing gas from the ISM of galaxies is the primary mode of intergalactic transfer (we return to this point in \S\ref{sec:windtrans}); we refer to this component as wind transfer.
Outflowing gas participating in intergalactic transfer can be either (1) temporally retained in the CGM of the source galaxy, i.e. within its virial radius $R_{\rm vir}$ (e.g. $z=1.9$ and $z=0.5$), then smoothly accreting onto the central galaxy after the parent halos merge, or (2) can be pushed to larger IGM scales before transferring to the central galaxy (e.g. $z=2.9$ and $z=2.4$). 
Gas stripping can in principle also contribute intergalactic transfer material, but we find this to be a subdominant contribution.  We discuss this possibility in \S\ref{sec:windtrans}, where we analyze the transfer of gas from dwarf satellites onto or Milky Way-mass galaxy {\bf m12i}.

Figure~\ref{fig:flows} also illustrates the trajectories of outflowing gas particles ejected at each redshift that recycle back onto the central galaxy at later times (blue lines).
Ejected gas can reach IGM scales before recycling back onto the central galaxy, especially at high redshift (e.g. $z=2.4$, $z=2$), but most recycling occurs well within $R_{\rm vir}$ of the host dark matter halo (we quantify recycling distances and timescales in \S\ref{sec:rec_dist}).
Similar intergalactic transfer and wind recycling trajectories are seen in all simulated galaxies.

\begin{figure*}
\begin{center}
\includegraphics[scale=0.46]{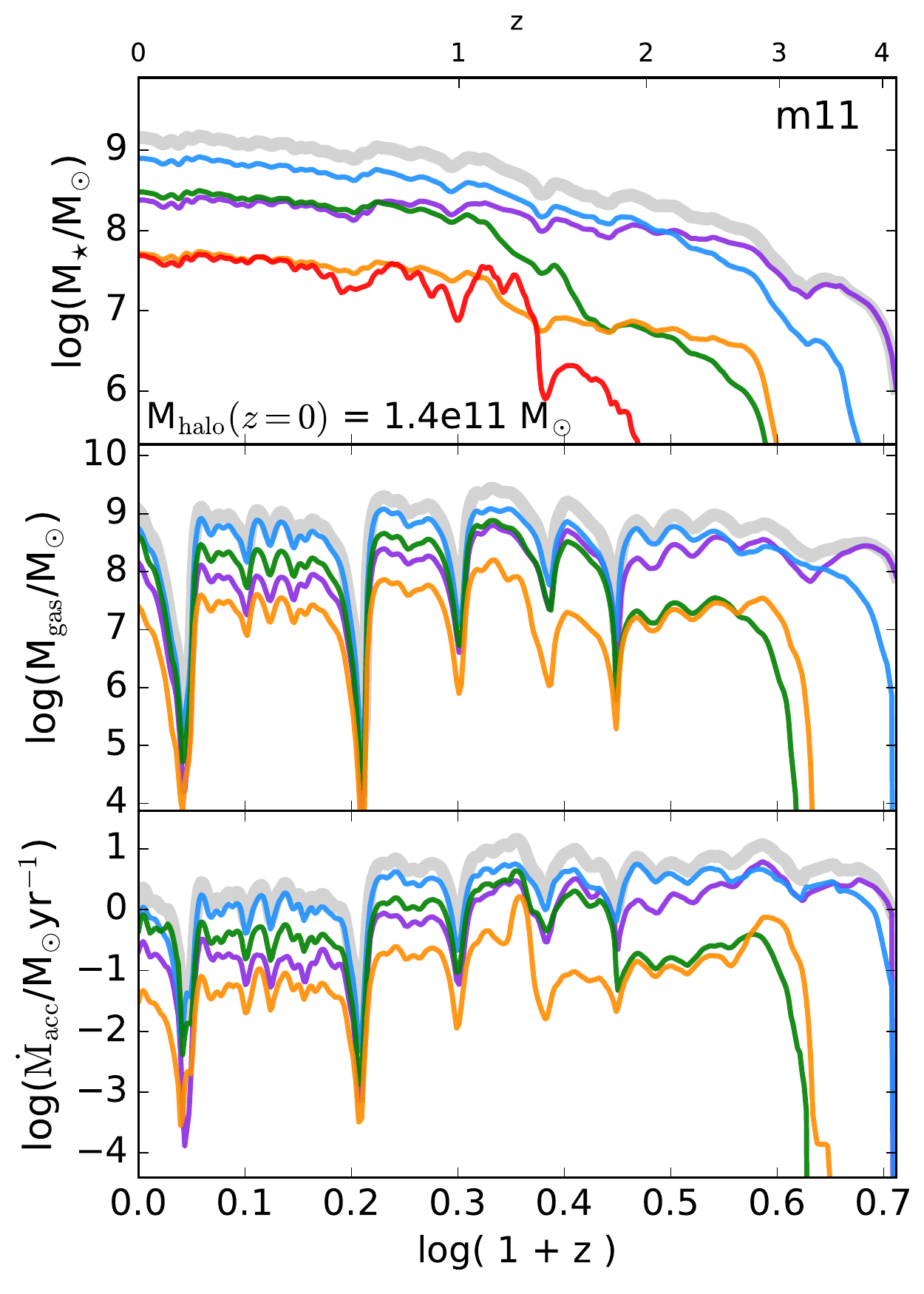}
\includegraphics[scale=0.46]{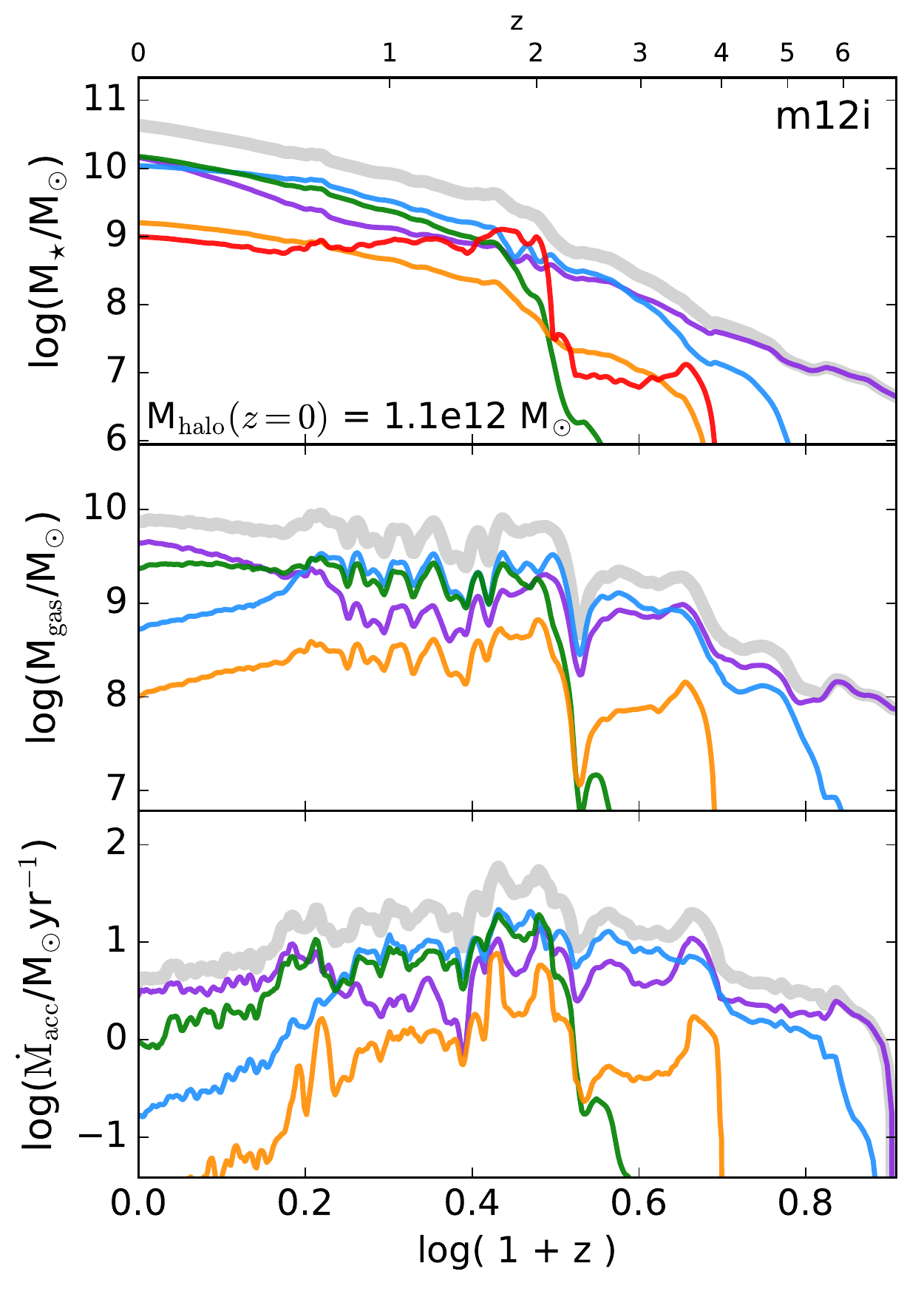}
\includegraphics[scale=0.46]{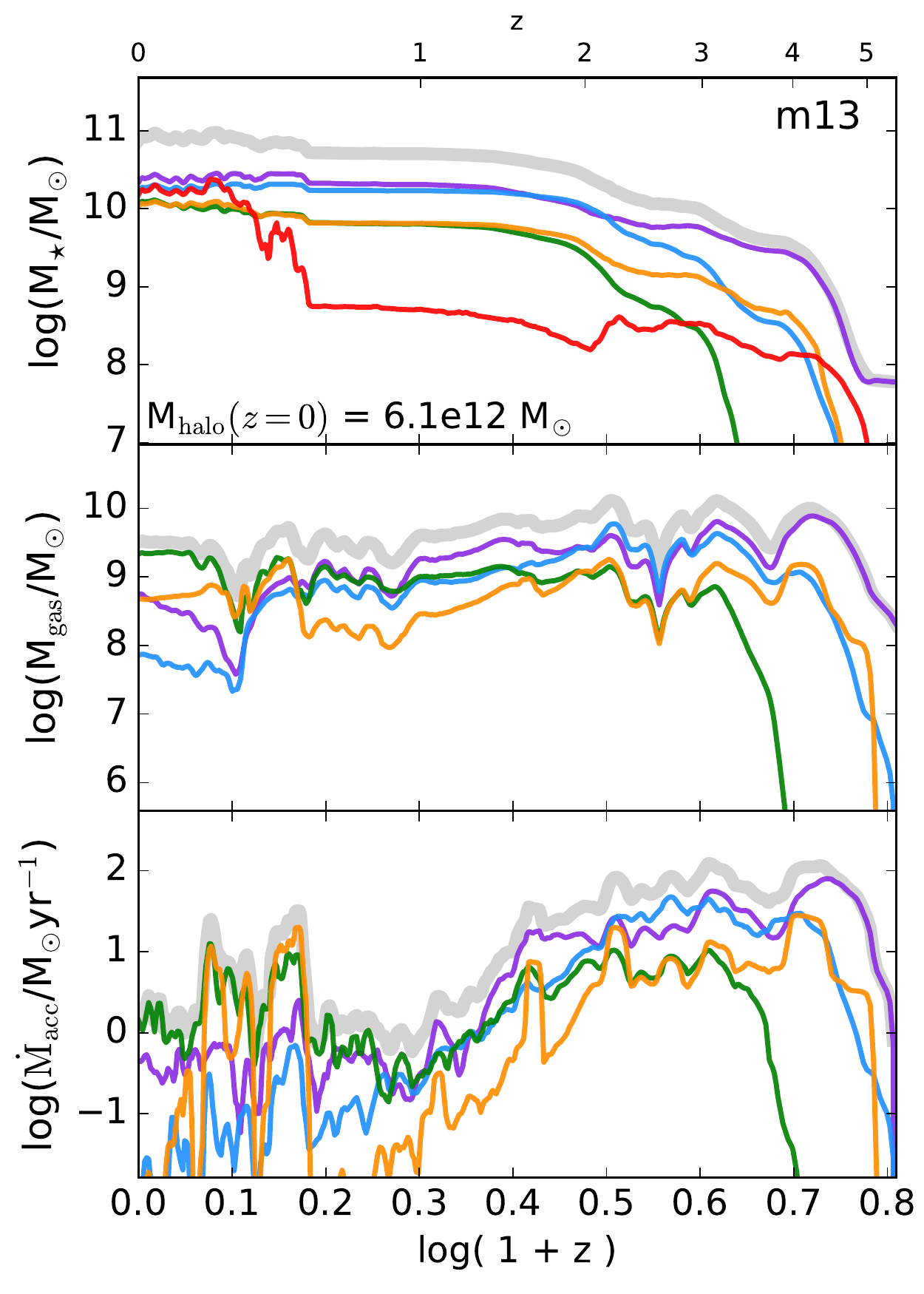}

\includegraphics[scale=0.3]{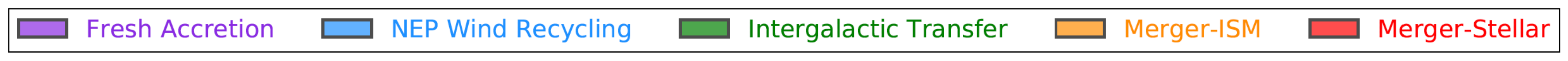}

\includegraphics[scale=0.46]{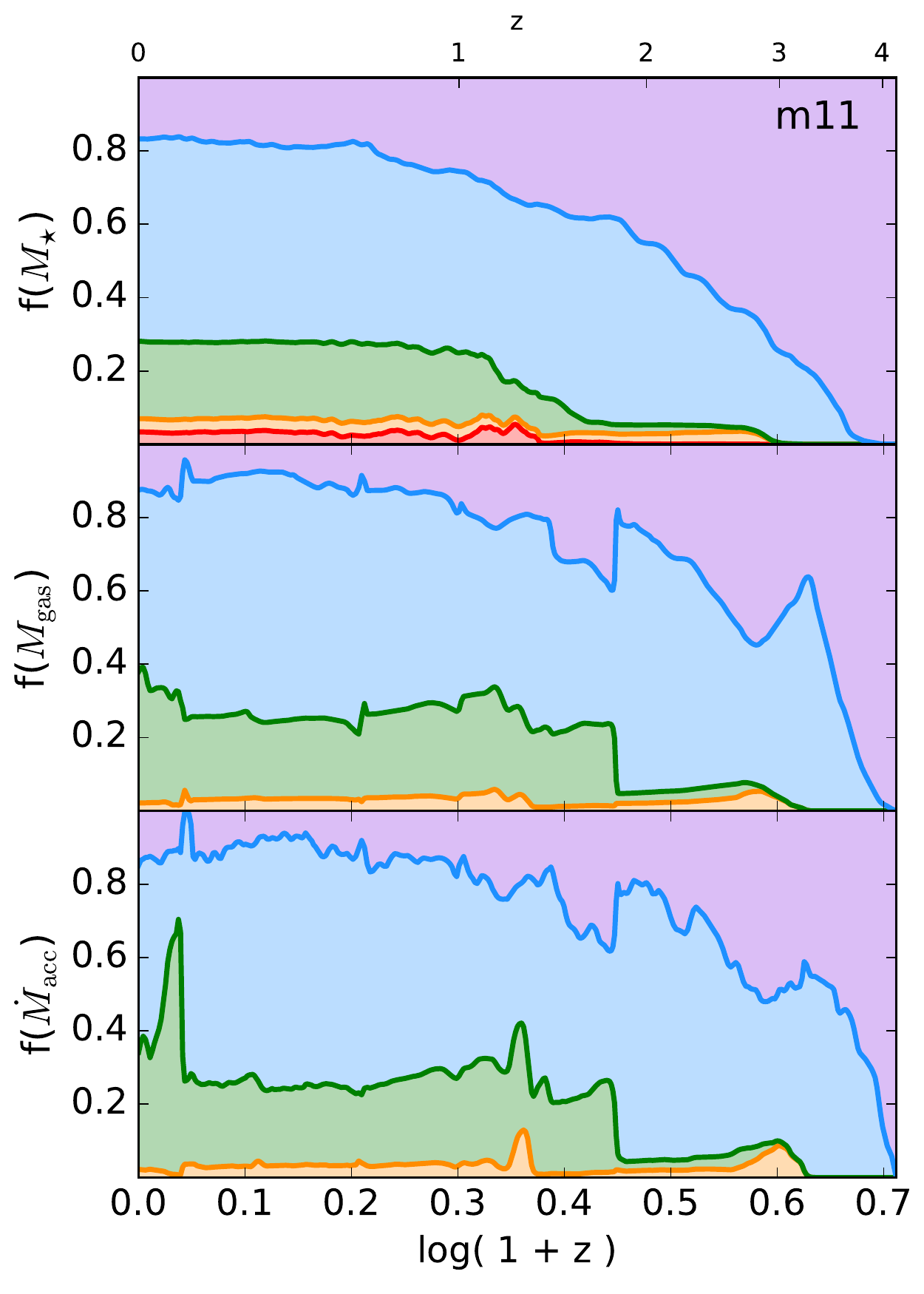}
\includegraphics[scale=0.46]{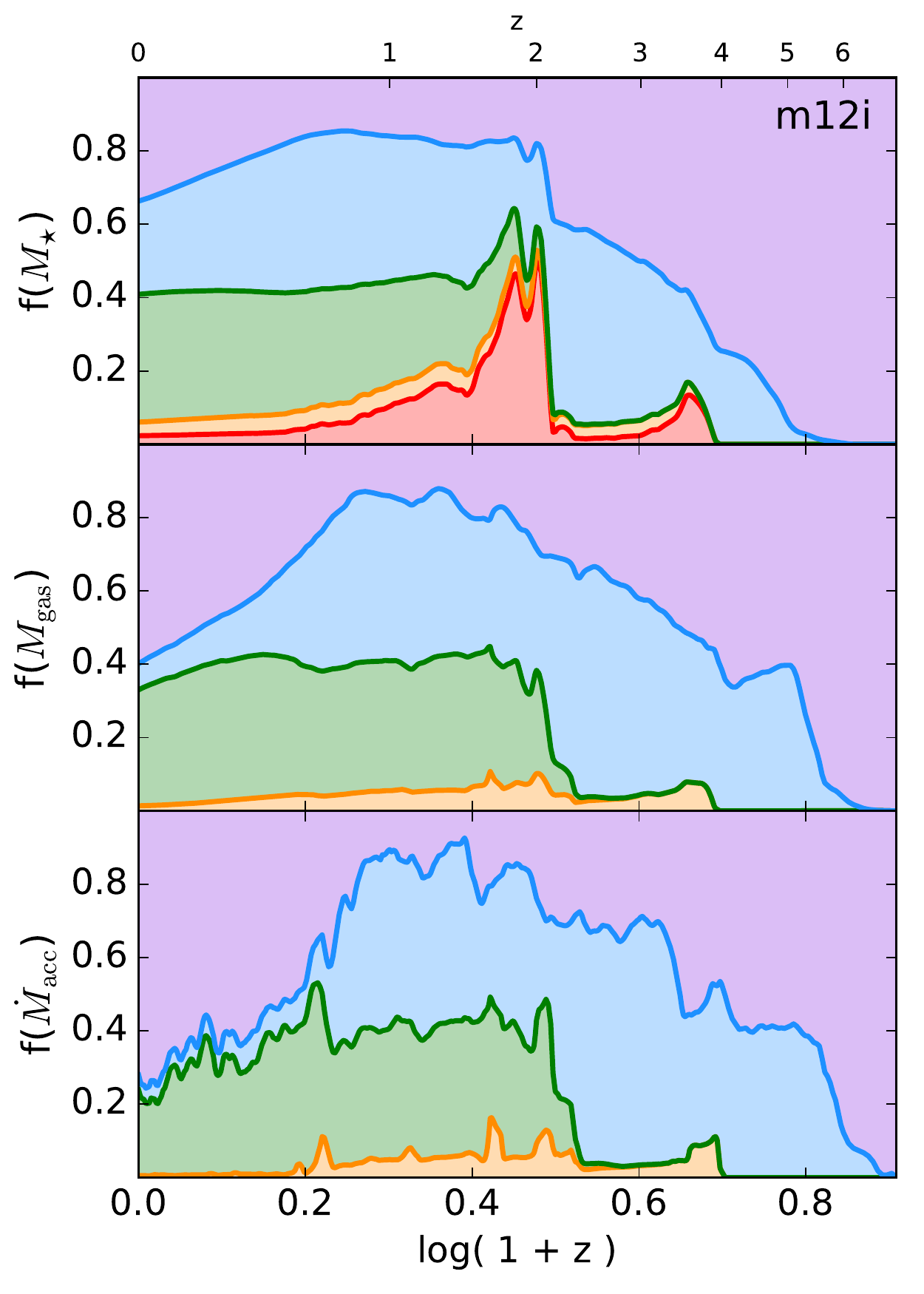}
\includegraphics[scale=0.46]{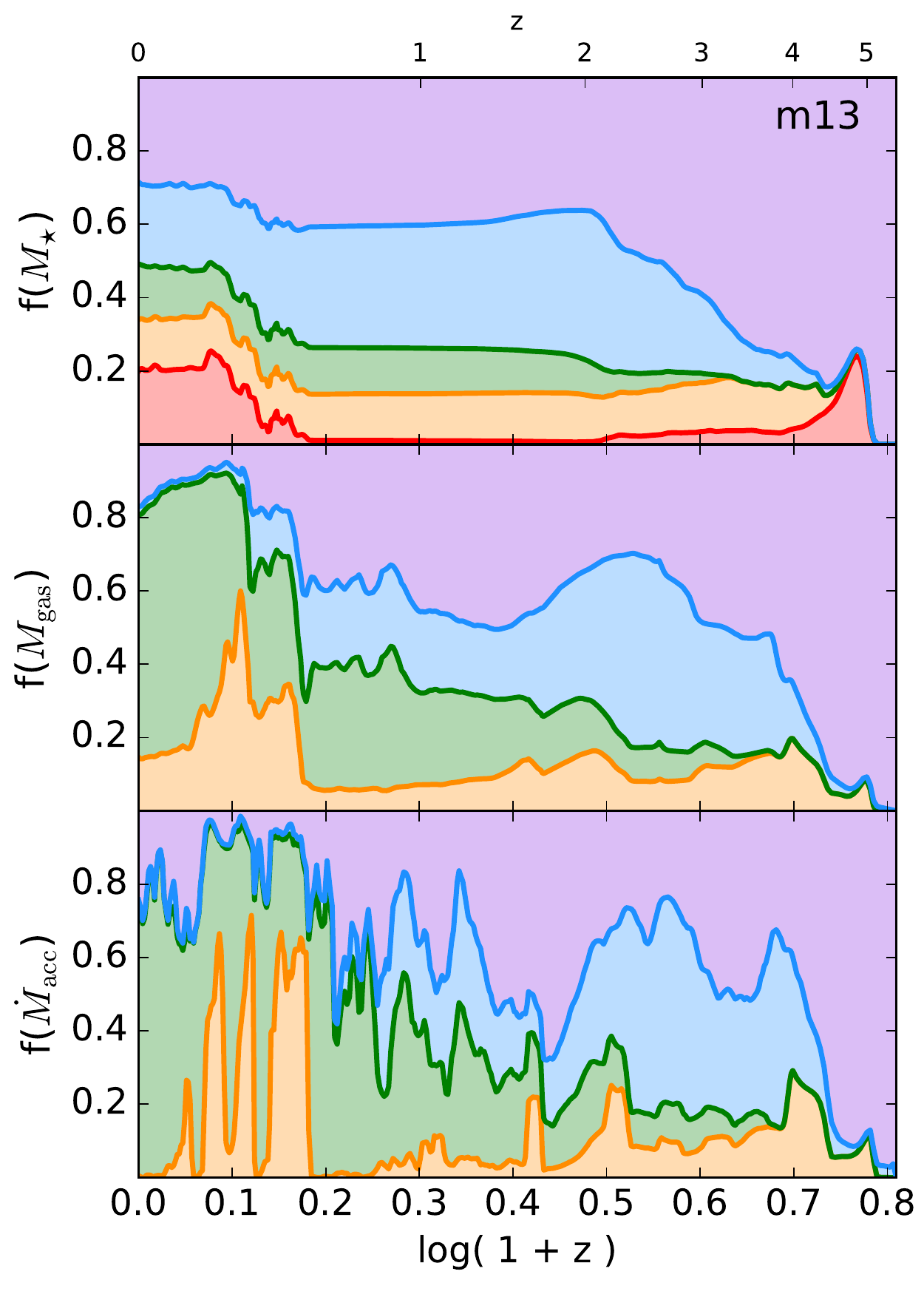}
\end{center}
\caption{Top panels: contribution of different processes to the evolution of galaxies {\bf m11} (left), {\bf m12i} (middle), and {\bf m13} (right) from early times down to $z=0$.  
Total stellar mass (top), ISM gas mass (middle), and gas accretion rate onto the galaxy (bottom) are indicated by the thick gray lines for each galaxy.
Lines of different colors show the contributions from fresh accretion (purple), NEP wind recycling (blue), intergalactic transfer of gas (green), and galaxy mergers (ISM gas: orange; stars: red).
All quantities represent average values over a timescale of $\sim 200$\,Myr.
Bottom panels: fraction of stellar mass, ISM gas mass, and gas accretion rate contributed by each process as a function of redshift.
Fractions are represented in terms of ranges, so that 
the contribution of each component at any redshift is given by the vertical extent of the corresponding color at that redshift.
}
\label{fig:gmode}
\end{figure*}

\subsection{Overview of representative galaxy growth histories}\label{sec:paths}

The top panels of Figure~\ref{fig:gmode} show the evolution of the total stellar mass ($M_*$), ISM gas mass ($M_{\rm gas}$), and gas accretion rate ($\dot{M}_{\rm acc}$) onto three representative simulated galaxies ({\bf m11}, {\bf m12i}, and {\bf m13}) from early times down to $z=0$ (gray solid lines). 
We separate the time-dependent $M_*$, $M_{\rm gas}$, and $\dot{M}_{\rm acc}$ into mutually exclusive components that add up to the total: (1) fresh accretion, (2) NEP wind recycling, (3) intergalactic transfer of gas, (4) ISM gas from mergers, and (5) stars from mergers.  
The bottom panels show the fraction of $M_*$, $M_{\rm gas}$, and $\dot{M}_{\rm acc}$ contributed by each process for the same simulated galaxies.

Note that fresh accretion and NEP recycling add up to the total supply of non-externally processed gas, while the intergalactic transfer and merger contributions correspond to the total supply of externally processed material (EP recycling is omitted for clarity; see Figure~\ref{fig:blocks}).
We have omitted the stellar stripping component as well as stars accreted from unresolved mergers (both contributions are minimal and thus neglected throughout the paper).
The gas content and SFR of galaxies are highly variable, on timescales shorter than the galaxy dynamical time, owing to the effects of explicit stellar feedback \citep{Hopkins2014_FIRE,Muratov2015,Sparre2015_SFbursts}.  For clarity, all quantities shown in Figure~\ref{fig:gmode} represent values averaged over $\sim 200$\,Myr.

\subsubsection{Massive dwarf spheroidal: {\bf m11}}

The early evolution of {\bf m11} begins with the rapid build up of a large ($> 10^8$\,\Msun) reservoir of gas due to the vigorous accretion of fresh gas from the IGM at rates 1--$10$\,\Msunyr.  As expected, early stellar growth is dominated by in situ star formation from freshly accreted, non-externally processed gas.  Interestingly, after the stellar component has grown to $M_* \gtrsim 10^8$\,\Msun~($z\sim 3$), stellar feedback has ejected enough gas from the galaxy that wind recycling begins to dominate gas accretion.  The contribution of recycled gas to in situ star formation increases further as the ISM cycles through periods of gas ejection and re-accretion.  Periods of intense star formation occur simultaneously with the presence of a large reservoir of gas, driving powerful outflows that sweep up a large fraction of gas in the ISM.  Wind recycling restitutes a substantial fraction of the ejected gas (\S\ref{sec:load}), producing distinct cycles in $M_{\rm gas}$ and $\dot{M}_{\rm acc}$.  This effect is most striking in {\bf m11} due to its extremely bursty star formation history, but this also occurs in more massive galaxies\footnote{Our lower mass dwarf {\bf m10} experiences frequent bursts of star formation decreasing $M_{\rm gas}$ by factors $\sim 3$--5, but these are not powerful enough to evacuate the ISM of the galaxy for extended periods of time (see Appendix~\ref{sec:appendix:m10}).}.  Indeed, a significant portion of every galaxy's evolution is dominated by intense bursts of star formation followed by powerful galactic outflows \citep{Muratov2015}.   

While the growth of {\bf m11} is clearly dominated by non-externally processed material, contributing $> 70$\,\% of $M_*$ at $z=0$, the stellar mass originated from gas preprocessed by other galaxies increases from $z \sim 2.5$ down to the present day.  Galaxy mergers contribute a total of $\sim 10^{7.5}$\,\Msun~directly in stars, corresponding to ex situ star formation, while ISM gas accreted through mergers fuels in situ star formation for a similar amount of stellar mass at $z=0$.  The total contribution of mergers represents only $\sim 6$\,\% of $M_*$ at $z=0$.  
By the time mergers begin to occur, the intergalactic transfer of gas via winds from other galaxies appears as a significant contribution to gas accretion onto {\bf m11} ($\sim 25$\,\%), persistently above the fresh accretion contribution\footnote{Note that gas provided by intergalactic transfer can also be ejected and recycled back onto {\bf m11} but we do not show here explicitly the EP wind recycling component (see Figure~\ref{fig:blocks}).} below $z \sim 1$.  Despite its quiescent merger history, intergalactic transfer fuels in situ star formation for $> 10^{8}$\,\Msun~in stars at $z=0$, about eight times more than the stellar mass provided by mergers in ex situ stars.
Nonetheless, wind recycling dominates the late time growth of {\bf m11}, providing $\sim 55$\,\% of its $z=0$ stellar mass.

\begin{figure*}
\begin{center}
\includegraphics[scale=0.5]{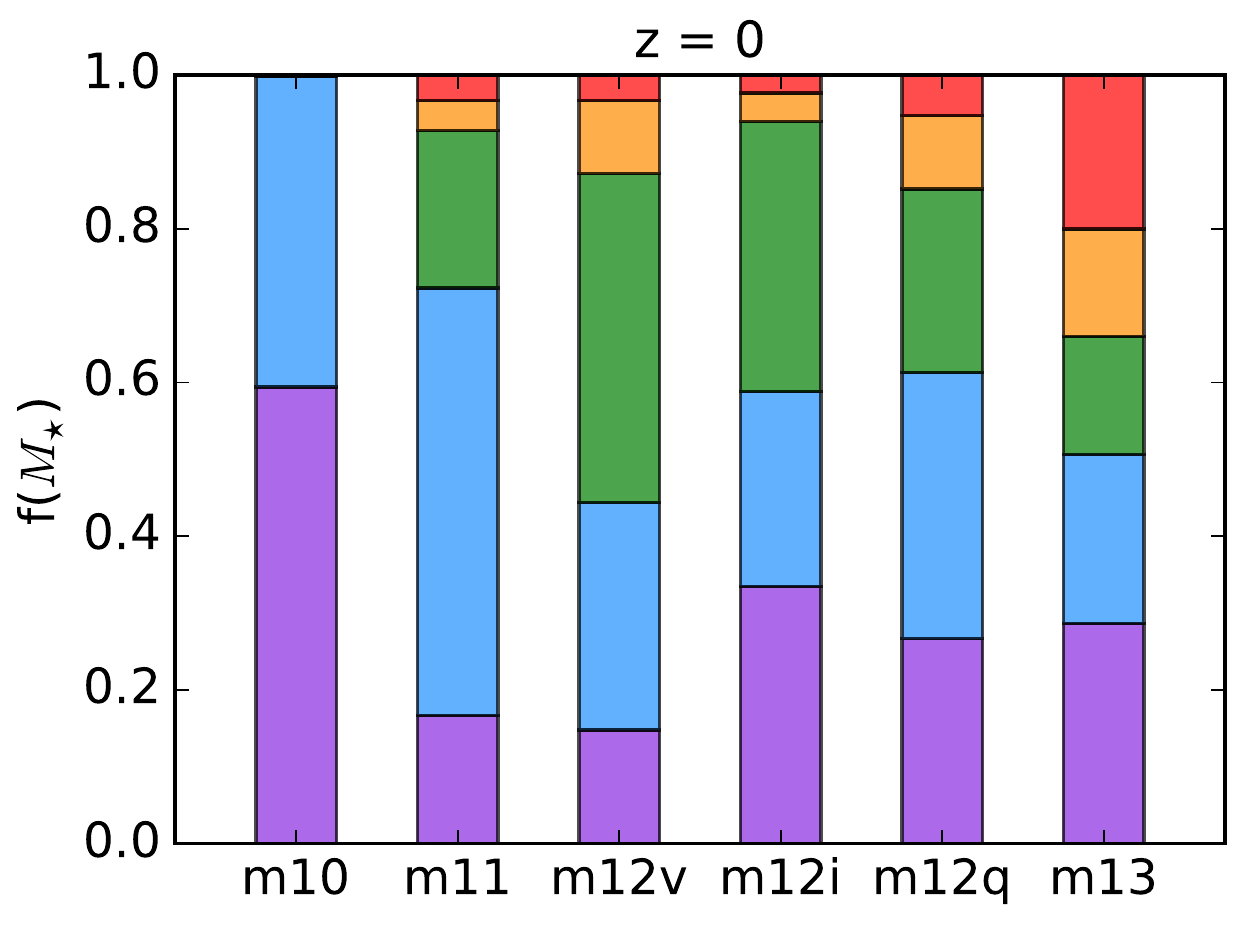}
\includegraphics[scale=0.5]{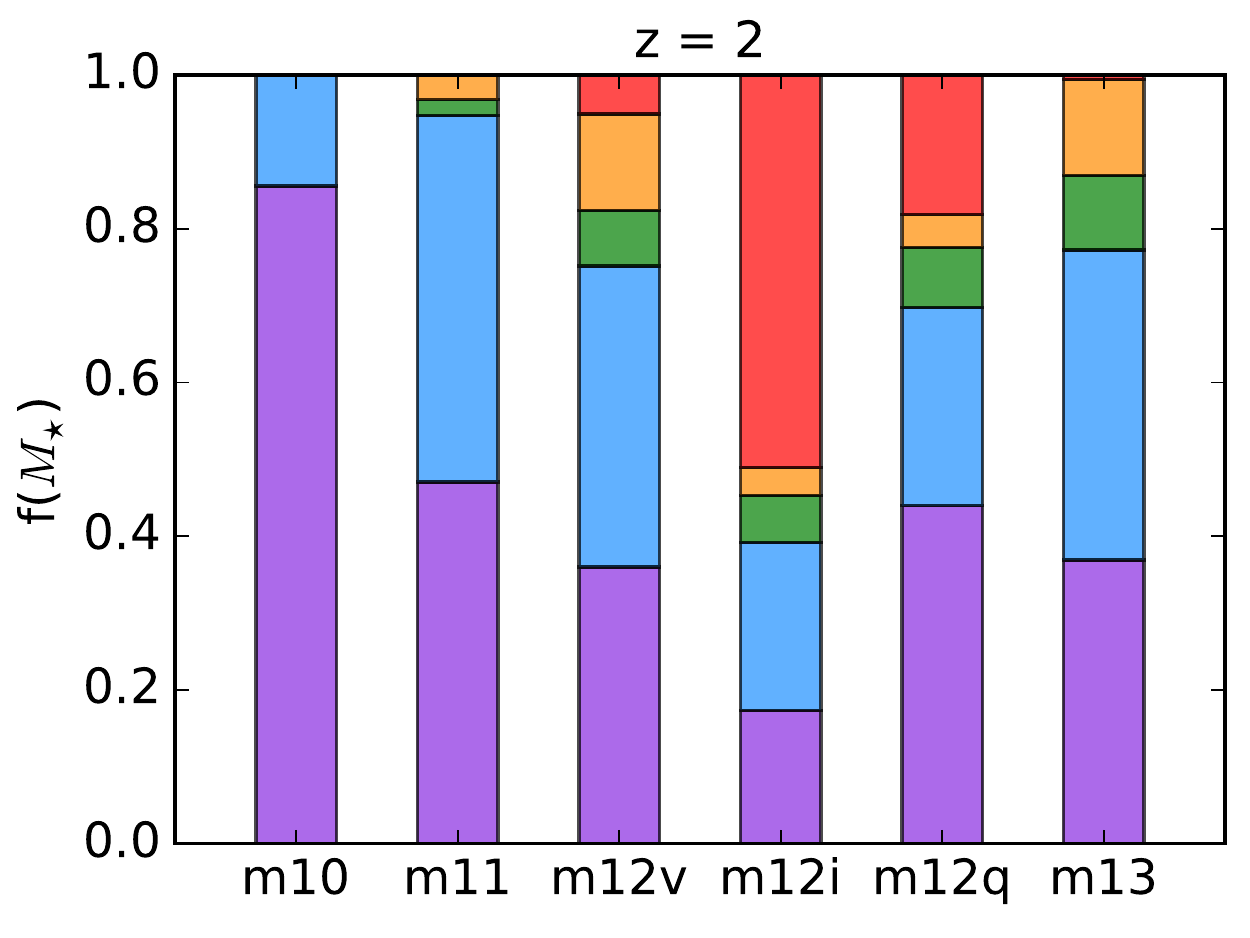}
\includegraphics[scale=0.3]{\pathI/legend.pdf}
\end{center}
\caption{Fraction of stellar mass at $z=0$ (left) and $z=2$ (right) contributed by fresh accretion (purple), wind recycling (blue), intergalactic transfer of gas (green), and galaxy mergers (ISM gas: orange; stars: red) for all simulated galaxies in order of increasing halo mass at $z=0$:
{\bf m10} ($M_{\rm halo} \approx 7.8\times10^{9}$\,\Msun), 
{\bf m11} ($M_{\rm halo} \approx 1.4\times10^{11}$\,\Msun), 
{\bf m12i} ($M_{\rm halo} \approx 1.1\times10^{12}$\,\Msun), 
{\bf m12q} ($M_{\rm halo} \approx 1.2\times10^{12}$\,\Msun), 
{\bf m12v} ($M_{\rm halo} \approx 6.3\times10^{11}$\,\Msun), and  
{\bf m13} ($M_{\rm halo} \approx 6.1\times 10^{12}$\,\Msun).
Wind recycling contributes more to the stellar component of lower mass systems.  The contribution from externally processed material (often dominated by intergalactic transfer) is higher in more massive systems.
The stellar mass of {\bf m12i} is dominated by ex situ stars from an ongoing merger at $z=2$.
}
\label{fig:bars}
\end{figure*}

\subsubsection{Milky Way-mass disk: {\bf m12i}}

Fresh accretion from the IGM also dominates the early stellar growth of our Milky Way-mass system {\bf m12i} (middle panel of Figure~\ref{fig:gmode}), a disk dominated galaxy at $z=0$. 
Recycling of gas previously ejected in massive outflows becomes comparable to fresh accretion at $z \sim 4$.  The contribution of wind recycling to $\dot{M}_{\rm acc}$ peaks at $z \sim 2$, feeding the ISM with gas at rates $\sim 20$\,\Msunyr.  Sufficiently massive galaxies ($M_{*} \gtrsim 10^{10}$\,\Msun) develop stable gaseous disks at late times and transition into a continuous and quiescent mode of star formation that does not drive large scale winds efficiently \citep{Muratov2015,Hayward2016}.  As a result, wind recycling steadily declines at lower redshifts, while the significant late time accretion of fresh gas from the IGM provides $\sim 3$\,\Msunyr~to {\bf m12i} at $z=0$. The total accretion rate onto the ISM ($\dot{M}_{\rm acc} \approx 4$\,\Msunyr~) is somewhat lower than the inflow rate at $0.25\,R_{\rm vir}$ \citep[$\sim 5$\,\Msunyr;][]{Muratov2015}.

The relatively quiescent history of this galaxy, disrupted only by one major merger ($z \sim 2$), illustrates the different contributions of externally processed material to the overall galaxy growth.  After an early period dominated by non-externally processed gas, the roughly equal-mass merger at $z \sim 2$ sharply rises the contribution of the merger-stellar component, providing $\sim 10^{9}$\,\Msun~directly in stars. Ex situ star formation briefly dominates $M_*$ at $z \sim 2$ but the merger-stellar component does not increase further due to the absence of prominent mergers at lower redshifts.  This major merger brings in about an equal amount of mass in ISM gas, decreasing afterwards due to consumption by star formation and the lack of substantial replenishment down to $z=0$. 

The transfer of gas via galactic winds provided $> 10^9$\,\Msun~of gas to the ISM of the galaxy around the time of the merger (about $1/3$ of $M_{\rm gas}$ at $z\sim2$), suggesting that a significant amount of gas ejected by the merging galaxy was retained in its parent halo before accreting onto {\bf m12i}.  Intergalactic transfer represents a major contribution to the gas and stellar content of {\bf m12i} at $z<2$, owing to significant accretion of gas ejected by other galaxies extending to late times, maintaining a gas supply of $\sim 1$\,\Msunyr~at $z=0$.  The stellar mass formed in situ from intergalactic transfer represents $\sim 35$\,\% of $M_*$ at $z=0$, while the total mass brought in by mergers represents $< 5$\,\% of the total stellar mass.

\subsubsection{Massive early-type galaxy: {\bf m13}}

The more massive, early-type galaxy {\bf m13} undergoes rapid growth at earlier times relative to our Milky Way-mass systems (right panel of Figure~\ref{fig:gmode}). By $z \sim 4$, fresh gas accreting at rates up to 200\,\Msunyr~had accumulated $\sim 10^{10}$\,\Msun~of gas in the ISM while feeding in situ star formation corresponding to $M_* \sim 10^{9.5}$\,\Msun.  By $z \lesssim 2$, the wind recycling contribution to $M_*$ becomes comparable to the fresh accretion component, and the total stellar mass from non-externally processed material remains roughly unchanged due to the decline in gas accretion rate at late times. 

Gas-rich galaxies merging with {\bf m13} at $z\sim [2.3, 3.0, 4.5]$ supply a fair amount of ISM gas, feeding in situ star formation that contributes $10\times$ more stellar mass than the ex situ stars provided by the merging galaxies.  Despite intergalactic transfer providing up to 10\,\Msunyr~of gas at $z \lesssim 3$, the stellar mass originated from gas preprocessed by other galaxies remains roughly $3\times$ below the non-externally processed contribution at $z>0.5$.
Subsequent mergers at $z\sim [0.2, 0.3, 0.4]$ modify the mass budget by providing $> 10^{10}$\,\Msun~in ex situ stars, which represent $\sim 20$\,\% of $M_*$ by $z=0$.  The intergalactic transfer component represents $\sim 15$\,\% of the total stellar mass of {\bf m13} at $z=0$, dominating the late time ($z < 0.5$) gas accretion.

\subsection{Galaxy growth components as a function of halo mass}\label{sec:all}

Figure~\ref{fig:bars} summarizes the relative contribution of different processes to the stellar mass of all simulated galaxies at $z=0$ (left) and $z=2$ (right), where we also include the isolated dwarf galaxy {\bf m10} and two additional Milky Way-mass galaxies, {\bf m12q} and {\bf m12v}.
With total stellar mass $M_* \approx 2\times10^6$\,\Msun~at $z=0$, the evolution of {\bf m10} is dominated by in situ star formation from non-externally processed gas, with negligible contribution of gas preprocessed by other galaxies.
The full redshift evolution of {\bf m10} is described in Appendix~\ref{sec:appendix:m10}, where we show that its late time gas accretion of $10^{-2}$--$10^{-3}$\,\Msunyr~is dominated by wind recycling.
Galaxy {\bf m12v} experiences a violent merger history resulting in the slight predominance of externally processed material at late times, owing primarily to persistent intergalactic transfer.  The stellar mass of {\bf m12q} is dominated by fresh accretion and wind recycling at all times, even after a late ($z < 0.5$) major merger that destroys its disk component. 

Despite their different merger histories, all three Milky Way-mass galaxies show a small contribution of ex situ stars from mergers at $z=0$ ($\lesssim 5$\,\%), clearly below the intergalactic transfer component ($> 25$\,\%).  However, note that while this seems a generic result, significant variations in the mass budget of galaxies may occur given the stochasticity of mergers. The $z=2$ contributions to $M_*$ show some clear differences relative to present day values: the stellar mass of {\bf m12i} is dominated by ex situ stars from an ongoing merger, while this contribution is negligible at $z=2$ for our higher mass galaxy {\bf m13}.  Moreover, the wind transfer component appears to be systematically lower at $z=2$ relative to $z=0$.  
We explore trends with halo mass and redshift further in the reminder of this section.

\subsubsection{Non-externally processed gas}\label{sec:ext}

\begin{figure}
\begin{center}
\includegraphics[scale=0.65]{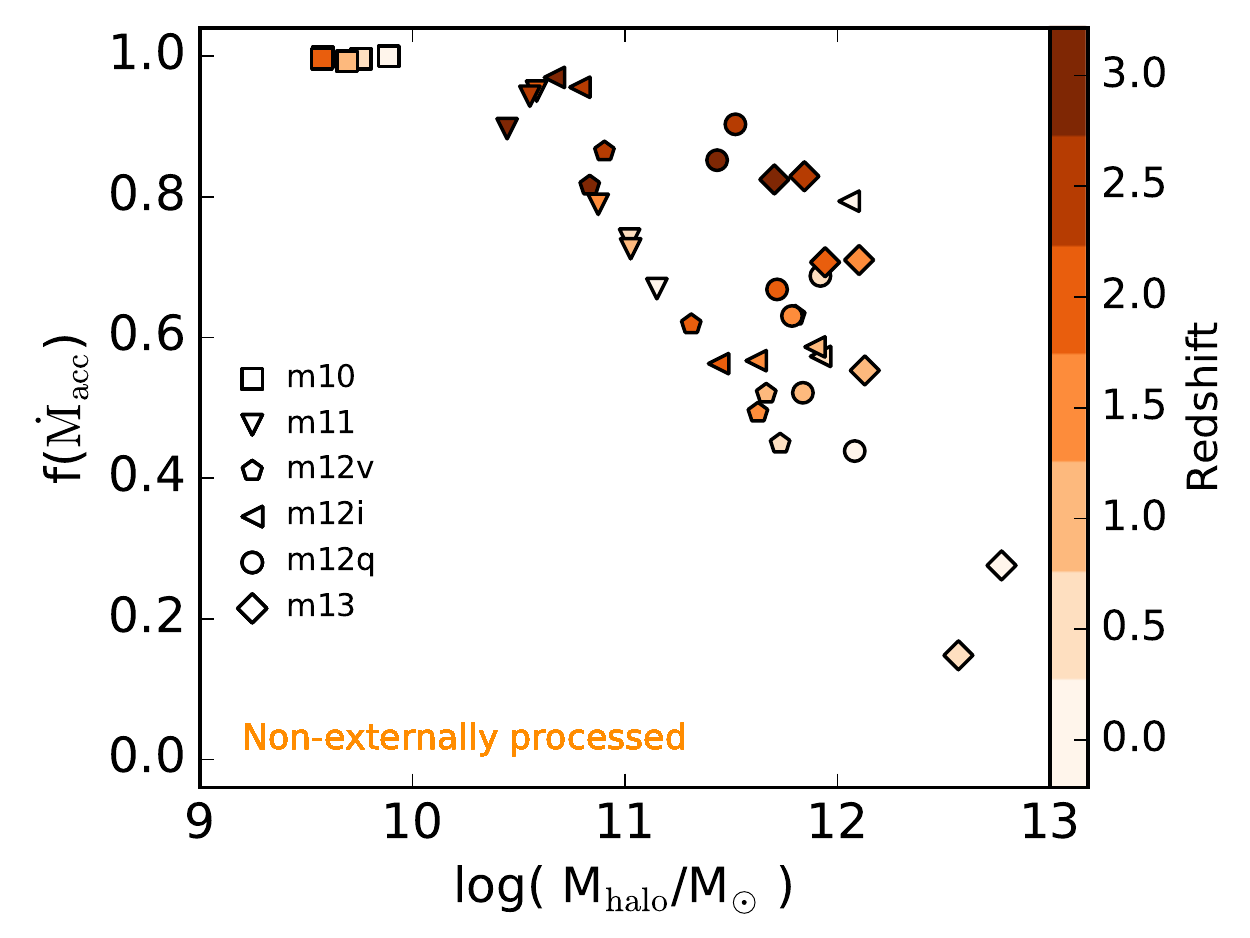}
\end{center}
\caption{Fraction of gas accretion rate onto the galaxy contributed by non-externally processed material as a function of halo mass.  The color scale indicates redshift evolution while different symbols correspond to each simulated galaxy.  
Non-externally processed gas dominates accretion onto sub-$L^*$ ($M_{\rm halo} < 10^{12}$\,\Msun) systems at all redshifts.
}
\label{fig:ext}
\end{figure}

\begin{figure}
\begin{center}
\includegraphics[scale=0.65]{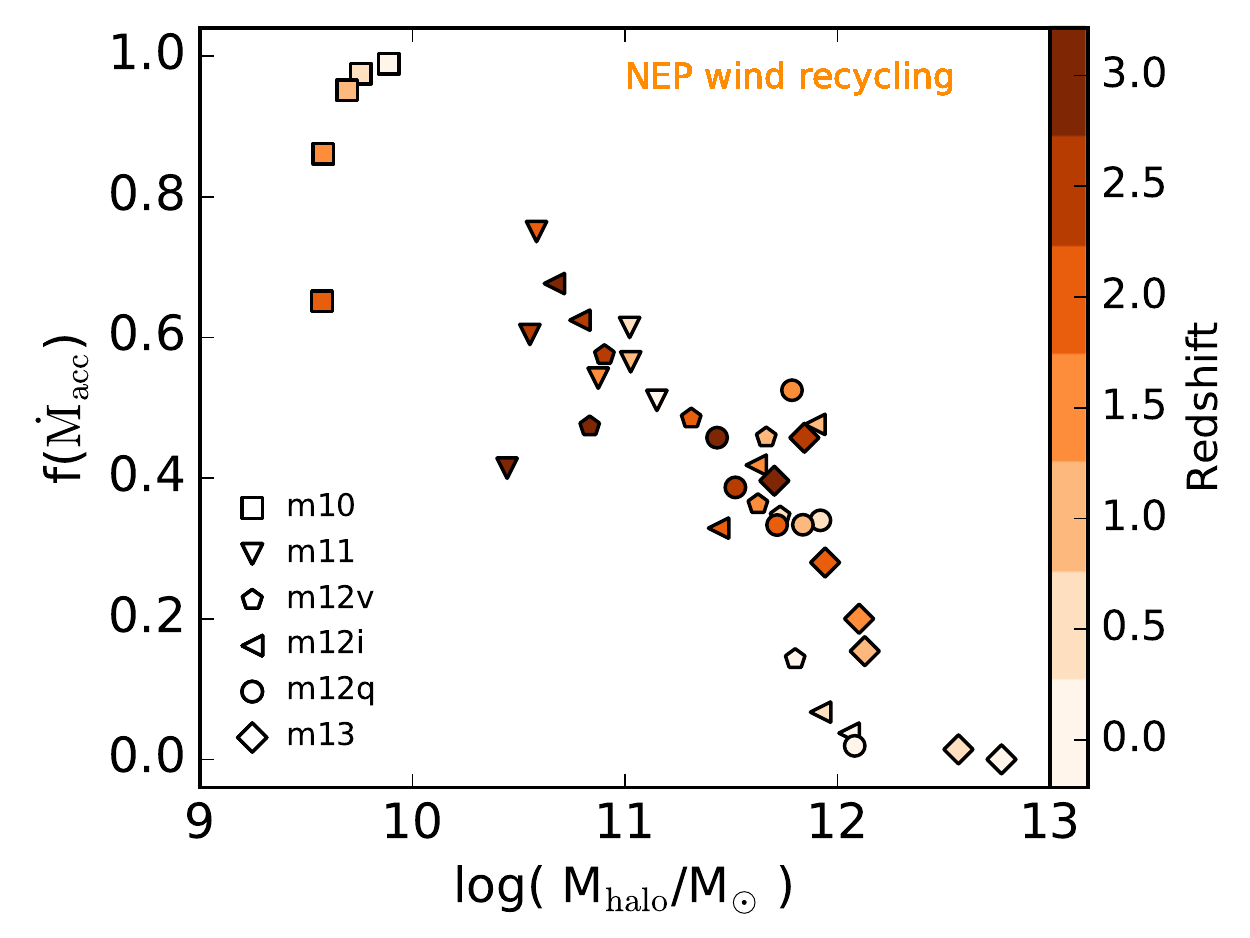}
\includegraphics[scale=0.65]{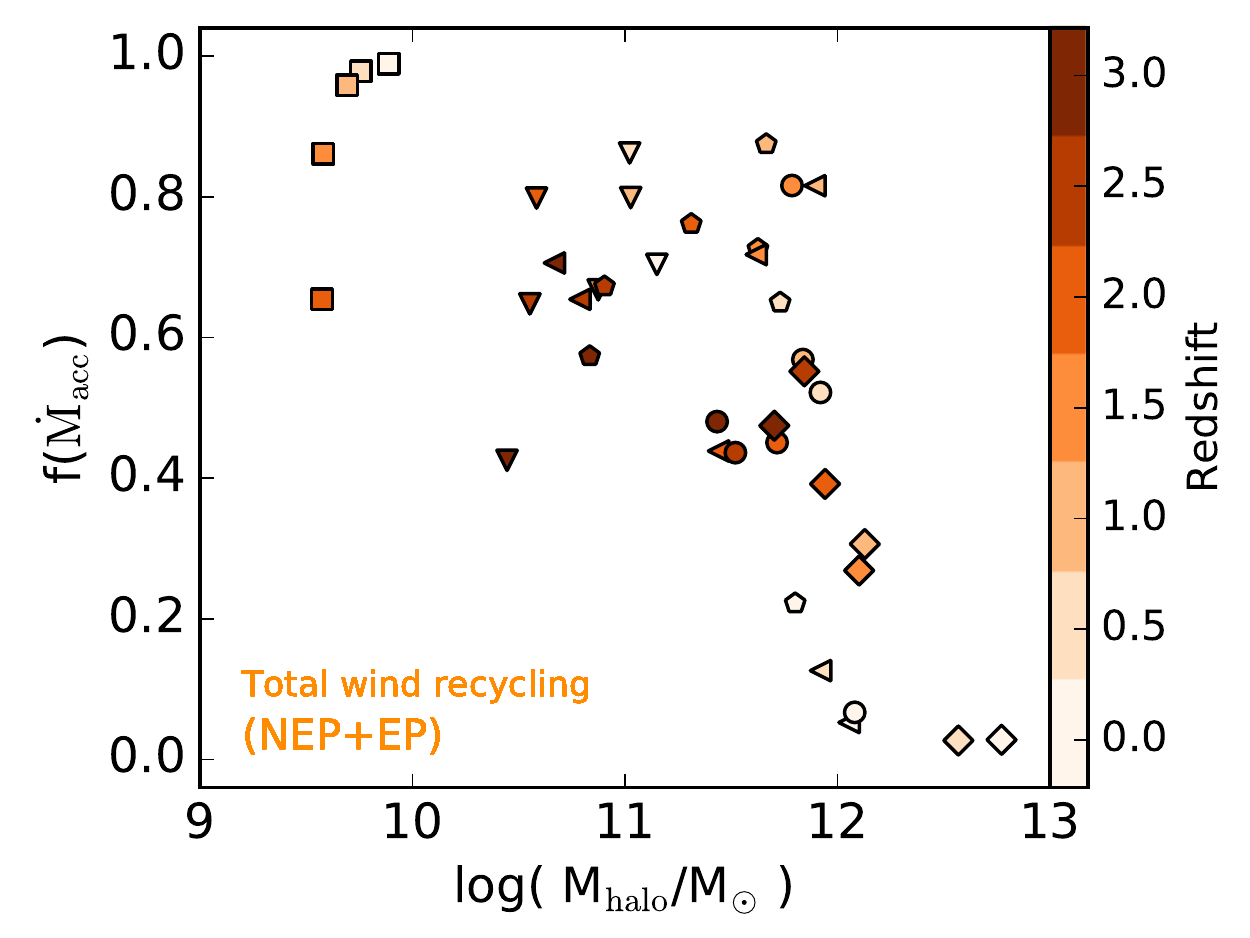}
\end{center}
\caption{Top: fraction of total gas accretion rate onto the galaxy (including ISM gas from mergers) contributed by the NEP wind recycling component as a function of halo mass and redshift (indicated by the color scale).  Bottom: same as top for total (NEP+EP) wind recycling, i.e. re-accretion of winds ejected from the central galaxy regardless of the original source of gas.    
Each symbol corresponds to a different simulated galaxy. 
Wind recycling contributes more in lower mass systems owing to larger mass loading factors and similar recycled-to-ejected ratios (see \S\ref{sec:load}).}
\label{fig:windrec}
\end{figure}

Figure~\ref{fig:ext} shows the contribution of non-externally processed material to the gas accretion rate onto galaxies as a function of halo mass, evaluated at different redshifts.  This includes fresh gas accretion directly from the IGM as well as NEP wind recycling (i.e. originally fresh gas ejected in winds recycling back onto the central galaxy). Despite significant scatter, the fraction of $\dot{M}_{\rm acc}$ contributed by non-externally processed gas decreases with (1) increasing halo mass at fixed redshift, and possibly also with (2) decreasing redshift at fixed halo mass.  This suggests that higher mass systems typically living in higher density regions receive more gas previously processed by other galaxies, increasingly so as the cosmological infall rate drops at lower redshifts.

\subsubsection{Wind recycling}\label{sec:windrec}

\begin{figure}
\begin{center}
\includegraphics[scale=0.65]{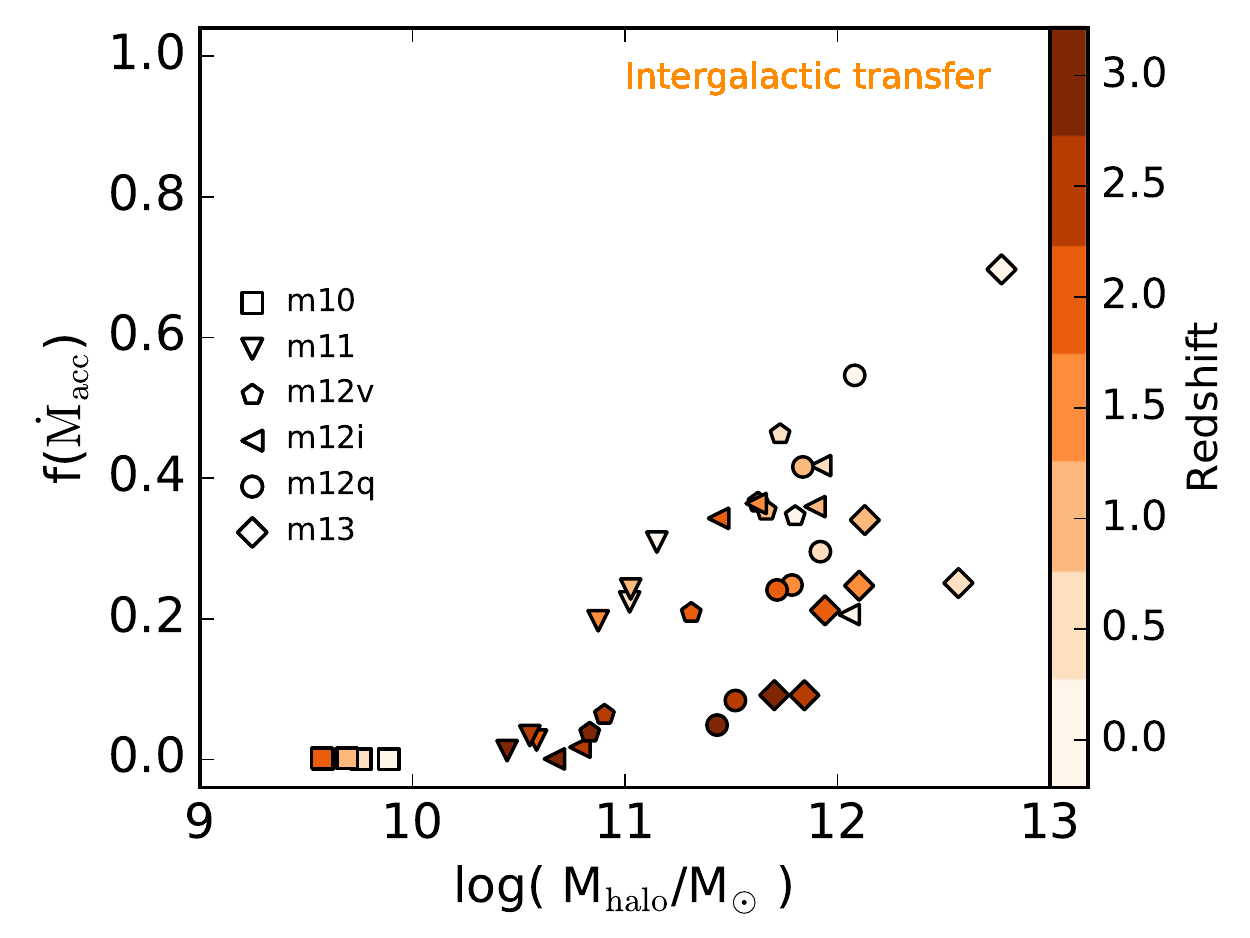}
\end{center}
\caption{Fraction of total gas accretion rate onto the galaxy (including ISM gas from mergers) contributed by the intergalactic transfer component as a function of halo mass and redshift (indicated by the color scale).  Each symbol corresponds to a different simulated galaxy. 
Intergalactic transfer contributes increasingly to higher mass systems at low redshift.}
\label{fig:windtrans}
\end{figure}

Figure~\ref{fig:windrec} shows the fraction of $\dot{M}_{\rm acc}$ provided by recycling of galactic winds back to the central galaxy as function of halo mass, evaluated at different redshifts for our six simulated galaxies.  
The top panel corresponds to the NEP wind recycling component, defined throughout this paper as the portion of non-externally processed gas recycled in winds (\S\ref{sec:windef}), while the bottom panel represents the total re-accretion of winds ejected from the central galaxy regardless of the original source of gas (including externally processed material).  In either case, the contribution of wind recycling to gas accretion decreases significantly with increasing halo mass, with the fraction of $\dot{M}_{\rm acc}$ at $z=0$ varying from order unity to zero across the halo mass range $10^{10}$--$10^{13}$\,\Msun.  This trend arises from the systematic decrease in wind mass loading factors in higher mass galaxies (\S\ref{sec:load}).  Moreover, wind recycling follows a non-trivial redshift evolution that depends on the star formation history of the central galaxy: (1) in higher mass systems, the peak of recycling is reached at earlier times, decreasing at lower redshifts owing to a transition into a quiescent mode of star formation that does not drive large scale winds efficiently \citep{Muratov2015,Hayward2016}, while (2) lower mass galaxies extend their bursty star formation histories down to lower redshift, continuously driving gas outflows that increase the recycling component. 
Figure~\ref{fig:param_Acc} shows that the fraction of $\dot{M}_{\rm acc}$ provided by recycling increases slightly when decreasing the minimum wind velocity by a factor of two, while the redshift and halo mass trends described here remain unaffected.

\subsubsection{Intergalactic transfer}\label{sec:trans}

\begin{figure}
\begin{center}
\includegraphics[scale=0.65]{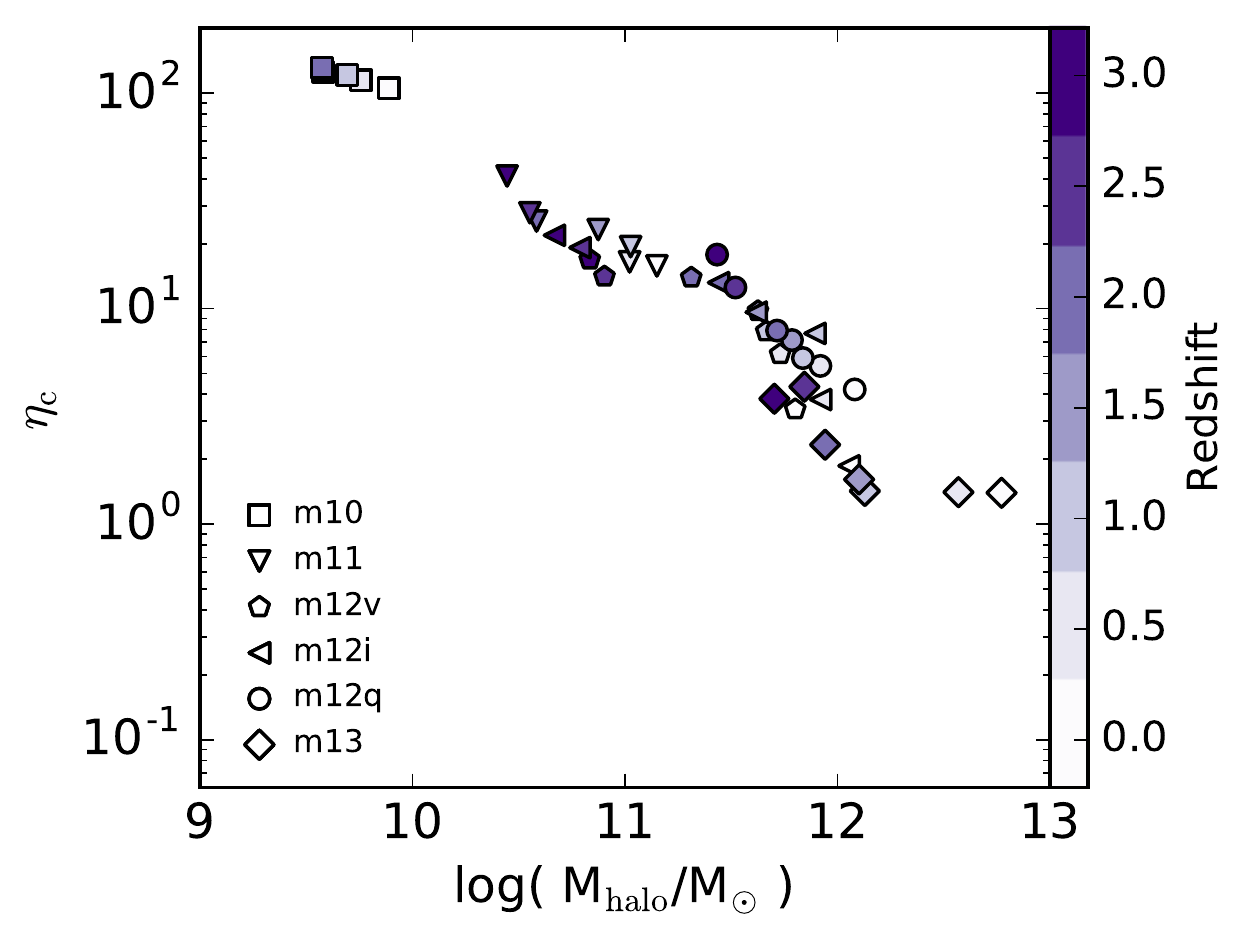}
\includegraphics[scale=0.65]{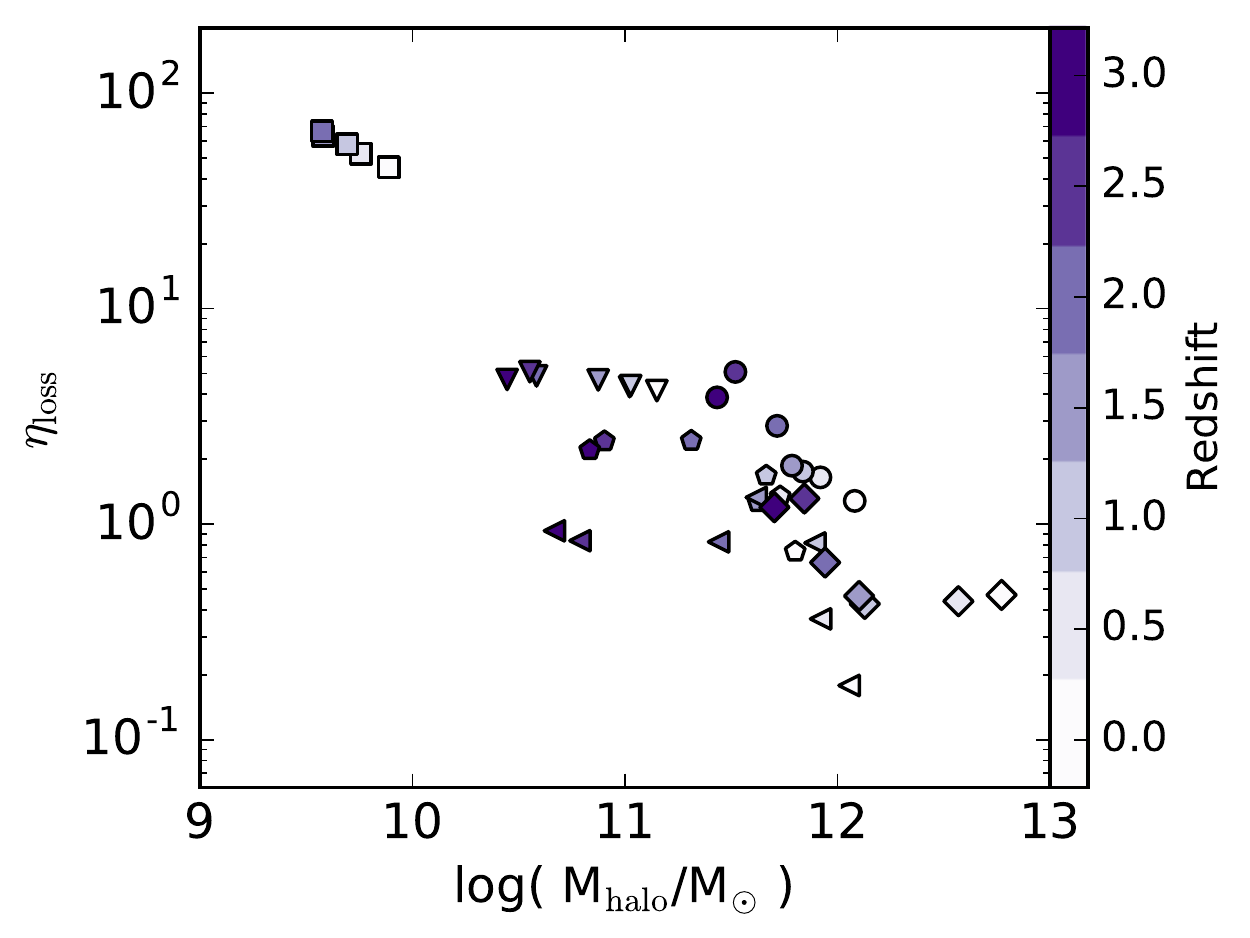}
\end{center}
\caption{Wind loading and mass loss from central galaxies.  
Top: cumulative wind loading factor $\eta_{\rm c}$ as a function of halo mass, evaluated at different redshifts as indicated by the color scale.  Each symbol corresponds to a different simulated galaxy.  For a given redshift, $\eta_{\rm c}$ is computed as the total gas mass ejected in winds divided by the total stellar mass formed.
Bottom: cumulative wind-loss factor as a function of halo mass and redshift, $\eta_{\rm loss}$, defined as the total gas mass lost in winds (i.e. gas ejected and not recycled prior to $z=0$) divided by the total stellar mass formed.}
\label{fig:windload}
\end{figure}

Figure~\ref{fig:windtrans} shows the contribution of intergalactic transfer to the overall gas accretion rate onto central galaxies as a function of halo mass and redshift.
The fraction of $\dot{M}_{\rm acc}$ contributed by intergalactic transfer correlates with halo mass, suggesting increasing exchange of mass between galaxies in higher density environments. This correlation is weak at high redshift, where intergalactic transfer represents $\lesssim 10\,\%$ of $\dot{M}_{\rm acc}$ for all galaxies, but becomes stronger at lower redshifts. 
The intergalactic transfer component increases at low redshift for a given halo mass, concurrently with the decrease in the overall contribution of non-externally processed gas.
Gas transferred by winds from other galaxies (previously largely neglected) is actually a major contributor to the ISM gas of all simulated galaxies at late times, with the exception of our isolated dwarf {\bf m10}, where the overall externally processed contribution is negligible.

\subsection{Statistics of galactic winds}\label{sec:winds}

\subsubsection{Wind loading factor and mass loss}\label{sec:load}

The top panel of Figure~\ref{fig:windload} examines the efficiency of stellar feedback at driving galactic winds by showing the cumulative wind loading factor ($\eta_{\rm c}$) as a function of halo mass and redshift.  For each galaxy, $\eta_{\rm c}(z_0)$ is computed as the ratio of the total gas mass ejected in winds from the ISM (regardless of its fate after ejection) to the total stellar mass formed in situ from early times down to $z = z_0$.
Cumulative mass loading factors are above unity for all simulated galaxies, ranging from $\eta_{\rm c} \gtrsim 1$ for {\bf m13} and $\eta_{\rm c} \gtrsim 3$ for the average of our three {\bf m12} galaxies at $z=0$, to $\eta_{\rm c} \approx 15$ and $\eta_{\rm c} \approx 100$ for our {\bf m11} and {\bf m10} galaxies, respectively. 
Despite the limited number of simulated galaxies, we find a clear trend of decreasing wind loading factors in higher mass halos.

The bottom panel of Figure~\ref{fig:windload} shows the wind-loss factor, $\eta_{\rm loss}(z_0)$, defined as the total gas mass lost in winds down to $z = z_0$ (i.e. gas ejected and never recycled) divided by the total stellar mass formed by that redshift.
With this definition, $\eta_{\rm loss}(z_0)$ represents the total gas mass deposited in the CGM and/or the IGM per unit stellar mass formed down to $z = z_0$.  Despite the large scatter relative to $\eta_{\rm c}$, we find a similar trend of decreasing $\eta_{\rm loss}$ with increasing halo mass.  Our dwarf galaxies {\bf m10} and {\bf m11} deposited $\sim 50$--5 times their $z=0$ stellar mass worth of gas in the CGM/IGM, while the mass loss in winds in higher mass galaxies is comparable to their stellar mass ($\eta_{\rm loss} \approx 0.6$ on average for our {\bf m12} and {\bf m13} galaxies).

\begin{figure}
\begin{center}
\includegraphics[scale=0.65]{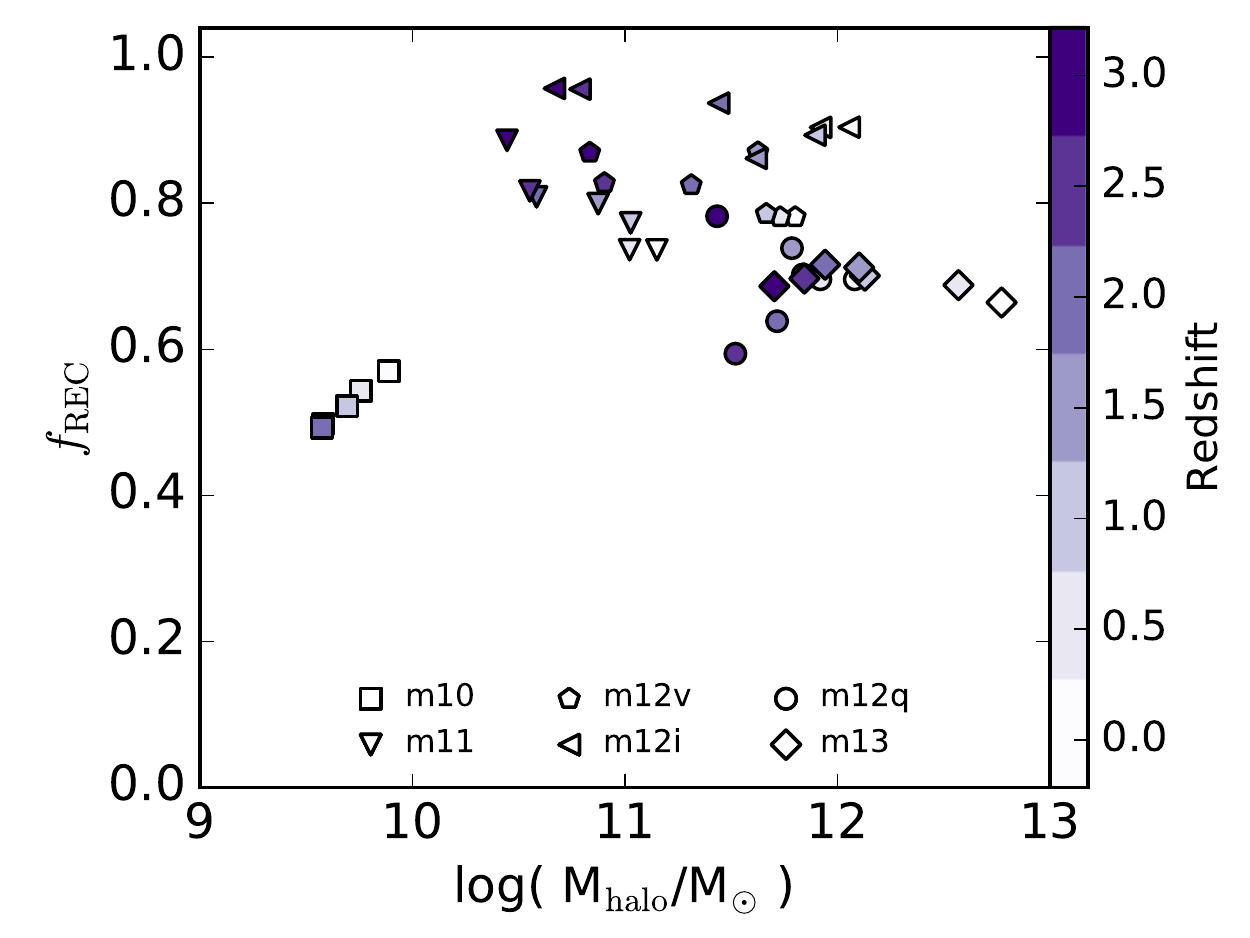}
\end{center}
\caption{Gas recycled to ejected ratio as a function of halo mass, $f_{\rm REC}(z_0)$, defined as the cumulative gas mass recycled divided by the cumulative gas mass ejected from early times down to $z=z_0$ (evaluated at different redshifts as indicated by the color scale).  Each symbol corresponds to a different simulated galaxy as in Figure~\ref{fig:windload}.}
\label{fig:recfrac}
\end{figure}

Figure~\ref{fig:recfrac} shows the cumulative wind recycled-to-ejected ratio as a function of halo mass and redshift, which we compute as $f_{\rm REC} \equiv (\eta_{\rm c} - \eta_{\rm loss})/\eta_{\rm c}$.  With this definition, $f_{\rm REC}(z_0)$ represents the fraction of total mass ejected in winds from early times down to $z=z_0$ that recycles prior to $z=0$, with the remaining mass deposited outside of the central galaxy.  We find that simulated galaxies recycle a significant fraction of the total mass ejected in galactic winds, with $f_{\rm REC} \sim 50$--95\,\% and no clear trend with either halo mass or redshift. Thus, while lower mass galaxies lose more mass in winds relative to their stellar growth (Figure~\ref{fig:windload}, bottom panel), their ability to recycle winds relative to the ejected mass is similar to higher mass galaxies.  As a result, wind recycling represents a larger contribution to the overall gas accretion onto lower mass galaxies (Figure~\ref{fig:windrec}).

\subsubsection{Recurrent wind recycling}\label{sec:wind_rec}

\begin{figure}
\begin{center}
\includegraphics[scale=0.45]{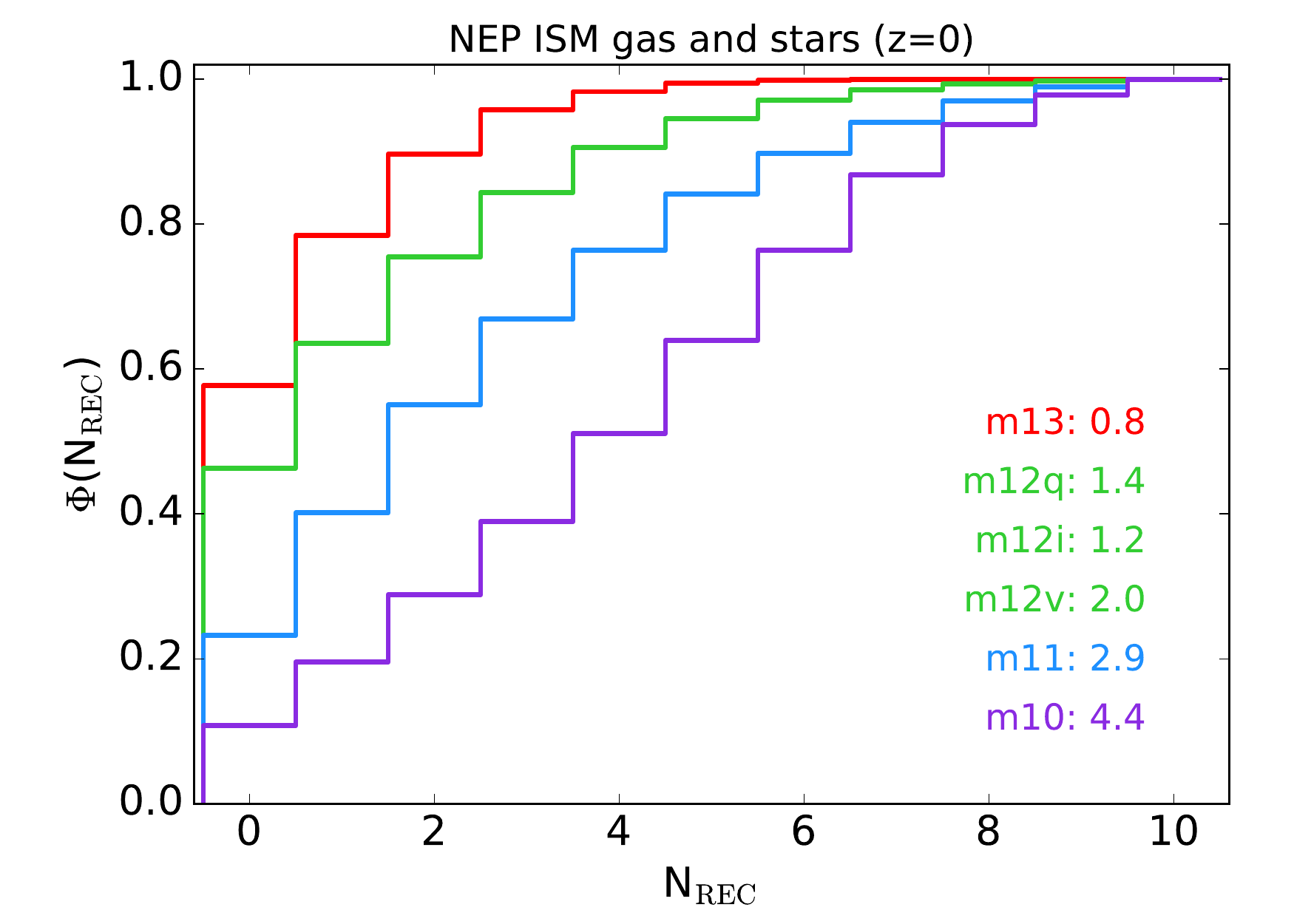}
\end{center}
\caption{Normalized cumulative distribution of the number of wind recycling events, N$_{\rm REC}$, experienced by the non-externally processed material (gas and stars) of simulated galaxies from early times down to $z=0$. N$_{\rm REC} = 0$ corresponds to gas (or stars formed from gas) never ejected from the central galaxy.
Different colors indicate galaxies {\bf m13} (red), {\bf m12q, m12i, m12v} (green), {\bf m11} (cyan), and {\bf m10} (blue), where only the average distribution is shown for the three {\bf m12} halos.  The average number of recycling times, indicated for each galaxy, increases in lower mass halos.
}
\label{fig:Nrec}
\end{figure}

\begin{figure}
\begin{center}
\includegraphics[scale=0.45]{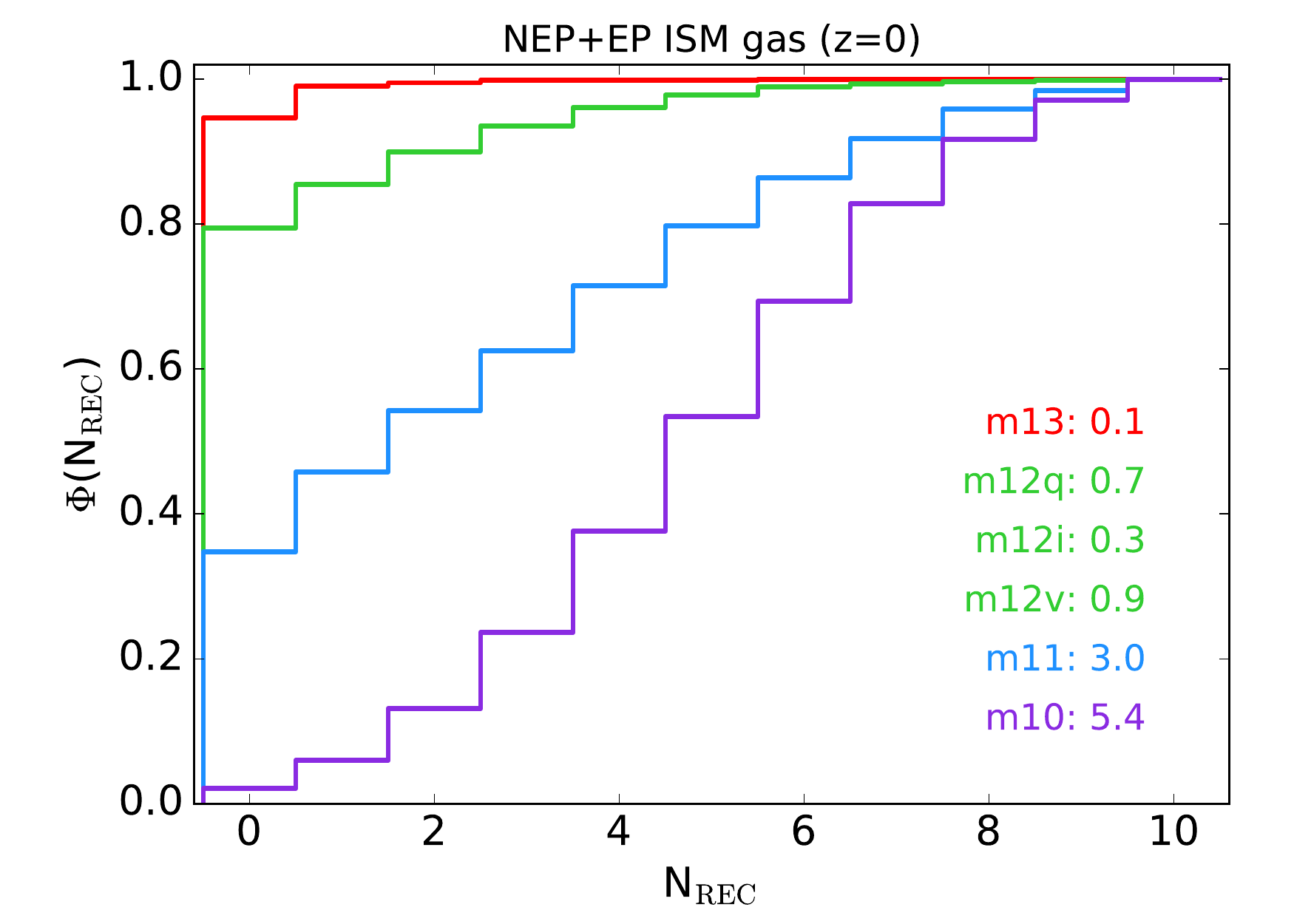}
\includegraphics[scale=0.45]{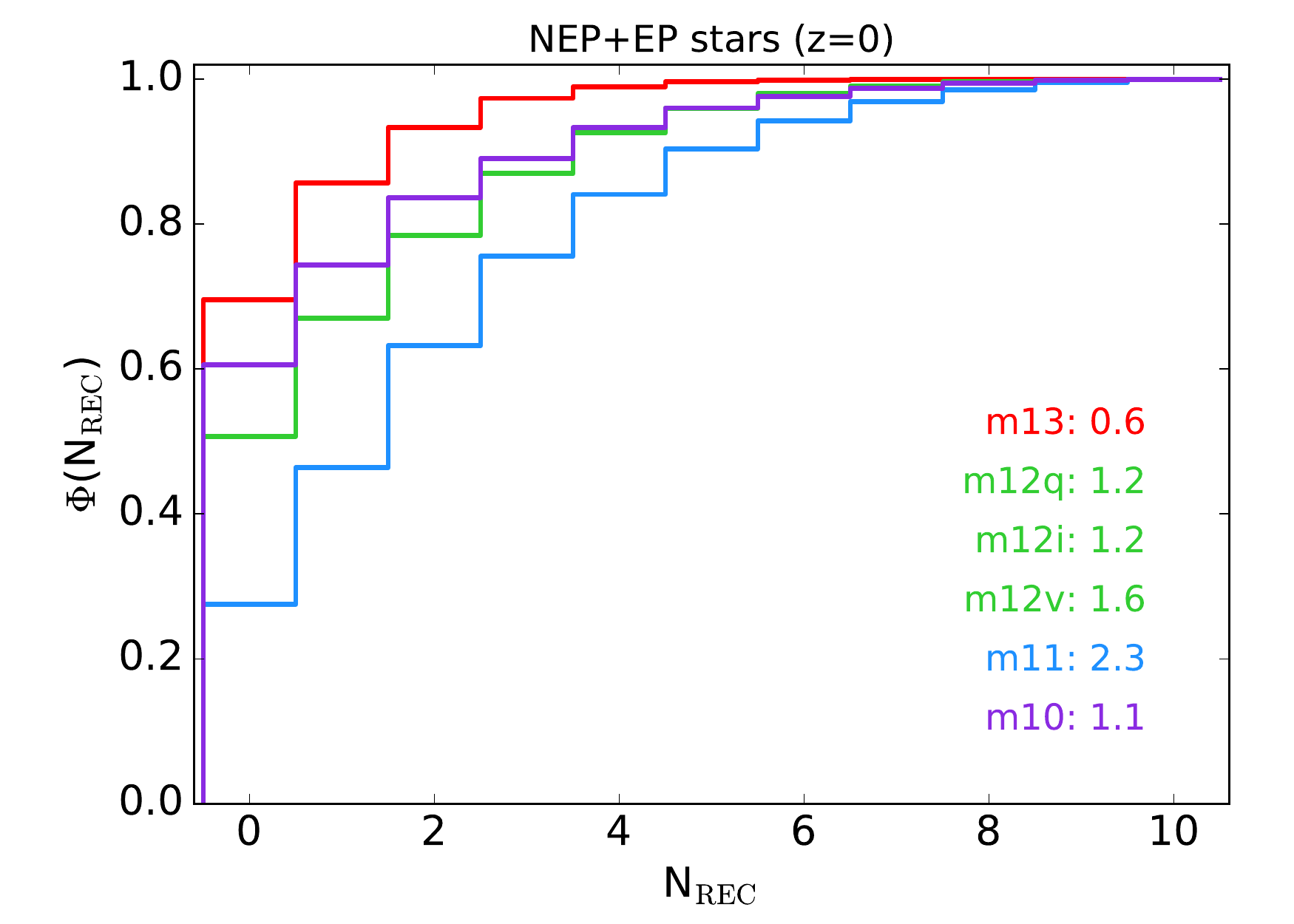}
\includegraphics[scale=0.45]{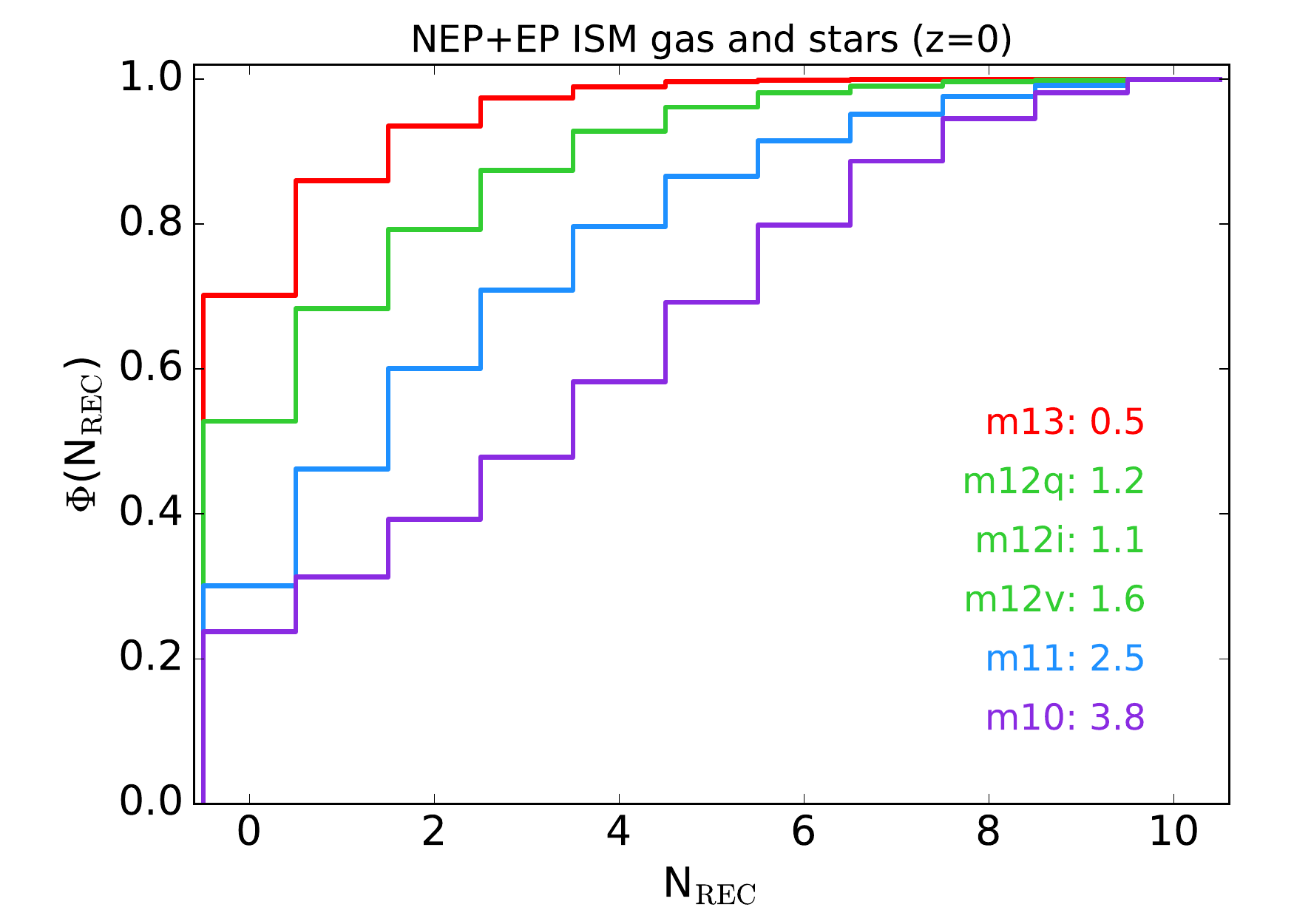}
\end{center}
\caption{Normalized cumulative mass in ISM gas (top), stars (middle), and gas+stars (bottom) in simulated galaxies at $z=0$ as a function of the number of recycling times, N$_{\rm REC}$. We include externally processed and non-externally processed material. 
Different colors indicate galaxies {\bf m13} (red), {\bf m12q, m12i, m12v} (green), {\bf m11} (cyan), and {\bf m10} (blue), where only the average distribution is shown for the three {\bf m12} halos.  Average N$_{\rm REC}$ values for the ISM and stellar components are indicated for each galaxy.
}
\label{fig:NrecM}
\end{figure}

Figure~\ref{fig:Nrec} explores the importance of recurrent wind recycling by showing the $z=0$ normalized cumulative mass of non-externally processed material (including gas and stars) as a function of the number of recycling events, N$_{\rm REC}$.  
Given that recycling occurring in other galaxies is neglected, the distribution of N$_{\rm REC}$ in terms of non-externally processed material provides a fair representation of the intrinsic efficiency of wind recycling in central galaxies.  We find that recurrent wind recycling is common in all simulated galaxies.  The average N$_{\rm REC}$ for non-externally processed material is systematically higher in lower mass systems, increasing from $\langle {\rm N}_{\rm REC} \rangle \sim 1 \rightarrow 4$ across the halo mass range $M_{\rm halo} \sim 10^{13}$\,\Msun\,$\rightarrow 10^{10}$\,\Msun.  
This can be explained by the bursty star formation histories of lower mass galaxies extending down to lower redshifts, driving frequent cycles of gas ejection and re-accretion events.

\begin{figure*}
\begin{center}
\includegraphics[scale=0.45]{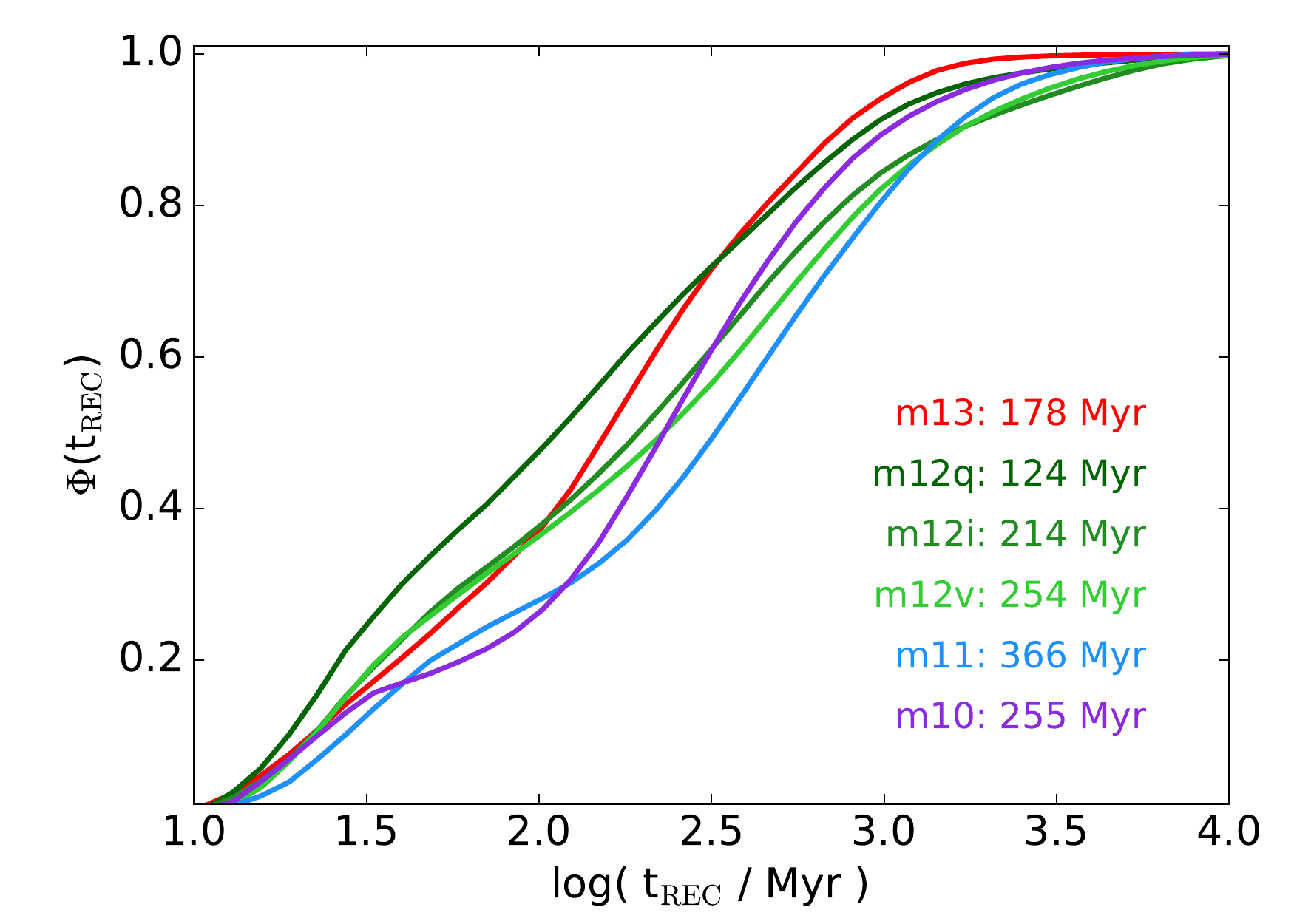}
\includegraphics[scale=0.45]{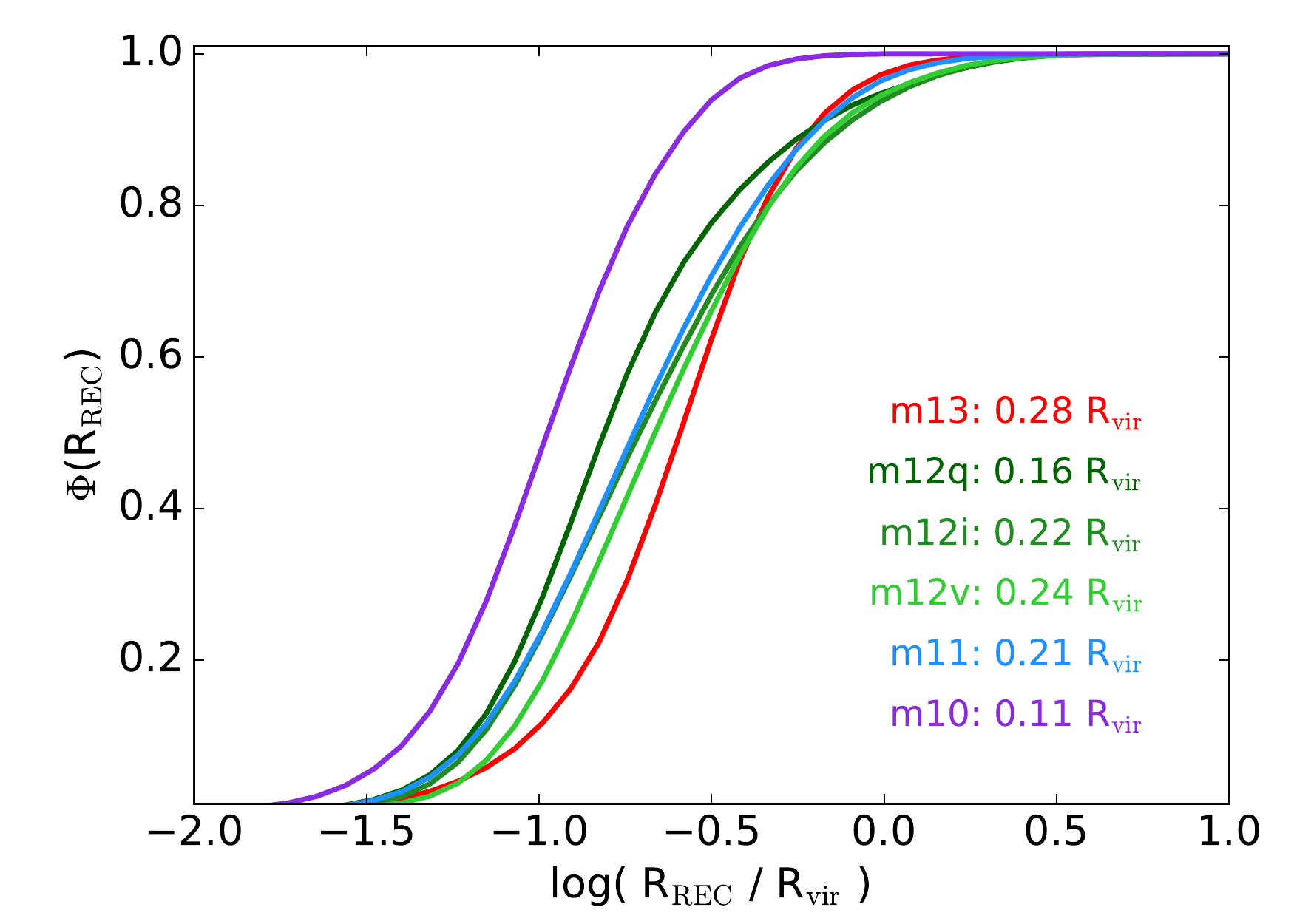}
\end{center}
\caption{Normalized cumulative distribution of wind recycling time (left) and distance (in units of $R_{\rm vir}$ at the time of ejection; right) for all recycling events occurring from $z=3$ down to $z=0$.  Each color corresponds to a different simulated galaxy.  Median recycling time and distance are indicated for each galaxy.
}
\label{fig:rec_hist}
\end{figure*}

Figure~\ref{fig:NrecM} examines the global incidence of recurrent wind recycling on the present day gas and stellar contents of galaxies.  We show the cumulative distribution of mass in ISM gas (top), stars (middle), and gas+stars (bottom) as a function of N$_{\rm REC}$, now including externally processed material as well.
The distribution of N$_{\rm REC}$ for ISM gas at $z=0$ is qualitatively similar to that of the non-externally processed material in Figure~\ref{fig:Nrec}, but shows a stronger trend with halo mass.  Most of the existing ISM gas in our dwarf galaxies {\bf m11} and {\bf m10} has been ejected and recycled back multiple times, with e.g. $\gtrsim 40$\,\% of the gas in {\bf m10} having cycled in and out of the galaxy six times or more.  In contrast, only $\sim 5$--20\,\% of the ISM gas content of our higher mass systems at $z=0$ has been ejected from the central galaxy at least once and recycled back.  
This is expected, owing to the rapid decline in gas accretion rate from wind recycling at late times in our {\bf m12} and {\bf m13} galaxies (Figure~\ref{fig:windrec}), caused by the lower efficiency of winds driven in high-mass galaxies at low redshift \citep[see also][]{Muratov2015}. Most present day gas content of our {\bf m12} and {\bf m13} galaxies comes from fresh gas accretion and intergalactic transfer at late times (Figure~\ref{fig:gmode}), well after the end of the epoch of powerful outflows, while our dwarf galaxies {\bf m11} and {\bf m10} continue to eject gas in winds promoting recycling and increasing N$_{\rm REC}$ down to $z=0$.

The middle panel of Figure~\ref{fig:NrecM} shows that $\sim 30$\,\%, 50\,\%, and 70\,\% of the total stellar mass of {\bf m13}, {\bf m12}, and {\bf m11} formed from gas recycled at least once by the central galaxy, respectively.  Interestingly, only $\sim 40$\,\% of the $z=0$ stellar mass of {\bf m10} comes from wind recycling while its ISM gas content has cycled through the galaxy more than five times on average.  As we show in Figure~\ref{fig:m10}, most of the stellar mass growth of {\bf m10} occurs at $z>2$ fueled by fresh accretion. The low level of star formation at $z<2$ generates winds continuously while growing the stellar content very little. 
When we consider the total baryonic content of galaxies at $z=0$ (bottom panel), we find an $N_{\rm REC}$--$M_{\rm halo}$ anticorrelation similar to that of the non-externally processed material in Figure~\ref{fig:Nrec}.

\subsubsection{Recycling distance and timescale}\label{sec:rec_dist}

Figure~\ref{fig:rec_hist} quantifies the recycling timescales ($t_{\rm REC}$; left) and distances ($R_{\rm REC}$; right) in our simulations.  For each gas recycling event, we define $t_{\rm REC}$ as the time interval between ejection from the central galaxy and subsequent re-accretion onto its ISM, while $R_{\rm REC}$ is defined as the maximum radial distance from the galaxy center reached by outflowing gas prior to recycling.  For each simulated galaxy, we compute $t_{\rm REC}$ and $R_{\rm REC}$ for all recycling events from $z=3$ down to $z=0$ and show their cumulative distributions in Figure~\ref{fig:rec_hist}. Gas ejected and not recycled prior to $z=0$ does not enter into the analysis, i.e. the distributions shown for $t_{\rm REC}$ and $R_{\rm REC}$ correspond specifically to the fraction $f_{\rm REC}$ (\S\ref{sec:wind_rec}) of outflowing material that does recycle.

We find a broad range of recycling times, extending from $\sim 10$\,Myr to $\sim 1$\,Gyr with roughly constant number of recycling events per logarithmic interval in $t_{\rm REC}$.
Note that the lower end of the distribution may be affected by the available number and distribution of output snapshots, which imposes a redshift-dependent limit to the minimum $t_{\rm REC}$ that can be measured from the simulations.  Recycling events with $t_{\rm REC} > 100$\,Myr are captured at all redshifts, while recycling on shorter timescales may be missed at low redshift where the snapshot frequency is lower ($\sim 20$--70\,Myr at $z < 1$).  Median recycling times range from $\sim 100$--350\,Myr, with no clear halo mass dependence. There is some indication for somewhat longer recycling times in our dwarf galaxies {\bf m11} and {\bf m10} relative to higher mass systems, but note the scatter in $t_{\rm REC}$ for our {\bf m12} galaxies.  At the upper end of the distribution, we find that $\sim 10$\,\% of recycling events may last $>1$\,Gyr.  The right panel of Figure~\ref{fig:rec_hist} indicates that most recycling occurs within the virial radius of the host dark matter halo, with the median recycling distance ranging from $\sim 0.3\,R_{\rm vir}$ for {\bf m13} to $\sim 0.1\,R_{\rm vir}$ for {\bf m10} (where $R_{\rm vir}$ is measured at the time of ejection for each recycling event).  We find that $\lesssim 5$\,\% of the gas recycled by our simulated galaxies crossed $R_{\rm vir}$ (at the time of ejection) before returning to the central galaxy.

\begin{figure}
\begin{center}
\includegraphics[scale=0.6]{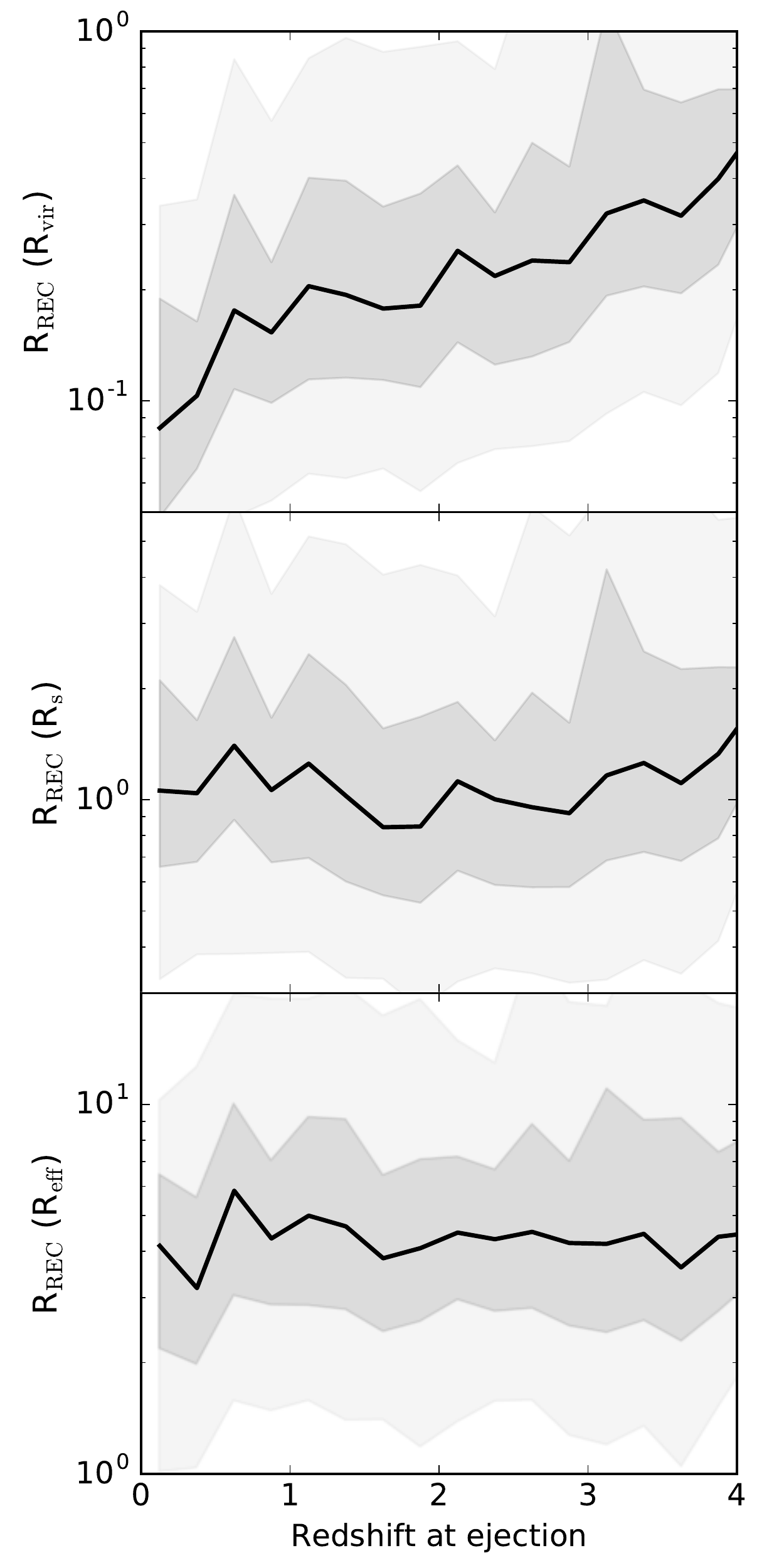}
\end{center}
\caption{Recycling distance as a function of redshift at ejection, expressed in units of the virial radius ($R_{\rm vir}$; top), the halo scale radius ($R_{\rm s}$; middle), and the galaxy stellar effective radius ($R_{\rm eff}$; bottom) at the time of ejection. Black solid lines show median values within equally-spaced redshift bins, including recycling events from all simulated galaxies.  Gray shaded areas indicate the 25\%--75\% and 5\%--95\% percentile ranges.
}
\label{fig:Rrec}
\end{figure}

Figure~\ref{fig:Rrec} explores the redshift dependence of the maximum distance reached by outflows prior to recycling.
The top panel shows that the median recycling distance decreases systematically with redshift relative to $R_{\rm vir}$, from $R_{\rm REC} \sim 0.6\,R_{\rm vir}$ to $\sim 0.1\,R_{\rm vir}$ in the redshift range $z=4\rightarrow0$.  This motivates comparing $R_{\rm REC}$ to other characteristic scales.
The middle and bottom panels of Figure~\ref{fig:Rrec} show the evolution of $R_{\rm REC}$ relative to the halo scale radius and the galaxy stellar effective radius at the time of ejection.  We compute the scale radius $R_{\rm s}\equiv R_{\rm vir}/c_{\rm vir}$ (where $c_{\rm vir}$ is the concentration parameter) using the halo-concentration model of  \citet{Diemer2015}, implemented in {\sc Colossus}\footnote{See \url{http://bdiemer.bitbucket.org}.}.
Most recycling occurs within 1--20\,$R_{\rm eff}$, where $R_{\rm eff}$ is the stellar effective radius, emphasizing that gas that will recycle typically remains close to the central galaxy after ejection.  Despite the large scatter, we find a typical recycling distance $R_{\rm REC} \sim R_{\rm s} \sim 5\,R_{\rm eff}$ roughly independent of halo mass and redshift, suggesting a characteristic {\it recycling zone} around galaxies that scales with the size of the inner halo and the galaxy's stellar component.

\section{Discussion}\label{sec:dis}

\subsection{Milky Way-mass galaxies}

We begin by summarizing the relative importance of different processes for the present day stellar content of Milky Way-mass galaxies, which we obtain by averaging over our three {\bf m12} galaxies.
The non-externally processed contribution represents $\sim 50$\,\% of the stellar mass of Milky Way-mass galaxies at $z=0$; this represents the portion of galaxy growth occurring in isolation, from material unaffected by other galaxies.  
Star formation fueled by fresh accretion accounts for $\sim 20$\,\% of the stellar mass of Milky Way size galaxies at $z=0$, while the remaining  $\sim 30$\,\% of non-externally processed material corresponds to recycling of their own wind ejecta.  

The remaining $\sim 50$\,\% of the stellar mass originates from externally processed material, i.e. gas and stars that belonged to a different galaxy at earlier times.
Mergers represent a small contribution to the growth of Milky Way-mass galaxies, providing $\sim 10$\,\% of the $z=0$ stellar mass, where $\lesssim 5$\,\% corresponds to direct contribution in stars while the remaining mass is provided in the form of ISM gas that fuels star formation.  
The externally processed contribution is dominated by intergalactic transfer, which may represent up to $\sim 40$\,\% of the stellar mass at $z=0$.  

Regardless of the original source of gas (i.e. externally processed or not), Milky Way-mass galaxies eject in winds and recycle $\sim 50$\,\% of the gas that makes up their stellar content at $z=0$.  Considering also the gas transferred from other galaxies via winds, $\gtrsim 70$\,\% of the stellar mass of Milky Way size galaxies today may have been in the form of galactic winds in the past.  Meanwhile, Milky Way-mass galaxies deposit as much gas in their CGM and the IGM via winds as their present day stellar content.

\newcommand{\sfig}{0.325}

\begin{figure*}
\begin{minipage}{0.495\textwidth}
\raggedright
    \includegraphics[width=1.\textwidth]{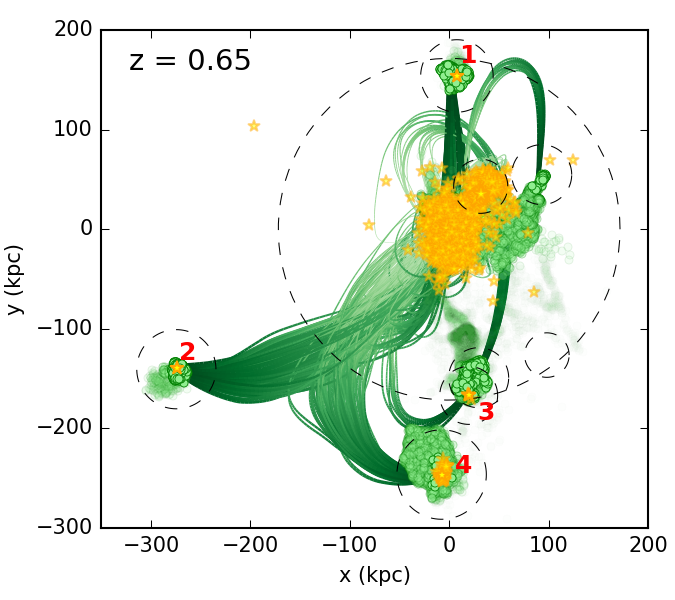}
\end{minipage}
\begin{minipage}{0.5\textwidth}
\raggedleft
\includegraphics[width=\sfig\textwidth]{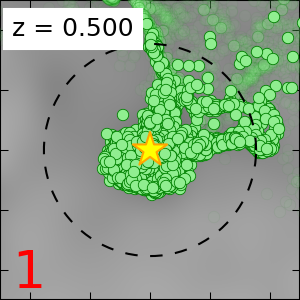}
\includegraphics[width=\sfig\textwidth]{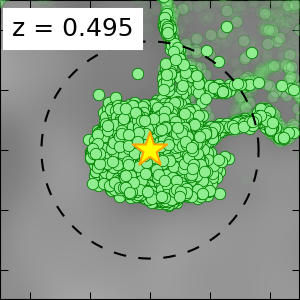}
\includegraphics[width=\sfig\textwidth]{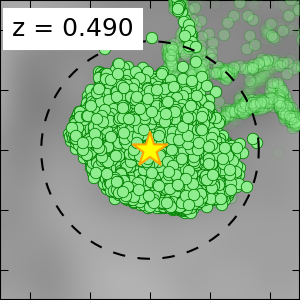}
\includegraphics[width=\sfig\textwidth]{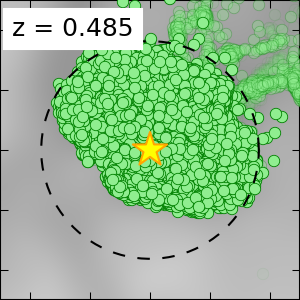}
\includegraphics[width=\sfig\textwidth]{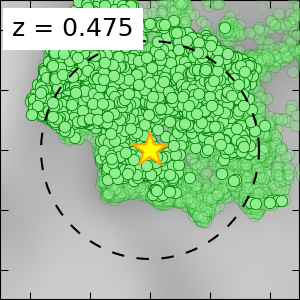}
\includegraphics[width=\sfig\textwidth]{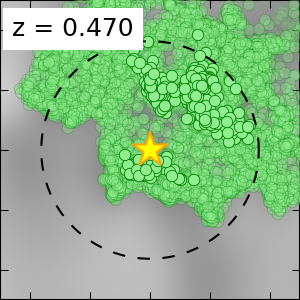}\\
\end{minipage}

\renewcommand{\sfig}{0.162}
\vspace{0.1cm}
\includegraphics[width=\sfig\textwidth]{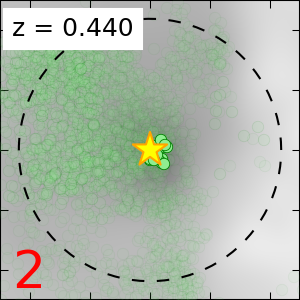}
\includegraphics[width=\sfig\textwidth]{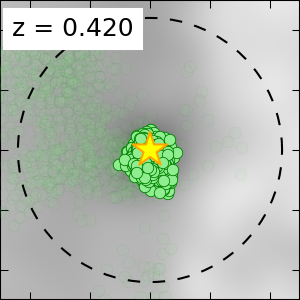}
\includegraphics[width=\sfig\textwidth]{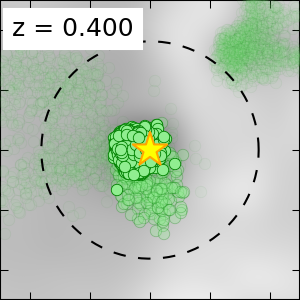}
\includegraphics[width=\sfig\textwidth]{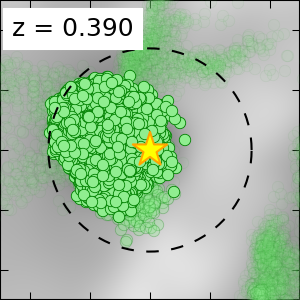}
\includegraphics[width=\sfig\textwidth]{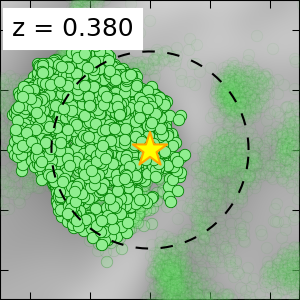}
\includegraphics[width=\sfig\textwidth]{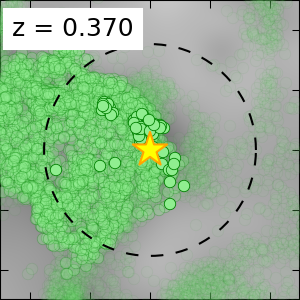}\\
\vspace{0.4cm}
\includegraphics[width=\sfig\textwidth]{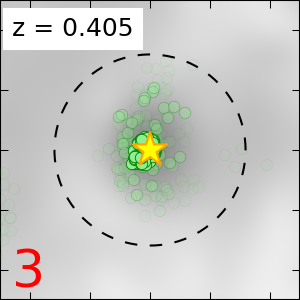}
\includegraphics[width=\sfig\textwidth]{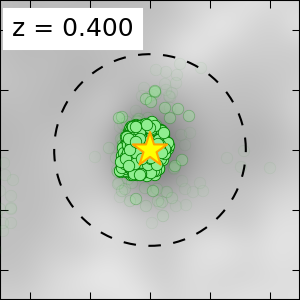}
\includegraphics[width=\sfig\textwidth]{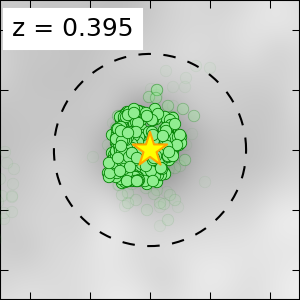}
\includegraphics[width=\sfig\textwidth]{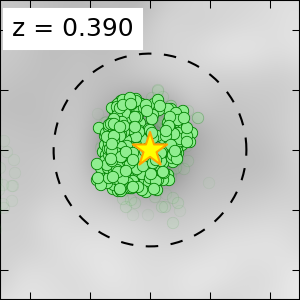}
\includegraphics[width=\sfig\textwidth]{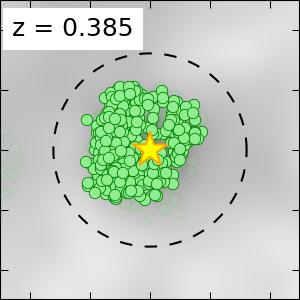}
\includegraphics[width=\sfig\textwidth]{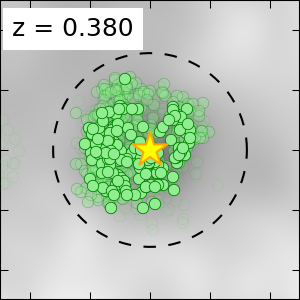}\\
\vspace{0.4cm}
\includegraphics[width=\sfig\textwidth]{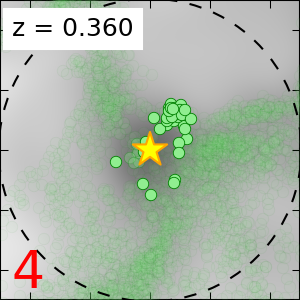}
\includegraphics[width=\sfig\textwidth]{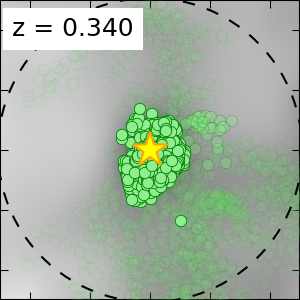}
\includegraphics[width=\sfig\textwidth]{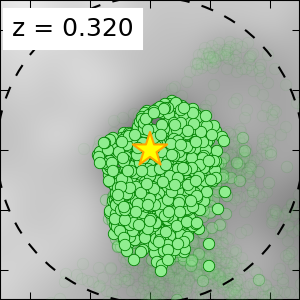}
\includegraphics[width=\sfig\textwidth]{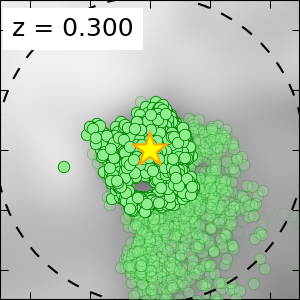}
\includegraphics[width=\sfig\textwidth]{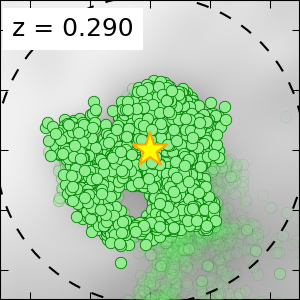}
\includegraphics[width=\sfig\textwidth]{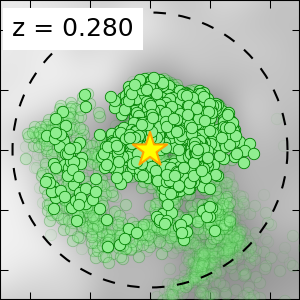}
\caption{Intergalactic transfer of gas from dwarf satellites onto galaxy {\bf m12i}  ($M_{\rm halo}(z=0) \approx 1.2\times10^{12}$\,\Msun).  Green lines in the top left panel show the trajectory of gas removed from the ISM of satellites after $z=0.65$ as it smoothly accretes onto the central galaxy (as in Figure~\ref{fig:flows}). 
Orange star symbols represent the stellar distribution and black dashed lines show the virial radius of identified (sub-)halos.
Thumbnails show a representative portion of the evolution of the four individual satellites numbered in the top left panel, each representing a [50\,kpc]$^3$ volume (physical) centered on the satellite.  
The gray scale shows projected gas density distributions (logarithmically scaled).
Green circles indicate the location of transfer gas particles in the vicinity of each satellite, with their color gradually vanishing after being removed from the source galaxy for a time $\sim 200$\,Myr. The star symbol indicates the center of the satellite galaxy and the black dashed line indicates $0.5\,R_{\rm vir}$ of the corresponding sub-halo at each redshift.  
Satellites experience multiple bursts of star formation as they orbit around the central galaxy, driving quasi-spherical outflows that dominate the intergalactic transfer component.}
\label{fig:thumbs}
\end{figure*}

\subsection{Intergalactic transfer}{\label{sec:windtrans}}

A significant result of this study is the identification of intergalactic gas transfer as a primary contributor to galaxy mass assembly.  Our analysis of gas particle trajectories reveals that the smooth exchange of gas between galaxies originates primarily from galactic winds, which we refer to as wind transfer.  
Gas ejected from the ISM of another galaxy can be retained in its CGM or reach the IGM before smoothly accreting onto the central galaxy (see Figure~\ref{fig:flows}).  
Massive galaxies are built hierarchically with large contributions of externally processed material. 
Stellar feedback removes material from lower mass galaxies, reducing the overall contribution of mergers in higher mass systems but increasing the amount of intergalactic gas transfer available for star formation  \citep{Oppenheimer2008,Keres2009_ObsProp,Oppenheimer2010}. 
Even if AGN feedback prevents late time star formation in high mass systems \citep[e.g.][]{Sijacki2007,Angles-Alcazar2016}, stellar feedback may greatly influence the evolution of massive galaxies.

Intergalactic transfer was implicitly included into the wind recycling mode in the analysis of \citet{Oppenheimer2010}, since they did not differentiate between galaxies accreting their own wind material versus accretion of winds from other galaxies. 
In a different set of large-volume simulations with the Illustris feedback model, \citet{Nelson2015} considered a ``stripped" growth mode for the smooth accretion of gas previously identified in a gravitationally bound substructure. 
While this is qualitatively similar to our definition of intergalactic transfer, \citet{Nelson2015} did not separate satellites' ISM material from gas in sub-halos. In their analysis, winds that remain within satellite halos are included in a ``clumpy" (merger) component.
Indeed, the relative importance of stripped accretion was similar between runs with and without feedback, suggesting that their stripped accretion mode was unrelated to wind transfer.

Tidal and ram pressure stripping can in principle remove gas from the ISM of satellites orbiting central galaxies, contributing also intergalactic transfer material.  Figure~\ref{fig:thumbs} explores in more detail the transfer of gas from dwarf satellites onto our Milky Way-mass galaxy {\bf m12i} at low redshift ($z<0.65$).  We show the trajectory of gas removed from the ISM of each satellite as it accretes onto the central galaxy in the top left panel.  In each case, the average transfer trajectory roughly corresponds to the path followed by the satellite on its way to merge with the central galaxy.  Projected trajectories relative to the size of sub-halos indicate that transfer gas is initially retained within the CGM of satellites.  

Thumbnails follow the evolution of the four most massive satellites at $z=0.65$ ($M_{*} \approx 2\times10^{7}$--$4\times10^{8}$\,\Msun), where we show the distribution of gas identified as intergalactic transfer recently removed from each satellite. In all cases, satellites experience multiple bursts of star formation as they orbit around the central galaxy, driving quasi-spherical outflows traced by the intergalactic transfer component.  Satellite winds interact with halo gas and eventually accrete onto the central galaxy.  This shows that galactic winds are indeed the primary mechanism driving intergalactic transfer also in the case of orbiting dwarf satellites, while direct stripping of ISM gas at more advanced merger stages is included in the merger-ISM component. Interestingly, intergalactic transfer from dwarf satellites provides a few \Msunyr~of gas to {\bf m12i} at $z \sim 0.5$ (Figure~\ref{fig:gmode}), similar to the inflow rates of metal-enriched gas onto galaxies of similar mass and redshift inferred by \citet{Rubin2012}.  As discussed in \S \ref{sec:flows} and illustrated in Figure \ref{fig:flows}, gas contributing to wind transfer can be first pushed all the way into the IGM before accreting onto another galaxy.
Intergalactic transfer may follow trajectories similar to wind recycling and fresh accretion. Further analysis is needed in order to separate these three smooth accretion modes by e.g. metallicity or kinematics.

\subsection{In situ versus ex situ star formation}

Galaxy mass assembly is commonly analyzed in simulations in terms of the contributions from in situ star formation and ex situ stellar growth \citep{Oser2010,Lackner2012,Tissera2012,Naab2014,Hirschmann2015,Pillepich2015,Qu2016,Rodriguez-Gomez2016}.  According to the definitions adopted here (\S\ref{sec:cwork}), in situ star formation clearly dominates the stellar mass growth of our simulated galaxies, while the ex situ component contributes a negligible amount to {\bf m10}, $\lesssim 5$\,\% to our {\bf m11} and {\bf m12} galaxies, and $\sim 20$\,\% to {\bf m13} at $z=0$. In all cases, stellar stripping represents $<1$\,\% of the stellar mass growth.  

Despite our limited sample of simulated galaxies, our results suggest a slow increase in the ex situ contribution from dwarfs to Milky Way-mass galaxies and a rapid increase for more massive systems, in good agreement with recent results from the Illustris \citep{Rodriguez-Gomez2016} and EAGLE \citep{Qu2016} simulations.  Abundance matching models also infer an 
increased importance of ex situ star formation in higher mass galaxies \citep{Behroozi2013,Moster2013}. 
The total contribution of externally processed material to $M_*$ is indeed lower in our isolated dwarf galaxies ($< 30$\,\% in {\bf m11} and $< 1$\,\% in {\bf m10}), while more massive systems typically living in higher density regions obtain an increasing amount of mass from material preprocessed by other galaxies. This is expected given that massive galaxies undergo more frequent mergers \citep[e.g.][]{Rodriguez-Gomez2015,Wellons2016}.  Nonetheless, our results emphasize that intergalactic transfer may significantly exceed the growth by ex situ stars from mergers, dominating the externally processed contribution for Milky Way-mass galaxies.

\subsection{Wind loading and mass loss}

The FIRE simulations predict that the efficiency of stellar feedback at driving large scale winds depends strongly on galaxy mass \citep{Muratov2015}.  Phenomenological wind models used in simulations that do not self-consistently predict wind properties typically use mass loading factors that follow momentum-conserving \citep[$\eta \propto V_{\rm c}^{-1}$; e.g.][]{Oppenheimer2010,Angles-Alcazar2014} or energy-conserving \citep[$\eta \propto V_{\rm c}^{-2}$; e.g.][]{Ford2013,Vogelsberger2014} scalings relative to the circular velocity $V_{\rm c}$.  \citet{Christensen2016} finds $\eta \propto V_{\rm c}^{-2}$ based on the analysis of zoom-in simulations including blastwave SN feedback, while the FIRE simulations predict a rough transition $\eta \propto V_{\rm c}^{-3}  \rightarrow \eta \propto V_{\rm c}^{-1}$ at $V_{\rm c} > 60$\,km\,s$^{-1}$ \citep{Muratov2015}.   
At $z=0$, the cumulative wind loading factor decreases from $\eta_{\rm c} \sim 100 \rightarrow 1$ across the halo mass range $M_{\rm halo} \sim 10^{10}$\,\Msun\,$\rightarrow 10^{13}$\,\Msun.  
This is consistent with the range of mass loading factors found in \citet{Muratov2015}.  However, note that here we consider the cumulative gas mass ejected from the ISM of the galaxy with velocity $v_{\rm out} > 2\,V_{\rm c}$, 
while \citet{Muratov2015} computed instantaneous mass fluxes ($v_{\rm out} > 0$) across a radial boundary at $0.25\,R_{\rm vir}$.  In principle, the mass flux at $0.25\,R_{\rm vir}$ could be (1) higher owing to mass loading of winds propagating from the ISM, or (2) lower owing to wind recycling at scales $<0.25\,R_{\rm vir}$.
Comparing mass loading factors and other wind properties across studies is not trivial given the different techniques applied in the analysis.

Recycling is quite common in our simulations, in qualitative agreement with previous work \citep{Oppenheimer2010,Ubler2014,Christensen2016,Nelson2015}. Our simulated galaxies recycle $f_{\rm REC} \approx 60$--90\,\% of the total ejected mass in winds down to $z = 0$ \citep[compared to $\sim 20$--70\,\% in][]{Christensen2016}, with no clear dependence on halo mass. 
The bulk of wind recycling occurs within the virial radius (see \S\ref{sec:reczone} below), indicating that a significant fraction ($f_{\rm REC}$) of the material ejected in winds over time never leaves the halo. 
Analyzing the distribution of baryons ejected and not present in the galaxy at $z=0$, we find that $\sim 75$\,\%  of the mass is in the IGM (i.e. outside $R_{\rm vir}$) while only $\sim 25$\,\% is located in the CGM (i.e. within $R_{\rm vir}$). Thus, while the CGM receives more mass in winds, efficient recycling within $R_{\rm vir}$ yields a higher fraction of wind material in the IGM at $z=0$. 
The net mass loss from the ISM relative to the stellar mass formed ($\eta_{\rm loss}$) depends strongly on halo mass, but the fraction of mass deposited in the IGM versus the CGM at $z=0$ is roughly independent of halo mass.

The halo baryon fraction of our simulated Milky Way-mass galaxies reaches $f_{\rm b} \approx 40$--90\,\% at $z=0$ \citep{Muratov2015}, where $f_{\rm b} \equiv (M_{\rm b}/M_{\rm halo})/f_{\rm b}^0$, $M_{\rm b}$ is the total halo baryonic mass, and $f_{\rm b}^0 \equiv \Omega_{\rm b}/\Omega_{\rm M}$ is the universal baryon fraction.
We find that the total mass ejected from the ISM and deposited in the IGM at $z=0$ implies a decrease in the halo baryon fraction $\Delta f_{\rm b} \approx 0.75\,\eta_{\rm loss}\,M_{*}/(f_{\rm b}^{0}\,M_{\rm halo}) \approx 5$--15\,\% for our Milky Way-mass galaxies, which is not enough to account for their baryonic content relative to cosmic mean ($\Delta f_{\rm b} \rightarrow 10$--60\,\%).
This suggests that (1) material removed from halos contains not only ISM ejecta but also swept-up material from the outer parts of the halo and/or (2) outflowing gas prevents gas accretion onto halos.
In the FIRE simulations, galaxies are able to retain a large fraction of the metals produced inside the halo \citep{Ma2016_Metallicity,Muratov2016}. We will explicitly track metal ejection and recycling in future work.

\subsection{Differential recycling}

We find that the contribution of galaxies' own wind recycling to gas accretion decreases with increasing halo mass (Figure~\ref{fig:windrec}).  This trend may seem counterintuitive given that increased hydrodynamic slowing of winds along with deeper gravitational potential wells can in principle make recycling more efficient in higher mass halos, as found in \citet{Oppenheimer2010}.  In the FIRE simulations, the mass dependence of the wind loading factor and the typical wind velocity are such that recycled-to-ejected ratios are roughly independent of halo mass \citep[Figure~\ref{fig:recfrac}; see also][]{Christensen2016}.  Lower mass galaxies lose more mass in winds relative to their stellar content but also re-accrete more gas relative to their mass.  As a result, gas accretion from wind recycling represents a larger fraction of the overall gas supply in lower mass systems (see \S\ref{sec:windrec}).  
This is in contrast to \citet{Oppenheimer2010}, where recycling in low mass galaxies is suppressed owing to longer recycling timescales.
However, note that the fraction of stellar mass attributed to wind mode accretion in \citet{Oppenheimer2010} corresponds to gas ejected from any galaxy, while we make an explicit distinction between each galaxy's own recycling and wind transfer from other galaxies.  Adding intergalactic transfer (which increases with halo mass) to wind recycling reduces the discrepancies between the two models; we find that $\sim 70$\,\% of the stellar content of Milky Way-mass galaxies at $z=0$ comes from gas that was in a wind in the past, roughly in agreement with \citet{Oppenheimer2010}.

 A characteristic feature of galactic winds in the FIRE simulations, not captured by current parameterized wind models in large volume simulations, is that outflows are significantly suppressed at low redshift ($z\lesssim 1$) in sufficiently massive galaxies \citep[see Figures~\ref{fig:gmode}--\ref{fig:windrec} and][]{Muratov2015}.    
The peak of wind recycling in higher mass systems is reached at earlier times when outflows are common, decreasing at lower redshifts.  Lower mass galaxies, however, extend their bursty star formation histories down to lower redshift, continuously driving gas outflows that increase the wind recycling component.  In fact, the number of recycling times $N_{\rm REC}$ increases systematically in dwarf galaxies, i.e. gas retained and/or converted into stars in lower mass systems has a higher probability of having cycled through the galaxy a larger number of times. This $N_{\rm REC}$--$M_{\rm halo}$ anticorrelation is not as apparent in \citet{Christensen2016}.

\subsection{Re-accretion time}

Wind recycling occurs over a broad range of timescales, from our resolution limit ($\gtrsim 10$\,Myr; the time interval between snapshots) up to a few Gyr.  Much recycling occurs on short timescales, with median recycling times $\,t_{\rm REC}\sim 100$--350\,Myr (for winds in the redshift range $z=0$--4) comparable to galaxy dynamical times (\S\ref{sec:rec_dist}). 
\citet{Oppenheimer2010} found significantly longer recycling times for parameterized momentum-driven winds in large volume simulations, with e.g. $t_{\rm REC} \sim 1$\,Gyr for Milky Way-mass galaxies at $z=1$.  Moreover, $t_{\rm REC}$ is anticorrelated with galaxy mass in their simulations ($t_{\rm REC} \propto M_{\rm halo}^{-1/2} $; i.e. winds recycle back to more massive galaxies faster), which was interpreted as increased hydrodynamic slowing of winds in denser environments around more massive galaxies \citep{Oppenheimer2008}. We find some indication for slightly longer $t_{\rm REC}$ in lower mass systems (Figure~\ref{fig:rec_hist}), but this may be due to the non-trivial connection between star formation histories and outflow properties in our simulations: while $t_{\rm REC}$ increases at low redshift in low mass galaxies, recycling times appear to decrease with the suppression of large scale winds in massive galaxies (e.g. $\,t_{\rm REC}\sim 500$\,Myr for our {\bf m12} galaxies at $z=1$, decreasing to $\,t_{\rm REC}\sim 100$\,Myr at lower redshift).  \citet{Christensen2016} find no strong dependence of $t_{\rm REC}$ with halo mass, in better agreement with our results, but their typical recycling times are longer ($t_{\rm REC} \sim 1$\,Gyr).

Recycling times can depend on star formation histories, outflow properties, treatment of hydrodynamic interactions, and numerical resolution.  Moreover, alternative definitions of recycled material and $t_{\rm REC}$ can introduce systematic differences between studies.  For example, \citet{Oppenheimer2010} (1) include wind transfer as recycling while we evaluate $t_{\rm REC}$ only for winds ejected from and recycled to the same galaxy, (2) consider lower limits for gas not recycled prior to $z=0$ while we evaluate $t_{\rm REC}$ specifically for winds that do recycle, and (3) define $t_{\rm REC}$ as the time from ejection to re-ejection (or gas conversion into stars) while we compute the time interval from ejection to re-accretion.  All of these can potentially increase $t_{\rm REC}$ relative to our results, in addition to intrinsic differences in galactic wind modeling.  The higher resolution of our simulations relative to \citet{Oppenheimer2010} and the shorter time interval between snapshots compared to the $\sim 100$\,Myr time resolution in \citet{Christensen2016} can also contribute to explaining our lower $t_{\rm REC}$ values.
 
The timescale for the reincorporation of wind ejecta is a crucial parameter in semi-analytic calculations \citep[e.g.][]{Henriques2013,White2015} and galaxy equilibrium models \citep[e.g.][]{Dave2012}.
Even in recent SAMs, low mass galaxies tend to form too early and are overabundant at high redshift.
A possible solution to this problem is to delay the reincorporation of gas ejected by strong stellar feedback, shifting the mass assembly of dwarf galaxies to lower redshifts. 
\citet{Henriques2013,Henriques2015} require a strong halo mass dependence $t_{\rm REC}\propto M_{\rm halo}^{-1}$ independent of redshift, with $t_{\rm REC} \sim 200$\,Myr in Milky Way-mass galaxies.
However, the equilibrium model of \citet{Mitra2015} favors a weaker scaling with halo mass, $t_{\rm REC} \propto (1+z)^{-0.3} \, M_{\rm halo}^{-0.45}$, with $t_{\rm REC} \sim 500$\,Myr for Milky Way-mass galaxies at $z=0$. 
Our lower recycling timescales may be reconciled with these models by the fact that we find $N_{\rm REC} >> 1$, particularly in low mass systems. In our simulations, stellar feedback suppresses early star formation despite the relatively short $t_{\rm REC}$, extending the bursty star formation histories of dwarf galaxies down to the present day \citep{Hopkins2014_FIRE}.
In SAMs and galaxy equilibrium models, $t_{\rm REC}$ may be partially degenerate with the effective mass loading factor. Our simulations suggest a highly dynamic gas reservoir around galaxies, continuously replenished and depleted by outflows and re-accretion, which way have important consequences for the properties of the CGM.

\subsection{Recycling zone and the CGM}{\label{sec:reczone}}

The median recycling distance relative to the virial radius $R_{\rm REC}/R_{\rm vir}$ decreases with redshift by a factor of $\sim 5$ from $z = 4 \rightarrow 0$.  Interestingly, the typical recycling distance does not evolve when normalized by the halo scale radius $R_{\rm s}$, defined as the radius where the logarithmic slope of the dark matter density profile is $-2$, or the galaxy stellar effective radius $R_{\rm eff}$: we find $R_{\rm REC} \sim R_{\rm s} \sim 5\,R_{\rm eff}$ roughly independent of halo mass and redshift, suggesting a characteristic {\it recycling zone} around galaxies that scales with the size of the inner halo and the galaxy's stellar component.

Simulations show that halo density profiles scale well with the virial radius during epochs of rapid accretion, where $R_{\rm vir}$ tracks the outer region enclosing recently accreted matter, but the inner density profile remains static in physical units when the accretion rate decreases at lower redshifts \citep{Wechsler2002,Cuesta2008,More2015}.
The evolution of the inner density profile and thus the inner gravitational potential well is better characterized by $R_{\rm s}$, which is not affected by halo pseudo-evolution \citep{Diemer2013,More2015}. 
Given that the median wind velocity in our simulations is proportional to halo circular velocity \citep{Muratov2015}, our finding $R_{\rm REC} \propto R_{\rm s}$ (independent of halo mass) suggests that the gravitational potential well in the inner region dominates the extent of the recycling zone around galaxies.  
\citet{Liang2016} showed that absorbers around galaxies of a wide range of stellar masses and redshifts \citep{Chen2010,Steidel2010,Tumlinson2011,Bordoloi2014,Liang2014,Werk2014,Johnson2015}  
appear to trace radial column density profiles that scale better with $R_{\rm s}$ than $R_{\rm vir}$.
CGM absorbers may thus be more closely connected to the inner wind recycling zone around galaxies \citep[see also][]{Ford2016}.


\section{Conclusions}\label{sec:con}

We have performed a detailed particle tracking analysis on a suite of FIRE cosmological zoom-in simulations spanning the halo mass range $M_{\rm halo} \sim 10^{10}$--$10^{13}$\,\Msun~at $z=0$ \citep{Hopkins2014_FIRE}.  
The FIRE simulations implement local stellar feedback processes that shape the multiphase structure of the ISM in galaxies while driving large scale outflows self-consistently in a full cosmological setting.  They thus represent an ideal tool to perform a thorough analysis of multiple aspects of the baryon cycle in galaxy evolution. In this work, we have focused on quantifying the origin of baryons that end up as stars in central galaxies across redshifts, evaluating the efficiency of galactic winds at removing gas from galaxies, and characterizing the main statistical properties of wind recycling. 
Our main conclusions can be summarized as follows:

\begin{enumerate}

\item Non-externally processed material dominates the evolution of isolated dwarf galaxies at all times.  This includes accretion of fresh gas directly from the IGM as well as wind recycling, i.e. re-accretion of galaxies' own wind ejecta.
The early growth of higher mass systems is also dominated by non-externally processed gas, while material preprocessed by other galaxies contributes increasingly at lower redshifts.  Externally processed material represents $\sim 50$\,\% of the $z=0$ stellar content of Milky Way-mass galaxies. 

\item Mergers contribute to externally processed material in the form of ex situ stars as well as ISM gas that fuels in situ star formation at later times. The merger-stellar and merger-ISM components provide about an equal amount of mass in Milky Way-mass galaxies which, combined, represents a small ($\lesssim 10$\,\%) contribution to galaxy growth. 
On average, the contribution of mergers to galaxy growth by $z=0$ increases with halo mass. 
Stars stripped from satellites represent $< 1$\,\% of the $z=0$ stellar content of Milky Way-mass galaxies.

\item Intergalactic gas transfer dominates the externally processed contribution to the $z=0$ stellar mass of all but our most massive galaxy. This includes primarily wind transfer, i.e. gas ejected in winds from other galaxies smoothly accreting onto the central galaxy. This previously under-appreciated growth mode represents a significant contribution to late time gas accretion in sufficiently massive galaxies, providing $\sim 20$--60\,\% of the gas inflow rate onto our Milky Way-mass galaxies at $z=0$. 
The contribution of intergalactic transfer to late-time inflows can exceed fresh accretion and the recycling of winds from the same galaxy.

\item Wind recycling dominates gas accretion for a large portion of every galaxy's evolution. 
Outflows are significantly suppressed at low redshift ($z\lesssim 1$) in sufficiently massive galaxies ($M_* \gtrsim 10^{10}$\,\Msun), owing to a transition into a quiescent mode of star formation, while isolated dwarfs extend their bursty star formation histories efficiently driving outflows down to $z=0$ \citep{Muratov2015,Hayward2016}.  Hence, wind recycling decreases at low redshift in more massive systems (we do not include weaker galactic fountains) but dominates gas accretion onto dwarf galaxies at $z=0$.
This characteristic feature of galactic winds in the FIRE simulations is not captured by current parameterized wind models in large volume simulations.

\item The total gas mass ejected in winds from the ISM per unit stellar mass formed depends significantly on halo mass and redshift, decreasing from $\eta_{\rm c} \sim 100 \rightarrow 1$ across the halo mass range $M_{\rm halo} \sim 10^{10}$\,\Msun\,$\rightarrow 10^{13}$\,\Msun~at $z=0$. Nonetheless, galaxies recycle $\gtrsim 60$\,\% of the total ejected mass down to $z=0$, with no clear trend with halo mass in the modest sample of zoom-in simulations analyzed. The total amount of gas deposited outside of galaxies (i.e. ejected and never recycled) may exceed many times the present day stellar mass of dwarf galaxies ($\eta_{\rm loss} \sim 5$--50), while Milky Way-mass galaxies may deposit as much gas in the CGM/IGM as their present day stellar mass. At $z=0$, $\sim 75$\,\% of the mass lost from the ISM is located in the IGM while only $\sim 25$\,\% remains in the CGM of galaxies.

\item The ISM gas content of galaxies can be ejected and recycled multiple times before forming stars or being lost to the CGM. The number of recycling times $N_{\rm REC}$ decreases with halo mass, i.e. recurrent wind recycling is more common and represents a larger mass contribution in lower mass galaxies. 
This $N_{\rm REC}$--$M_{\rm halo}$ anticorrelation can be explained by the bursty star formation histories of lower mass galaxies extending down to lower redshifts.

\item Gas recycling occurs over a broad range of timescales, from $\gtrsim 10$\,Myr to a few Gyr, with median recycling times $t_{\rm REC}\sim 100$--350\,Myr significantly shorter than previous work.
At high redshift, outflowing gas may reach the virial radius before recycling, while in the low redshift universe most recycling occurs within $\sim0.3\,R_{\rm vir}$. The typical recycling distance is roughly independent of halo mass and redshift when normalized by halo scale radius $R_{\rm s}$ or galaxy stellar effective radius $R_{\rm eff}$: $R_{\rm REC} \sim R_{\rm s} \sim 5\,R_{\rm eff}$.  
This suggests that CGM absorbers may trace
a characteristic {\it recycling zone} around galaxies that scales with the size of the inner halo and the galaxy's stellar component. 

\end{enumerate}

Overall, our results highlight the role of galactic winds as a primary contributor to the baryonic mass budget of central galaxies, where recycling of their own wind ejecta and the intergalactic transfer of mass from other galaxies via winds dominate gas accretion for a significant portion of every galaxy's evolution. 
The particle tracking analysis developed in this work can be used to address many additional questions using the FIRE simulations.
In future work, we plan to address the implications of the cosmological cycling of baryons for the chemical evolution of galaxies, the acquisition of angular momentum, and the structural properties of galaxies, which will help developing observational diagnostics of the cosmological baryon cycle.

\acknowledgments 

We thank C.~R. Christensen, R. {Dav{\'e}}, S. Genel, Z. Hafen, A.~V. Kravtsov, M. Kriek, A.~L. Muratov, B.~D. {Oppenheimer}, A.~J. Richings, J. Rojas-Sandoval, R.~S. Somerville, and S. Veilleux for useful discussions. 
DAA acknowledges support by a CIERA Postdoctoral Fellowship.
CAFG was supported by NSF through grants AST-1412836 and AST-1517491, by NASA through grant NNX15AB22G, and by STScI through grants HST-AR- 14293.001-A and HST-GO-14268.022-A.
DK was supported by NSF grant AST-1412153 and a Cottrell Scholar Award.
Support for PFH was provided by an Alfred P. Sloan Research Fellowship, NASA ATP Grant NNX14AH35G, and NSF Collaborative Research Grant \#1411920 and CAREER grant \#1455342. 
EQ was supported by NASA ATP grant 12-ATP12-0183, a Simons Investigator award from the Simons Foundation, and the David and Lucile Packard Foundation.
The simulations analyzed in this paper were run using the Extreme Science and Engineering Discovery Environment (XSEDE), which is supported by NSF grant ACI-1053575 (allocations TG-AST120025, TG-AST130039, and TG-AST140023).
This work greatly benefited from the hospitality of the Aspen Center for Physics, supported by NSF grant PHY-1066293.

\vspace{0.5cm}

\bibliography{BaryonCycle}

\begin{appendix}

\section{Low mass dwarf}\label{sec:appendix:m10}

\begin{figure}
\begin{center}
\includegraphics[scale=0.46]{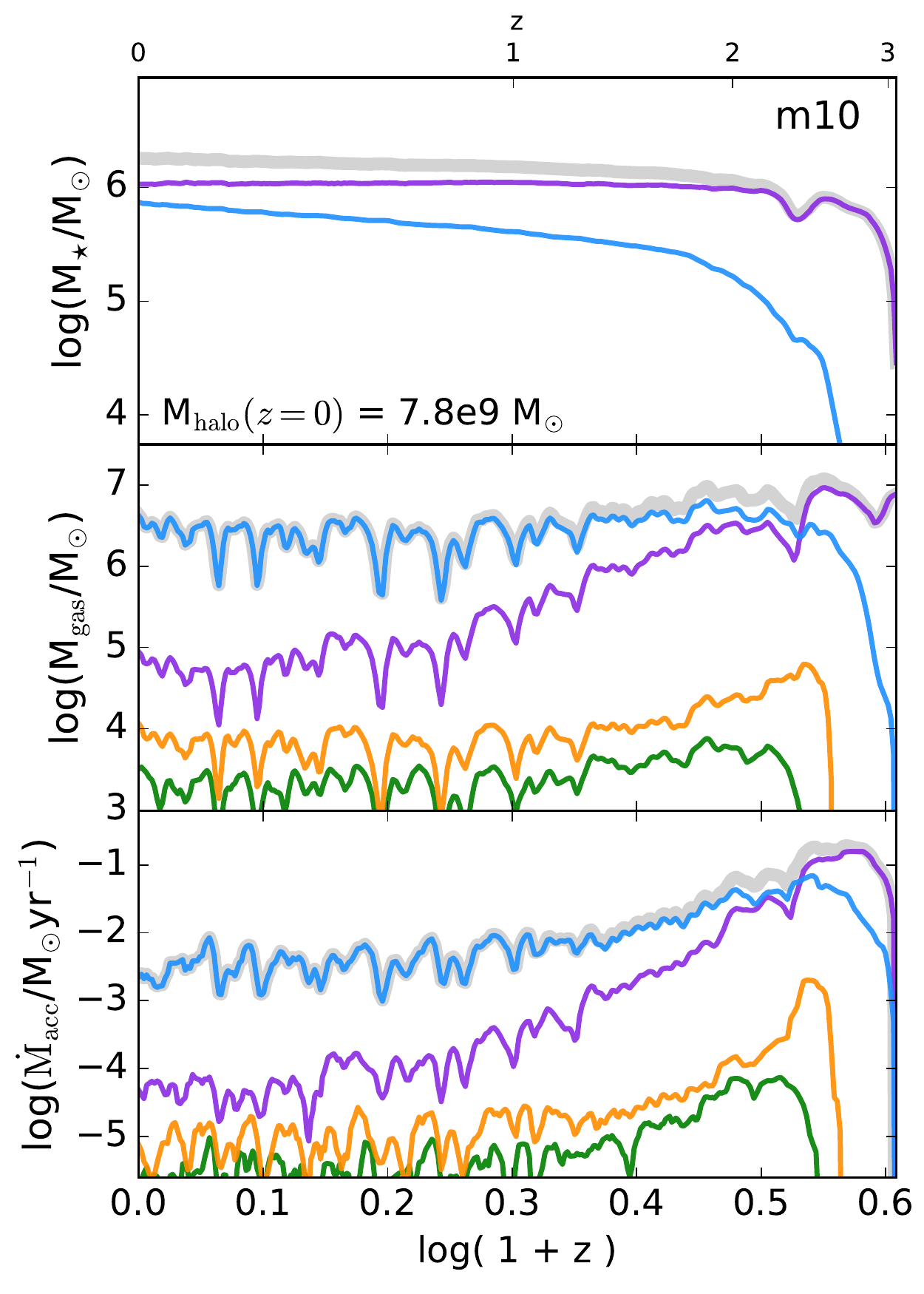}
\end{center}
\caption{Total stellar mass (top), ISM gas mass (middle), and gas accretion rate onto the ISM (bottom) as a function of redshift for our low mass dwarf {\bf m10} (thick gray lines).
Different colors indicate the contributions from fresh accretion (purple), NEP wind recycling (blue), intergalactic transfer of gas (green), and ISM gas from galaxy mergers (orange).  All quantities represent average values over a timescale of $\sim 200$\,Myr.}
\label{fig:m10}
\end{figure}

Figure~\ref{fig:m10} shows the evolution of our low mass dwarf {\bf m10} from early times down to $z=0$, where we indicate the contribution of fresh accretion, wind recycling, intergalactic transfer, and mergers to the total stellar mass, ISM gas mass, and gas accretion rate onto the galaxy.
Consistent with the trends found in the main text for more massive galaxies, the evolution of this isolated dwarf galaxy is dominated by in situ star formation from non-externally processed gas, with a very small contribution of gas preprocessed by other galaxies.
The stellar content of {\bf m10} grows rapidly at early times ($z \gtrsim 2.5$), fueled by the accretion of fresh gas at rates $> 0.1$\,\Msunyr.  The primary growth mode transitions to wind recycling at $z \lesssim 2$, dominating gas accretion as it declines to $10^{-2}$--$10^{-3}$\,\Msunyr~at late times.  
As our higher mass dwarf {\bf m11}, {\bf m10} exhibits a bursty star formation history down to $z=0$.  The frequent bursts drive winds that temporarily decrease the ISM gas content by factors $\sim 3$--5, until recycling replenishes a substantial fraction of the ejected gas (\S\ref{sec:wind_rec}).  
Nonetheless, the stellar content grows very little during each burst, maintaining an overall low level of star formation at late times. As a result, only $\sim 40$\,\% of the $z=0$ stellar mass of {\bf m10} comes from wind recycling, while most of its ISM gas content has been ejected in winds and recycled. Mergers and intergalactic transfer provide gas at rates $\lesssim 10^{-5}$\,\Msunyr~at $z=0$, while the contribution of ex-situ stars from mergers is negligible.

\section{Numerical robustness}\label{sec:appendix:num}

\begin{figure*}
\begin{center}
\includegraphics[scale=0.5]{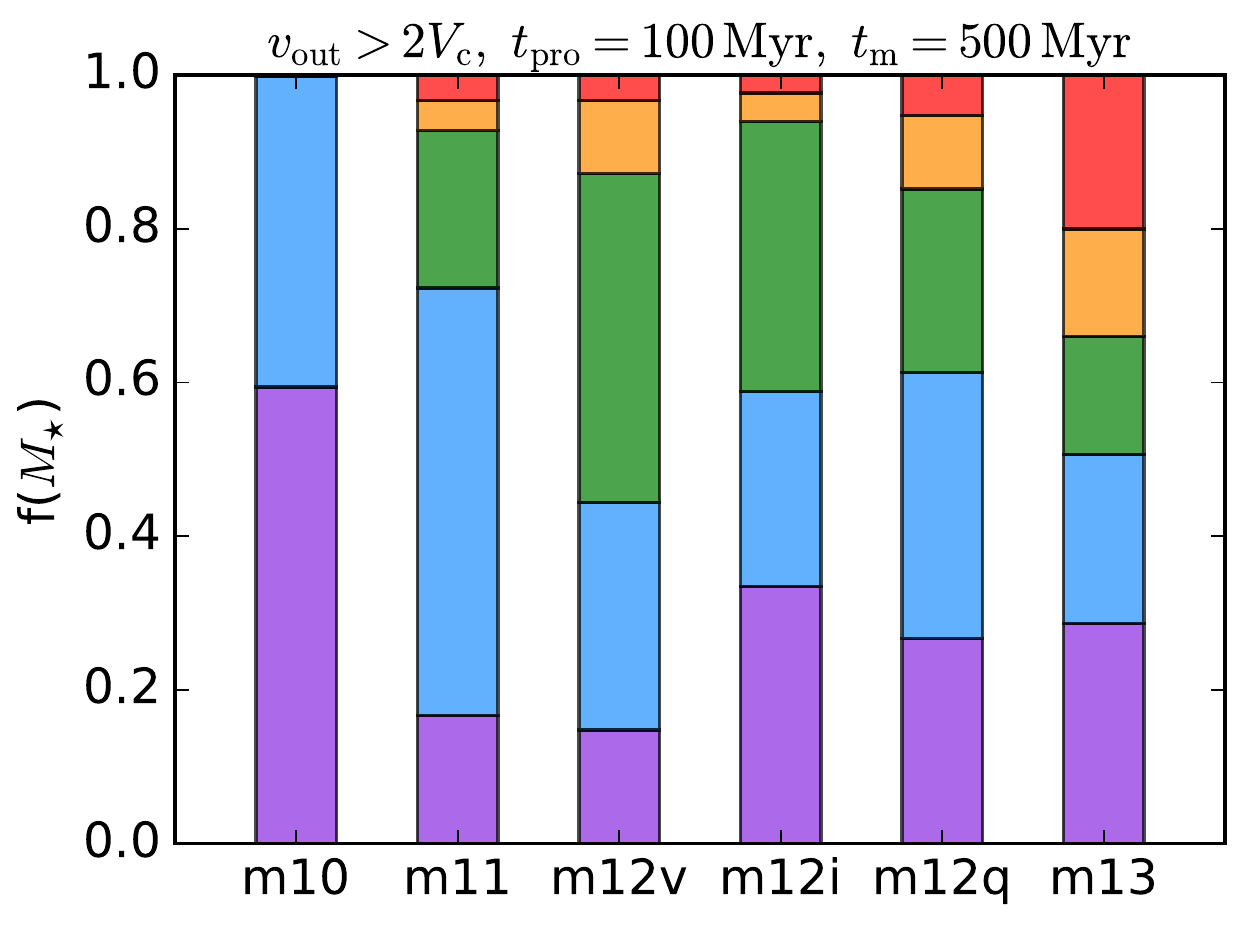}
\includegraphics[scale=0.5]{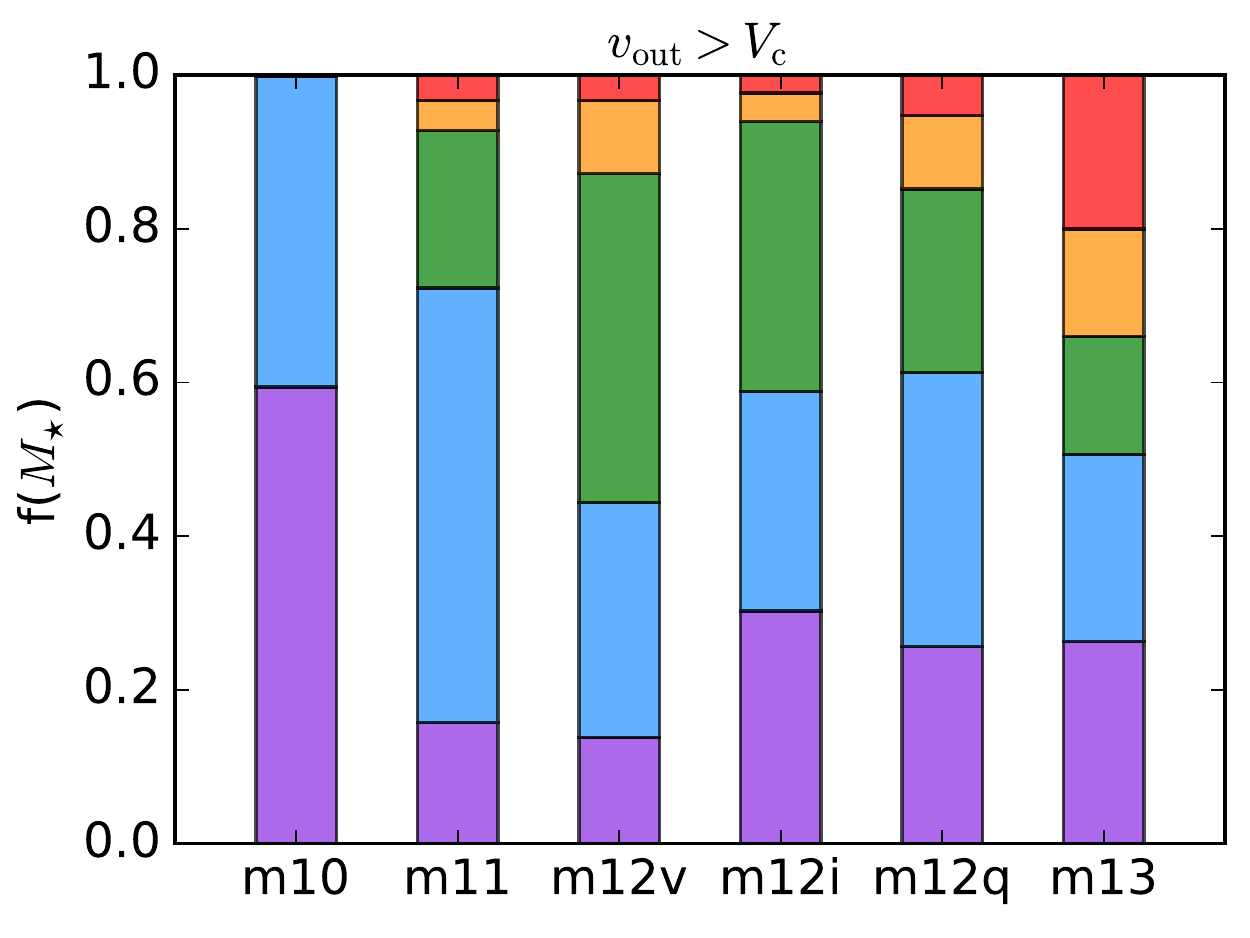}
\includegraphics[scale=0.5]{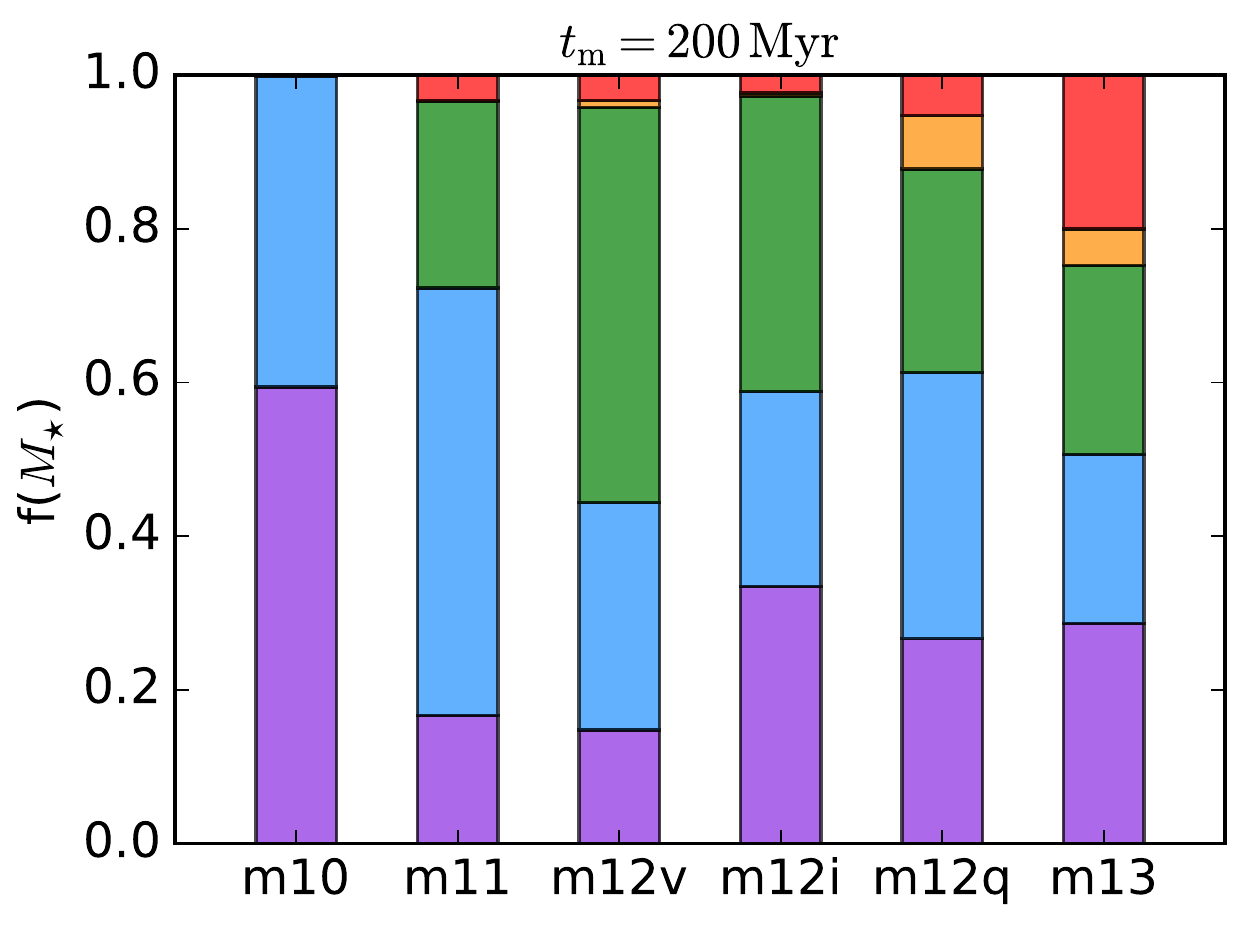}
\includegraphics[scale=0.5]{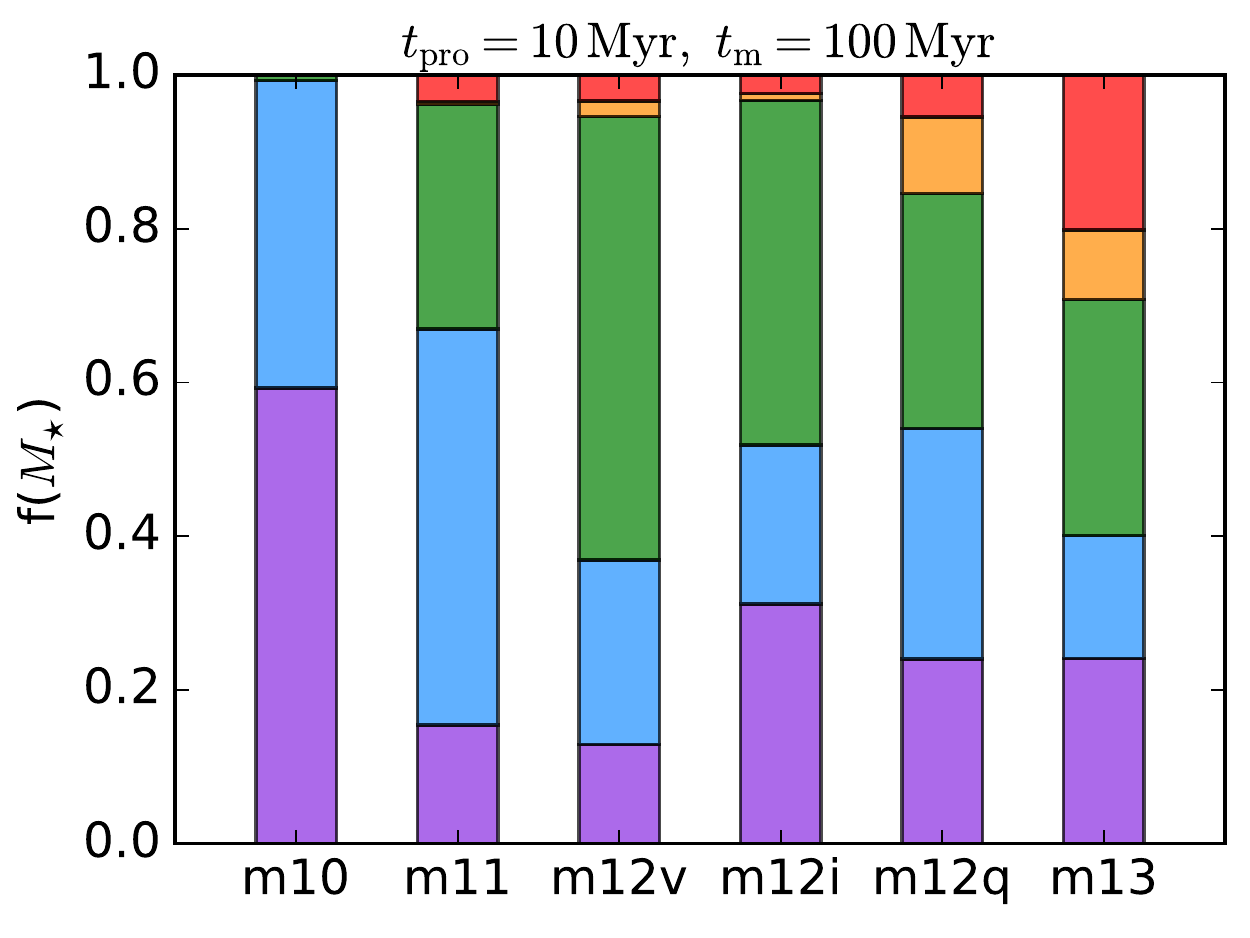}
\includegraphics[scale=0.3]{\pathI/legend.pdf}
\end{center}
\caption{
Fraction of stellar mass at $z=0$ contributed by fresh accretion (purple), wind recycling (blue), intergalactic transfer of gas (green), and galaxy mergers (ISM gas: orange; stars: red) for all simulated galaxies in order of increasing halo mass at $z=0$:  
{\bf m10} ($M_{\rm halo} \approx 7.8\times10^{9}$\,\Msun), 
{\bf m11} ($M_{\rm halo} \approx 1.4\times10^{11}$\,\Msun), 
{\bf m12i} ($M_{\rm halo} \approx 1.1\times10^{12}$\,\Msun), 
{\bf m12q} ($M_{\rm halo} \approx 1.2\times10^{12}$\,\Msun), 
{\bf m12v} ($M_{\rm halo} \approx 6.3\times10^{11}$\,\Msun), and  
{\bf m13} ($M_{\rm halo} \approx 6.1\times 10^{12}$\,\Msun).
We compare our fiducial analysis (upper left panel; see also Figure~\ref{fig:bars}) to the results obtained when we 
(1) decrease the minimum wind velocity to $v_{\rm out} > V_{\rm c}$ (upper right),
(2) decrease the time interval for merger identification to $t_{\rm m} = 200$\,Myr (lower left), and
(3) decrease the threshold preprocessing time and the time interval for merger identification to $t_{\rm pro} = 10$\,Myr and  $t_{\rm m} = 100$\,Myr, respectively (lower right).
While quantitative results depend slightly on the specific parameters adopted, our main conclusions remain unchanged.
}
\label{fig:param_bars}
\end{figure*}

\begin{figure*}
\begin{center}
\includegraphics[scale=0.65]{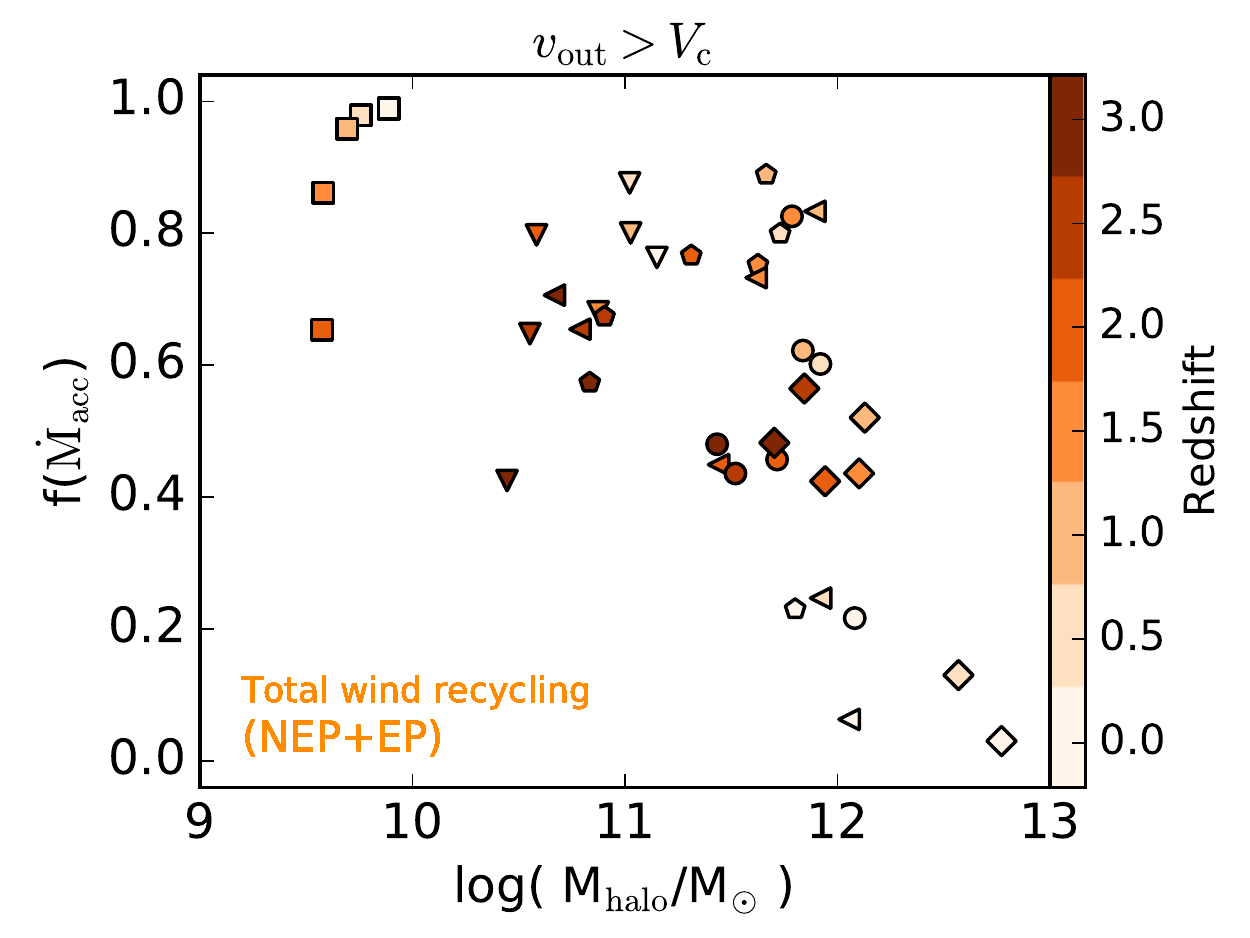}
\includegraphics[scale=0.65]{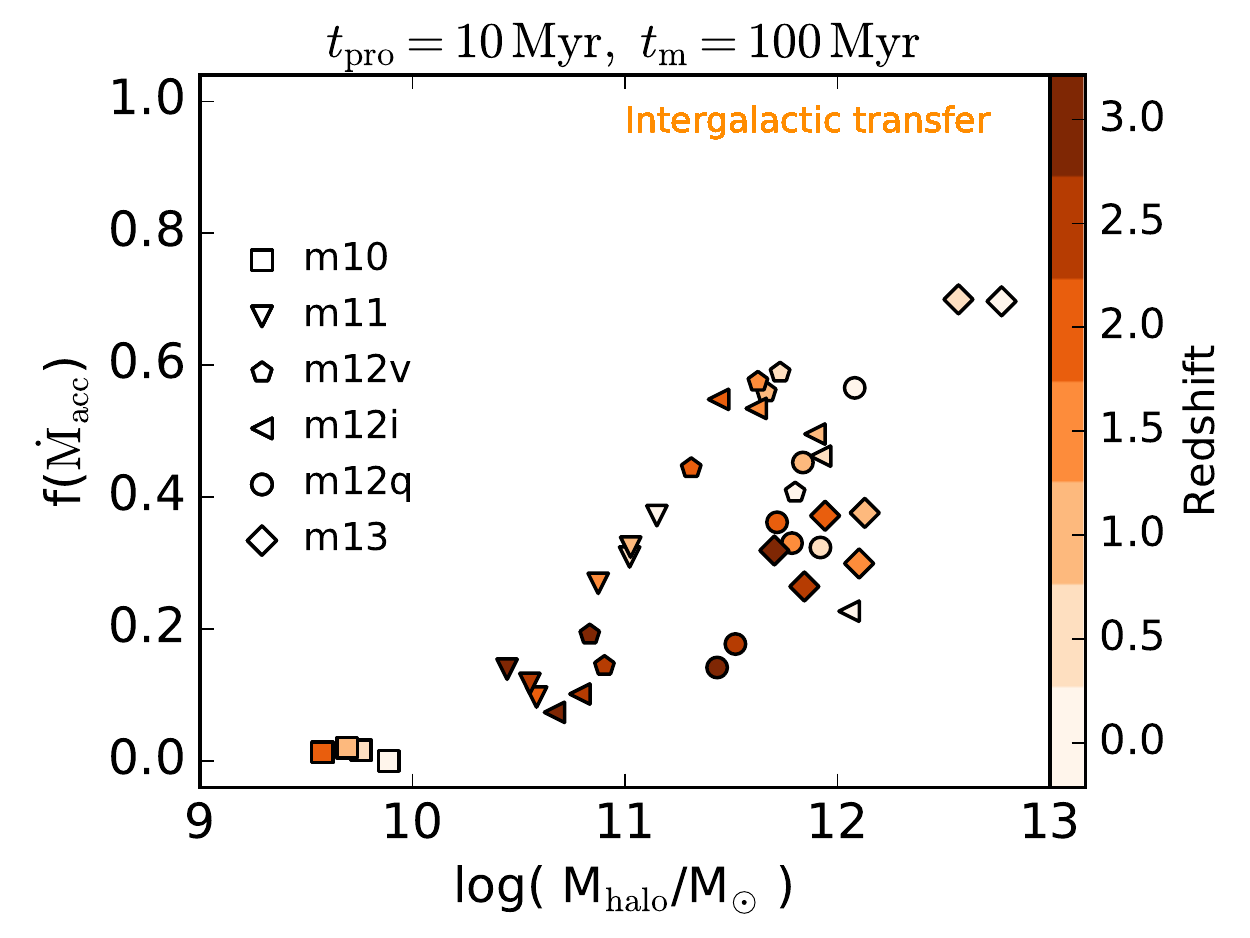}
\end{center}
\caption{
Left: fraction of gas accretion rate onto the galaxy contributed by the total (NEP+EP) wind recycling component as a function of halo mass and redshift, as in Figure~\ref{fig:windrec}, where we now reduce the minimum wind velocity by a factor of two (i.e. $v_{\rm out} > V_{\rm c}$). 
Right: fraction of gas accretion rate onto the galaxy contributed by the intergalactic transfer component as a function of halo mass and redshift, as in Figure~\ref{fig:windtrans}, where we now reduce the threshold preprocessing time and the time interval for merger identification to $t_{\rm pro} = 10$\,Myr and  $t_{\rm m} = 100$\,Myr, respectively.
}
\label{fig:param_Acc}
\end{figure*}

Our particle tracking analysis relies on three parameters that control the operational definitions of galaxy growth components used throughout the paper (\S\ref{sec:DissectGrowth}): 
(1) the threshold preprocessing time $t_{\rm pro} = 100$\,Myr, i.e. the minimum residence time in another galaxy prior to first accretion onto the central galaxy required for externally processed material, 
(2) the timescale for merger identification $t_{\rm m} = 500$\,Myr, i.e. the time interval prior to first accretion before which externally processed material is required to be removed from the source galaxy to classify as intergalactic transfer as opposed to mergers, and
(3) the minimum radial velocity $v_{\rm out} > 2\,V_{\rm c}$ imposed on outflowing gas to classify as galactic wind.
In this section, we evaluate how changes in these parameters affect our results.

Figure~\ref{fig:param_bars} shows the fraction of stellar mass at $z=0$ contributed by each process (as in Figure~\ref{fig:bars}) for different classification parameters.  
The top right panel corresponds to decreasing the threshold velocity for galactic winds by a factor of two (using fiducial values for $t_{\rm pro}$ and $t_{\rm m}$).  Imposing $v_{\rm out} > V_{\rm c}$ increases the number of identified wind ejection events and thus the amount of wind recycling relative to our fiducial analysis.  This yields an average increase of $\sim 5$\,\% in the fraction of $M_{*}(z=0)$ contributed by NEP recycling relative to fiducial values. The fresh accretion component decreases correspondingly, while other contributions remain unchanged.
The bottom left panel shows the effect of lowering $t_{\rm m} = 500 \rightarrow 200$\,Myr, which increases the amount of gas transfer relative to the merger-ISM contribution while leaving non-externally processed components unchanged.  In this case, the fraction of $M_{*}(z=0)$ contributed by intergalactic transfer increases by $\sim 10$--60\,\% relative to fiducial values (except in {\bf m10} where gas transfer is negligible).
The bottom right panel shows the effect changing $t_{\rm pro} = 100 \rightarrow 10$\,Myr and $t_{\rm m} = 500 \rightarrow 100$\,Myr simultaneously.  The short preprocessing time $t_{\rm pro} = 10$\,Myr (equivalent to a single snapshot) increases the overall externally processed contribution while decreasing the fresh accretion and NEP recycling components (the total wind recycling depends only on $v_{\rm out}$ and it is thus unaffected by $t_{\rm pro}$).  Moreover, $t_{\rm m} = 100$\,Myr favors intergalactic transfer over mergers relative to our more conservative fiducial value, since transfer material may leave the source galaxy only 100\,Myr prior to accreting onto the central galaxy.  In this case, intergalactic transfer contributes $\sim 30$--60\,\% of $M_{*}(z=0)$ for all galaxies but {\bf m10}, becoming the dominant process in our Milky Way-mass galaxies.

Figure~\ref{fig:param_Acc} shows the fraction of gas accretion rate onto the galaxy contributed by wind recycling as a function of halo mass and redshift when we decrease the threshold wind velocity to $v_{\rm out} > V_{\rm c}$ (left) and the fraction of gas accretion rate contributed by intergalactic transfer when we use $t_{\rm pro} = 10$\,Myr and $t_{\rm m} = 100$\,Myr (right).
As expected from Figure~\ref{fig:param_bars}, each parameter modification increases gas accretion from either wind recycling or intergalactic transfer relative to our fiducial analysis using more conservative parameters, emphasizing further the importance of these processes, while the redshift and halo mass trends identified for each component remain unaffected (see Figures~\ref{fig:windrec} and~\ref{fig:windtrans} for comparison).  
Overall, while quantitative results depend slightly on the specific parameters adopted, our main conclusions remain unchanged.

\end{appendix}

\end{document}